\documentclass[10pt, draftclsnofoot, onecolumn]{IEEEtran}

\usepackage{cite}
\usepackage{amsmath,amssymb,amsfonts}
\usepackage{algorithm}
\usepackage{algpseudocode}%
\usepackage{graphicx}
\usepackage{hyperref}
\usepackage{textcomp}
\usepackage{xcolor}
\usepackage{subcaption}
\usepackage{multirow}
\usepackage{siunitx}
\usepackage{lipsum}
\usepackage{mathtools}
\usepackage{cuted}
\usepackage{stfloats}
\usepackage{bbm} 
\usepackage{bm} 

\usepackage{enumitem}

\usepackage{amsthm}

\newtheorem{lemma}{Lemma}
\newtheorem{remark}{Remark}
\newtheorem{proposition}{Proposition}
\newtheorem{corollary}{Corollary}
\newtheorem{conjecture}{Conjecture}

\newenvironment{proofN3}{%
  \proof}{\endproof}

\newcommand{\rev}[1]{{{\textcolor{black}{#1}}}}

\usepackage{tabularx}
\makeatletter
\newcommand{\multiline}[1]{%
  \begin{tabularx}{\dimexpr\linewidth-\ALG@thistlm}[t]{@{}X@{}}
    #1
  \end{tabularx}
}
\makeatother

\usepackage[font=small]{caption}

\allowdisplaybreaks

\usepackage{cite}
\usepackage{amsmath,amssymb,amsfonts}
\usepackage{algorithm}

\usepackage{algpseudocode}%
\usepackage{graphicx}
\usepackage{hyperref}
\usepackage{textcomp}
\usepackage{xcolor}
\usepackage{subcaption}
\usepackage{multirow}
\usepackage{siunitx}

\usepackage{lipsum}
\usepackage{mathtools}
\usepackage{cuted}
\usepackage{stfloats}
\usepackage{bbm} 
\usepackage{bm} 

\usepackage{enumitem}

\usepackage{amsthm}

\def\bb0{{\mathbb{0}}}

\def\bb{{\mathbf{b}}}

\def\bg{{\mathbf{g}}}

\def\bm{{\mathbf{m}}}

\def\bx{{\mathbf{x}}}
\def\by{{\mathbf{y}}}

\def\b0{{\mathbf{0}}}

\def\bA{{\mathbf{A}}}
\def\bB{{\mathbf{B}}}

\def\bF{{\mathbf{F}}}

\def\bI{{\mathbf{I}}}

\def\bO{{\mathbf{O}}}

\def\sf0{{\mathsf{0}}}

\begin{document}


\title{ 
Coding for Gaussian Two-Way Channels: Linear and Learning-Based Approaches
}



\author{Junghoon~Kim,
        Taejoon~Kim,
        Anindya Bijoy Das,
        Seyyedali Hosseinalipour,
        David J. Love, 
        and
        Christopher G. Brinton
\thanks{This work was supported in part by ONR under grant N000142112472 and by NSF under grants ITE2226447, CNS2225577, CNS2212565, EEC1941529, and CNS2146171.
A part of this work was presented at 58th Annual Allerton Conference on Communication, Control, and Computing (Allerton) in 2022 \cite{kim2022linear}.     %
J. Kim is with Motorola Mobility, Chicago, IL 60654, USA
(e-mail: junghoon@motorola.com). This work was done while he was at Purdue University. 
T. Kim is 
with the School of Electrical, Computer and Energy Engineering, Arizona State University, Tempe, AZ 25281, USA (email: taejoonkim@asu.edu).
A. B. Das, D. J. Love, and C. G. Brinton are with the Department
of Electrical and Computer Engineering, Purdue University, West Lafayette, IN 47907, USA
(e-mail: \{das207, djlove, cgb\}@purdue.edu).
S. Hosseinalipour is with the Department of Electrical Engineering, University at Buffalo-SUNY, NY 14260, USA (email: alipour@buffalo.edu).}
}
\maketitle

\begin{abstract}

Although user cooperation cannot improve  the capacity of  Gaussian two-way channels (GTWCs) with independent noises, it can improve communication reliability.
In this work, 
we aim to enhance and balance the communication reliability in GTWCs by minimizing the sum of error probabilities via joint design of encoders and decoders at the users.
We first formulate general encoding/decoding functions, where the user cooperation is captured by the coupling of user encoding processes. 
The coupling effect renders the encoder/decoder design non-trivial, requiring effective decoding to capture this effect, as well as efficient power management at the encoders within power constraints.
%
To address these challenges,
we propose two different \textit{two-way coding} strategies: linear coding and learning-based  coding.
For linear coding, 
we propose optimal linear decoding and discuss  new
insights on encoding regarding user cooperation to balance reliability. 
We then propose an efficient algorithm 
for joint encoder/decoder design.
%
For learning-based coding, we introduce a novel recurrent neural network (RNN)-based coding architecture, where we propose \textit{interactive} RNNs and a power control layer for encoding, and  we incorporate bi-directional RNNs with an attention mechanism  for decoding.
%
Through simulations, we show that our two-way coding methodologies outperform conventional channel coding schemes (that do not utilize user cooperation) significantly in sum-error performance.  
We also demonstrate that our linear coding 
excels at high signal-to-noise ratios (SNRs), while 
our RNN-based coding performs best at low SNRs.
We further investigate our two-way coding strategies in terms of power distribution, two-way coding benefit, different coding rates, and block-length gain.






\end{abstract}


\section{Introduction}










Most of the modern communication systems, including cellular networks, WiFi networks, satellite communications, and social media platforms, facilitate two-way interaction, enabling users to exchange messages in both directions~\cite{zhang2016full,kim2015survey}.
This interactive capability promotes a seamless exchange of information and feedback, supports real-time communication, and fosters effective collaboration among users.
The input-output model that allows for the
bidirectional exchange of information is referred to as \textit{two-way channels}, which was studied by
Shannon in~\cite{shannon1961two}.
A practically relevant two-way channel model is the Gaussian two-way channel (GTWC),
where Gaussian-distributed  noise is added independently to each direction  of the two-way channel between the users~\cite{han1984general}.
%
Earlier studies on GTWCs have been mostly focused on analyzing the channel capacity region of GTWCs.
An important result was obtained by Han in \cite{han1984general}, which revealed that  incorporating the previously received symbols (i.e., feedback information) into generating transmit symbols at the users does not increase 
the capacity of GTWCs.
In other words, the channel capacity for GTWCs is achieved when the two-way channel is considered as two independent one-way channels, i.e., when  two users do not cooperate with each other.




In addition to channel capacity, 
communication reliability or error probability is another important metric 
in information/communication theory. 
Currently, some researchers are focusing on examining GTWCs in terms of communication reliability.
In \cite{palacio2021achievable}, Palacio-Baus and Devroye defined error exponents for GTWCs and showed
how cooperations between the two users, or using previously received symbols into creating transmit symbols, can improve the error exponents in comparison to the non-cooperative case.
In \cite{vasaldynamic}, Vasal suggested a dynamic programming (DP)-based methodology for encoding to improve the communication reliability for GTWCs. 
Despite ongoing efforts to improve the communication reliability for GTWCs, 
the existing works still lack in providing a finite block-length coding scheme and its performance evaluation.
%

To the best of our knowledge, there is no framework for designing practical codes in GTWCs, which is the main motivation behind this study.
A foundational coding strategy for GTWCs is to carry out a \textit{linear} processing  for encoding and decoding in order to simplify the system model of GTWCs and mitigate the coding complexity. 
It is important to note that GTWCs can be thought of as an expanded system model of feedback-enabled 
Gaussian one-way channels (GOWCs), where a linear  coding framework for GOWCs with feedback has been well developed.
The seminal work done by Schalkwijk and Kailath in \cite{schalkwijk1966coding} introduced a simple linear encoding for GOWCs that can achieve doubly exponential decay in the probability of error upon having noiseless feedback information. In \cite{butman1969general}, Butman introduced a general framework for linear coding, in which the noise may be colored, nonstationary, and correlated in GOWCs.
In \cite{chance2011concatenated}, Chance and Love proposed a linear encoding scheme for GOWCs with noisy feedback, which is further analyzed and revealed to be an optimal structure under some conditions for linear encoding by~\cite{agrawal2011iteratively}.
There have been attempts to view the linear code design in feedback-enabled GOWCs as 
feedback stabilization in control theory~\cite{elia2004bode} and
optimal DP~\cite{mishra2023linear}.
While the aforementioned studies focus on GOWCs with passive  feedback, where the receiver feeds back its received symbol without alteration, Ben-Yishai and Shayevitz proposed a linear coding scheme for active feedback, where the receiver is allowed to employ coding for feedback~\cite{ben2017interactive}.
Overall, despite the availability of well-developed frameworks of linear coding for feedback-enabled GOWCs, such a linear framework has not been developed for GTWCs, 
which is one of the main motivations of this work. 
Linear processing offers the significant advantage of low complexity for encoding/decoding with a simplified system model.
However, a significant limitation of linear coding is its inherent constraint on producing optimal codes because of the linearity.

To tackle the issues of linear processing,
we further develop \textit{non-linear} coding for GTWCs
to provide higher degrees of flexibility in designing the codes.
It is important to note that there have been research efforts on
designing non-linear codes in feedback-enabled GOWCs. Along these lines, Kim. \textit{et al.}~\cite{kim2020deepcode} proposed Deepcode, which exploits recurrent neural networks (RNNs) for non-linear coding in feedback-enabled GOWCs, and showed performance improvements in the error probability across many noise scenarios as compared to linear coding.
In \cite{safavi2021deep}, Safavi. \textit{et al.} proposed deep extended feedback (DEF) codes that generalizes Deepcode to improve the spectral efficiency and error performance. 
In \cite{ozfatura2022all}, Ozfatura. \textit{et al.} proposed generalized block attention feedback (GBAF) codes that exploit self-attention modules to  incorporate different neural network architectures.
They showed that GBAF codes can outperform
the existing solutions, especially in the noiseless feedback scenario.
Kim. \textit{et al.}~\cite{kim2023feedback} proposed a novel RNN-based feedback coding methodology through state propagation-based encoding, which incorporates efficient power control, an attention mechanism, and block-level classification of the messages.
The proposed methodology greatly enhances the error performance in practical noise regions as compared to the existing coding schemes.
While there have been many successful attempts to develop non-linear codes via deep learning for feedback-enabled GOWCs, to the best of our knowledge, 
no such efforts have been made yet for GTWCs.


Although GTWCs can be understood as an expanded system model of feedback-enabled GOWCs, it is not trivial to generalize the linear and learning-based coding methods of feedback-enabled GOWCs to GTWCs, since the objectives and system models are different from each other. 
While the objective of feedback-enabled GOWCs is to improve communication performance for the one-way transmission from the transmitter to the receiver~\cite{schalkwijk1966coding, chance2011concatenated,kim2020deepcode, kim2023feedback}, the objective of GTWCs is to improve communication performance for the two-way transmissions from User 1 to 2 and from User 2 to  1. 
The performance of these two-way transmissions can be balanced through the cooperation of the users.
We note that the cooperation incurs a coupling effect in their encoding processes, intertwining the encoding and decoding operations of the users.
To capture this two-way cooperation and the encoder's coupling, 
a novel system model is essential for GTWCs.
In contrast, in feedback-enabled GOWCs, the receiver only acts as a helper while the transmitter is the recipient of help, which does not capture the two-way cooperation and the encoder's coupling.
 
We specify the three challenges in the design of coding methods for GTWCs as below.

(C-1) \textit{Coupling of user encoding processes}. 
It is important to understand how to use receive symbols in generating transmit symbols (i.e., for encoding) at both users in order to improve the user's reliability.
However, designing these encoding methods is challenging, since using the received symbols in generating transmit symbols at both users across multiple channel uses in GTWCs will incur a coupling effect in their encoding processes.
This coupling intertwines the encoding and decoding operations of the users, ultimately affecting the overall system behavior.


(C-2) \textit{Requirement of effective decoding to capture the coupling effect.}
It is crucial for the decoding process at each user to effectively capture the coupling effect introduced by the encoders in GTWCs. This requirement necessitates a joint design of encoders and decoders for both users, adding complexity and posing a significant challenge in the design of decoders, compared to feedback-enabled GOWCs.
%

(C-3) \textit{Need for efficient power management within power constraints.}  
When designing encoding schemes that utilize feedback information, power control over the sequence of transmit symbols becomes crucial for achieving robust error performance.
In GTWCs, power management becomes more complicated, as
compared to feedback-enalbed GOWCs,
due to the coupled encoding processes requiring proper power allocation into the message and feedback transmissions at both users.
Therefore, an effective power control strategy should be investigated and incorporated
into the encoder design for both users under the average power constraints in GTWCs.



In this paper, we aim to bridge the gaps between the two pieces of literature on GOWCs and GTWCs by addressing the above challenges. 
In light of the aforementioned challenges,
we propose two distinct \textit{two-way coding} strategies for GTWCs: 
(i) linear coding, which is based on tractable mathematical derivations, and (ii) non-linear coding, which is inspired by recent advancements in deep learning.

The major contributions of the paper are summarized below. 
\begin{itemize}
    \item We first introduce a general functional form of encoding and decoding for two-way channels, where each user 
    generates the transmit symbols by encoding both its own message and the past received symbols from the other user.
    Using the defined encoding/decoding functions, we then formulate an optimization problem for minimizing the sum of the error probabilities of the users under each user's power constraint,
    aiming to improve and balance the communication reliability in two-way channels.
    \item We propose a \textit{linear} coding scheme for GTWCs. We first formulate a signal model for linear coding that captures the coupling of encoders to address (C-1). 
    We then derive a maximum likelihood decoding, i.e., an optimal form of linear decoding, at each user as a function of encoding schemes of the users, which addresses (C-2).
    To obtain a tractable solution, we convert the sum-error minimization problem to the weighted sum-power minimization problem.
    In weighted sum-power minimization, we provide new insights on user cooperation by exploring the relationship between the channel noise ratio and the weight coefficient imposed in weighted sum-power minimization. 
    Building upon the insights presented in \cite{palacio2021achievable} that a user with lower channel noise can act as a helper by providing feedback, our work extends this concept to address weighted sum problems. 
    In particular, 
    whether a user would act as a helper  is affected not only by the channel ratio but also by the weight coefficient.
    %
    We then develop an efficient algorithm for joint encoder-decoder design under power constraints for (C-3). 
    Lastly, we discuss how to extend our linear coding approach to medium/long block-lengths.
    %
    %
    %
    \item We propose a novel \textit{learning-based} coding architecture to capture the three aforementioned challenges of (C-1), (C-2), and (C-3).
    Specifically, to address (C-1), we introduce state propagation-based encoding to understand the encoding behavior in two-way transmissions and
    propose employing a pair of \textit{interactive} RNNs for encoding, where the previous output of one user's RNN is fed into the other user's RNN as a current input.
    This innovative use of RNNs in an interactive manner effectively captures user cooperation across many channel uses  in GTWCs, while in feedback-enabled GOWCs, only one-way encoding RNN has been designed at the transmitter~\cite{kim2020deepcode,kim2023feedback}, which does not  capture the two-way transmissions and the encoder's coupling.
    To address (C-3), we introduce
    a power control layer at each user's encoder, which we prove satisfies all power constraints asymptotically. 
    To address (C-2), we adopt bi-directional RNNs with an attention mechanism to  fully exploit correlation among receive symbols,
    in which the encoders' coupling behavior is implicitly captured as a form of symbols.
    To address all these three challenges jointly in the coding architecture, we train the overall encoders/decoders at the users via auto-encoder. 
    \item We analyze the computational complexity of the proposed linear and non-linear coding schemes.
    Through numerical simulations, we show that our two-way coding strategies outperform conventional channel coding schemes (that do not utilize user cooperation) by wide margins in terms of sum-error performance. 
    For linear coding, we introduce an enhancement strategy by transmitting messages in alternate channel uses, which is motivated by the solution behavior.
    Under asymmetric channels, linear coding performs best when the channel signal-to-noise ratios (SNRs) of both users are high, while RNN-based coding excels when either of the channel SNRs is low. 
    For RNN-based coding, we observe that when the difference between the channel SNRs of the users is larger than some threshold, the user with higher channel SNR sacrifices some of its error performance to improve the other user’s error performance, which in turn improves the sum-error performance. 
    This behavior is consistent with the understanding presented in \cite{palacio2021achievable} that a user with lower channel noise can function as a helper.
    %
    We further provide information-theoretic insights on power distribution at the users, where both linear and RNN-based coding schemes allocate more power to early channel uses in asymmetric channels. This behavior is aligned with power distribution for feedback-enabled GOWCs~\cite{chance2011concatenated,kim2023feedback}.
    Lastly, we demonstrate that our coding schemes support higher coding rates and  study the block-length gain of RNN-based coding.
    \item 
    We highlight the new contributions made in this work, in contrast to our previous study~\cite{kim2022linear}.
    Overall,  we introduce both linear and non-linear coding approaches, while our prior work focused only on linear coding.
    For non-linear coding, we propose a new neural coding methodology based on deep learning.
    Also, we shift our focus towards enhancing reliability by minimizing sum-errors within power limitations as opposed to
      weighted sum-power minimization in our prior study, and
    then 
    establish the connection between the optimization of the two minimization problems.
    Building upon the tractable solution framework for  weighted sum-power minimization in our previous work, we propose a new  linear coding solution for sum-error minimization.
    Furthermore, we add discussions on the extension to medium/long block-lengths, the effect of different modulation orders, and user cooperation under asymmetric channels.
    Also, we analyze the computational complexity of our linear and non-linear coding schemes.
    Finally, we conduct extensive 
    performance evaluations, including sum-error performance, two-way coding benefits, varying coding rates, and block-length gain, as well as  power distribution mainly discussed in our prior work.
\end{itemize}

\section{System Model and Optimization Problem for Coding}

In this section, we first present a transmission model 
for GTWCs in  Sec.~\ref{ssec:trans}.
Then, we provide a general functional form of encoding and decoding at the users in Sec.~\ref{ssec:enc/dec}. Lastly, we formulate an optimization problem for sum-error minimization in Sec.~\ref{ssec:opt}.

\begin{figure}[t]
    \centering
    \includegraphics[width=.8\linewidth]{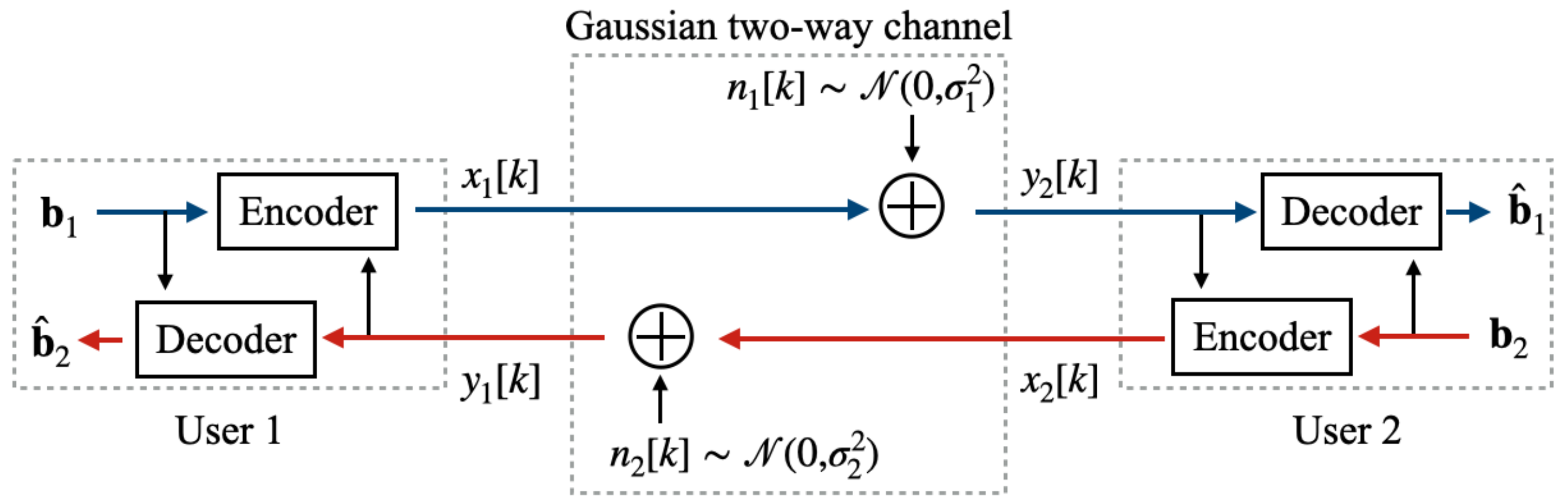}
    \caption{System model for Gaussian two-way channels. The goal is to successfully convey the messages ${\bf b}_i$ to each other by exchanging transmit symbols $x_i[k]$ between the users. 
    The users employ encoders and decoders to ensure successful message transmissions, where the two-way interaction between the users should be captured.}
    \label{fig:system}
\end{figure}

\subsection{Transmission Model}
\label{ssec:trans}

We consider a two-way channel between two users, User 1 and User 2, with additive white Gaussian noise (AWGN), as shown in Fig. \ref{fig:system}. This channel model is referred to as a GTWC. 
 We assume that  transmission occurs over $N$ channel uses (or timesteps).
Let $k\in\{1,\cdots, N\}$ denote the channel use index, and $x_1[k] \in \mathbb{R}$ and $x_2[k] \in \mathbb{R}$ represent the transmit symbols of User 1 and User 2, respectively, at time $k$.
Subsequently,
the receive symbols at User 2 and User 1, respectively,  at time $k$ are given by
\begin{align}
    {y}_2[k] = x_1[k] + n_1[k] \in \mathbb{R}, \quad
    {y}_1[k] = x_2[k] + {n}_2[k] \in \mathbb{R},
    \label{eq:y}
\end{align}
where ${n}_1[k] \sim \mathcal{N}(0,\sigma_1^2)$ and ${n}_2[k] \sim \mathcal{N}(0,\sigma_2^2)$ are \rev{Gaussian random variables representing the additive noises,} which are independent of each other and across the channel use index $k$. 
We consider an average power constraint for transmission at the users as
\begin{equation}
    \mathbb{E}\bigg[ \sum_{k=1}^N x^2_i[k]  \bigg] \le N P, \quad i \in \{1,2\},
    \label{eq:power:first}
\end{equation}
where $P$ denotes the average transmit power constraint per channel use at each user. The distribution of  variables over which the expectation in \eqref{eq:power:first} is taken will be specified in \eqref{opt:const:power}.

\subsection{Functional Form of Encoding and Decoding}
\label{ssec:enc/dec}

The goal of exchanging transmit symbols  among the users is to transmit a \textit{message} available at each user to the other. As shown in Fig. \ref{fig:system}, we consider that each User $i$, $i\in\{1,2\}$, aims to transmit a unique message 
 represented by a bit vector, ${\bf b}_i \in \{0,1\}^{K_i}$, to the other user, where $K_i$ is the number of source bits at User $i$. \rev{Note that ${\bf b}_i$ denotes a random variable.} \rev{In this work, we assume that all bits in ${\bf b}_1$ and ${\bf b}_2$
are independent and identically distributed (IID).}
We first provide a general functional form of encoding and decoding at the users.

\subsubsection{Encoding}
User $i$ encodes the bit vector ${\bf b}_i \in \{0,1\}^{K_i}$ to generate the $N$ transmit symbols, $\{{x}_i[k]\}_{k=1}^N$.
The rate at User $i$ is defined by $r_i=K_i/N$.
Rather than encoding the bit vector
${\bf b}_i$ solely to generate the transmit symbols $\{{x}_i[k]\}_{k=1}^N$,
we consider a joint encoding scheme that leverages both the bit vector and the receive symbols in GTWCs. 
The utilization of receive (or feedback) symbols in generating transmit symbols has been demonstrated to effectively enhance communication reliability in feedback-enabled GOWCs~\cite{schalkwijk1966coding, chance2011concatenated, kim2020deepcode, kim2023feedback}. This observation 
inspires us to utilize feedback in the framework of GTWC encoding.  
%
In particular, at time $k$,
the encoding at User $i$ is described as a function of the bit vector ${\bf b}_i$ and the $k-1$ receive symbols $\{y_i[j]\}_{j=1}^{k-1}$ in \eqref{eq:y}. 
We denote an encoding function of User $i$ at time $k$  as $f_{i,k}: \mathbb{R}^{K_i+k-1} \to \mathbb{R}$. 
We can subsequently represent the encoding of User $i$ at time $k$  as  
\begin{align}
    x_i[k] = f_{i,k}({\bf b}_i, y_i[1], ..., y_i[k-1]), \quad k=1,...,N.
    \label{eq:encoding}
\end{align}
The encoding functions $\{f_{i,k}\}_{i,k}$ can be linear or non-linear depending on design choices. In this work, we address both cases, linear coding in Sec.~\ref{sec:linear} and non-linear coding in Sec.~\ref{sec:RNN}.

\subsubsection{Decoding} 
Once the $N$ transmissions are completed, Users $1$ and $2$ compute  estimates of the bit vector of the other user, respectively, $\hat {\bf b}_2 \in \{0,1\}^{K_2}$ and $\hat {\bf b}_1 \in \{0,1\}^{K_1}$.
For decoding, User $i$ can utilize the $N$ receive symbols, $\{{y}_i[k]\}_{k=1}^N$ in \eqref{eq:y}, and its own bit vector ${\bf b}_i$.
Specifically, its own bit vector ${\bf b}_i$ can be used as side information for decoding at User $i$ since the receive symbols $\{{y}_i[k]\}_{k=1}^N$ contain not only the other user's bit vector information but also its own bit vector information due to the causal encoding processes in \eqref{eq:encoding} and the subsequent symbol exchange in \eqref{eq:y}.
We denote the decoding functions of User 1 and 2 as $g_1:\mathbb{R}^{N + K_1} \to \{0,1\}^{K_2}$ and $g_2:\mathbb{R}^{N + K_2} \to \{0,1\}^{K_1}$, respectively. We can then represent the decoding processes conducted by User 1 and User 2, respectively, as
\begin{align}
    \hat {\bf b}_2 = g_1({\bf b}_1, y_1[1], ..., y_1[N]), \quad
    \hat {\bf b}_1 = g_2({\bf b}_2, y_2[1], ..., y_2[N]).
    \label{eq:dec_func_g}
\end{align}
The decoding functions $\{g_i\}_i$ can be either linear or non-linear. We address linear coding in Sec.~\ref{sec:linear} and non-linear coding in Sec.~\ref{sec:RNN}.

\subsection{Optimization of Encoders and Decoders}
\label{ssec:opt}

Since our goal is to design practical codes with finite block-lengths, we consider 
error probability as our main performance indicator. 
We consider two different metrics of interest. 
The first metric is the block error rate (BLER), which represents the probability that the original bit vector from one user is incorrectly recovered by the other user.
Formally, the BLER of User $i$'s bit vector is defined as $\text{BLER}_i = \text{Pr}[\{{\bf b}_i \neq \hat {\bf b}_i\} ]$, $i\in\{1,2\}$.
The other metric is the bit error rate (BER), which represents the probability that each bit in the original bit vector from one user is incorrectly recovered by the other user.
We first represent the $\ell$-th entry of ${\bf b}_i$ and $\hat {\bf b}_i$ as ${b}_i[\ell] \in \{0,1\}$ and $\hat {b}_i[\ell] \in \{0,1\}$, respectively. Then, the BER of ${b}_i[\ell]$
is defined by
$\text{BER}_{i,\ell} = \text{Pr}[ \{{b}_i[\ell] \neq \hat {b}_i[\ell]\} ]$, $i\in\{1,2\}$ and $\ell \in\{1,...,K_i\}$.
It is worth noting that BER values can differ for each individual bit entry,
i.e., $\text{BER}_{i,\ell} \neq \text{BER}_{i,m}$ for $\ell \neq m$, depending on the design of encoding/decoding functions in \eqref{eq:encoding}-\eqref{eq:dec_func_g}.
To achieve the goal of minimizing BERs across all entries, our objective is to minimize the average BER over all entries.
Formally, we define the average BER of User $i$'s bit vector as $\text{BER}_i = (1/K_i) \sum_{\ell=1}^{K_i} \text{BER}_{i,\ell} = (1/K_i) \sum_{\ell=1}^{K_i} \text{Pr}[ \{{b}_i[\ell] \neq \hat {b}_i[\ell]\} ]$.





For generalization, we assume User $i$ uses an error probability metric denoted by $\mathcal{E}_i ( \{f_{1,k}\}_{k=1}^N,$ $\{f_{2,k}\}_{k=1}^N, g_{\bar i}, $ $ P, \sigma_1^2, \sigma_2^2, K_1, K_2, N)$, which can represent either $\text{BLER}_i$ or $\text{BER}_i$ depending on design choices.
Here, $\bar i$ denotes the index of the counterpart of User $i$, i.e., $\bar i=2$ if $i=1$ while $\bar i=1$ if $i=2$.
%
In this work, we aim to design the encoding and decoding functions when the values of $P, \sigma_1^2,$ $\sigma_2^2, K_1, K_2,$ and $N$ are given. We thus simplify the dependency as $\mathcal{E}_i ( \{f_{1,k}\}_{k=1}^N, \{f_{2,k}\}_{k=1}^N, g_{\bar i})$.
The objective in this work is to minimize the error probability of both users, while balancing the communication reliability in two-way channels.
Accordingly, we formulate the following optimization problem:
\begin{align}
&  \underset{ \{f_{1,k}\}_{k=1}^N, \; \{f_{2,k}\}_{k=1}^N, \; g_1, \; g_2} {\text{minimize}} & &
\mathcal{E}_1 (\{f_{1,k}\}_{k=1}^N, \{f_{2,k}\}_{k=1}^N,  g_{2})
+ \mathcal{E}_2 (\{f_{1,k}\}_{k=1}^N, \{f_{2,k}\}_{k=1}^N,  g_{1})
 \label{opt:obj}
\\
& \hspace{1cm} \text{subject to}
& &  
 \mathbb{E}_{ {\bf b}_1, {\bf b}_2, {\bf n}_1, {\bf n}_2 } \bigg[ \sum_{k=1}^N  x^2_i[k]   \bigg] \le N P, \quad i \in \{1,2\},
 \label{opt:const:power}
\end{align}
where ${\bf n}_i = [n_i[1], ..., n_i[N]]^\top$, $i \in \{1,2\}$.
The expectation in \eqref{opt:const:power} is  taken over the distributions of \rev{the random variables, i.e.,} the bit vectors, $ {\bf b}_1$ and $ {\bf b}_2$, and the noise terms, ${\bf n}_1$ and ${\bf n}_2$; the transmit symbol $x_i[k]$ in \eqref{opt:const:power} is the encoding output as a function of its local bit vector $ {\bf b}_i$ and the receive symbols, $y_i[1], ..., y_i[k-1]$, shown in \eqref{eq:encoding}. 
The receive symbols, $y_i[1], ..., y_i[k-1]$, are dependent on the noises ${\bf n}_1$ and ${\bf n}_2$ and the other user's encoding process, i.e., the bit vector of the other user $ {\bf b}_{\bar i}$, through the successive encoding operation of the both users as shown in \eqref{eq:y} and \eqref{eq:encoding}.

Solving \eqref{opt:obj}-\eqref{opt:const:power} is non-trivial  and challenging since the encoding processes of both the users (in \eqref{eq:encoding}) are coupled to each other in a causal manner, and the coupling effect should be incorporated into the design of the encoders/decoders of both users. 
To address these challenges, we propose two distinct \textit{two-way coding} approaches. First, we propose a \textit{linear} coding approach with low complexity for designing the encoders/decoders  in Sec.~\ref{sec:linear}.
Despite the low design complexity, the linear assumption on coding may limit the performance due to limited degrees of flexibility in the design of encoders/decoders. To allow higher degrees of freedom for coding, we propose a \textit{non-linear} coding approach based on deep learning in Sec.~\ref{sec:RNN}.

\section{Linear Coding: Framework and Solution}
\label{sec:linear}

In this section, we first formulate a signal model for linear coding in Sec.~\ref{ssec:linear:signal}. We then derive the optimal linear decoding scheme in Sec.~\ref{ssec:linear:decoding}. To obtain a tractable solution, we convert the original sum-error minimization problem to the weighted sum-power minimization problem  in Sec.~\ref{ssec:linear:conversion}.
We next analyze encoders' behaviors and further reduce the design complexity of encoders
in Sec.~\ref{ssec:linear:opt:weighted-sum}. 
Afterwards, we propose a solution method for the weighted sum-power minimization in Sec.~\ref{ssec:linear:opt:weighted-sum:sol},
based on which
we provide an overall algorithm for the sum-error minimization in Sec.~\ref{ssec:linear:opt:sum-error:sol}. 
Lastly, we discuss how to 
extend our linear coding approach to long block-lengths in Sec.~\ref{ssec:linear:long}.






\subsection{Signal Model for Linear Coding}
\label{ssec:linear:signal}

To conduct linear coding for GTWCs in Fig. \ref{fig:system}, we consider a message symbol $m_i \in \mathbb{R}$, where
User $i$, $i \in \{1,2\}$, converts the bit vector ${\bf b}_i  \in \{0,1\}^{K_i}$ to the message symbol $m_i$ through $2^{K_i}$-ary pulse amplitude modulation (PAM)~
\cite{proakis2008digital}. 
\rev{Note that  $m_i$ is a random variable.}
We assume that each message $m_i$ is chosen from a set of possible messages $\mathcal{M}_i$ according to a uniform distribution, given that all bits in ${\bf b}_1$ and ${\bf b}_2$
are assumed to be IID to one another. We note that the cardinality of $\mathcal{M}_i$ is $2^{K_i}$.
We consider $\mathbb{E}_{\rev{m_i}}[m_i]=0$ and $\mathbb{E}_{\rev{m_i}}[m^2_i]=1$.
We focus on the scenario, where Users 1 and 2 exchange a single pair of messages, i.e., User 1 conveys $m_1$ to User 2, and User 2 conveys $m_2$ to User 1.
We represent the $N$ receive symbols at User $i$ (in \eqref{eq:y}) in vector form as ${\bf y}_i = [y_i[1], ..., y_i[N]]^\top \in \mathbb{R}^{N \times 1}$.
Then, we can rewrite the expressions in \eqref{eq:y} as
\begin{align}
    {\bf y}_2 = {\bf x}_1 + {\bf n}_1, \quad
    {\bf y}_1 = {\bf x}_2 + {\bf n}_2,
    \label{eq:y_vec}
\end{align}
where ${\bf x}_i = [x_i[1], ..., x_i[N]]^\top$, $i\in\{1,2\}$. Note that we refer to the vector form of the symbols as signals, i.e., ${\bf x}_i$ as the transmit signal and ${\bf y}_i$ as the receive signal, $i \in \{1,2\}$.

We consider that User $i$, $i \in \{1,2\}$, employs the message encoding vector  $\tilde {\bf g}_i \in \mathbb{R}^{N \times 1}$ for encoding the message $m_i$ and the feedback encoding matrix $\tilde {\bf F}_i \in \mathbb{R}^{N \times N}$ for encoding the receive symbols.
Note that $\tilde {\bf F}_i$, $i \in \{1,2\}$, is strictly lower triangular (i.e., the matrix entries are zero on and above the diagonal) due to the causality of the system.
To avoid feeding back redundant information under  limited transmit power,
 we consider that each user removes the contribution of its known prior transmit symbols  from the receive symbols to generate its future transmit symbols, similar to linear coding for feedback-enabled GOWCs~\cite{butman1969general, schalkwijk1966coding, chance2011concatenated}.
For the case of User 1, the transmit signal ${\bf x}_1$ is transmitted across the channel to User 2, who encodes a noisy version of ${\bf x}_1$, i.e., ${\bf y}_2$, by using $\tilde {\bf F}_2$ to generate ${\bf x}_2$. User 2 then transmits ${\bf x}_2$ to User 1, who receives a noisy version of ${\bf x}_2$, i.e., ${\bf y}_1$. 
In short, ${\bf y}_1$ contains some information of ${\bf x}_1$ to which the feedback encoding matrix $\tilde {\bf F}_2$ is applied.
Therefore,
 User 1 subtracts its signal portion ${\bf x}_1$ from the receive signal ${\bf y}_1$ as ${\bf y}_1 - \tilde {\bf F}_2 {\bf x}_1$.
Similarly, User 2 subtracts its signal portion ${\bf x}_2$ from ${\bf y}_2$ and obtains the modified feedback information ${\bf y}_2 - \tilde {\bf F}_1 {\bf x}_2$. 
The transmit signals of the users are then given by
\begin{align}
    {\bf x}_1 & = \tilde {\bf g}_1 m_1 + \tilde {\bf F}_1 ({\bf y}_1 - \tilde {\bf F}_2 {\bf x}_1),
    \label{eq:x1:tilde}
    \\
    {\bf x}_2 & = \tilde {\bf g}_2 m_2 + \tilde {\bf F}_2 ({\bf y}_2 - \tilde {\bf F}_1 {\bf x}_2).
    \label{eq:x2:tilde}
\end{align}
%

\begin{remark}
    \label{remark:general_signal_model}
    There are various ways to represent the transmit signals, these representations can be converted into one another through variable changes due to linearity. Our signal model in \eqref{eq:x1:tilde}-\eqref{eq:x2:tilde} is one such representation.
    Other possible linear signal models and their conversions are discussed in detail in Appendix~\ref{app:linear:general_signal}.  
    Among various choices of general linear signal models, we chose \eqref{eq:x1:tilde}-\eqref{eq:x2:tilde} because this form does not result in any inverse matrix  appearing in the power expressions, as shown in \eqref{eq:power1}-\eqref{eq:power2}. 
    However, other linear signal models would lead to the introduction of an inverse matrix form in the power expressions, which are difficult to handle in optimization. The derivation of the power expressions when using other linear signal models is discussed in Appendix~\ref{app:linear:general_signal}.
\end{remark}

Since each user transmits the signals encapsulating the received signals  from the other user over multiple channel uses in a causal manner, a \textit{coupling} occurs between the transmit signals of the users. 
Since this coupling occurs successively across time, it intertwines the encoding/decoding operations and impacts the overall system behavior.
To mitigate the coupling effects in the signal representation,
we rewrite the signal model in
\eqref{eq:x1:tilde}-\eqref{eq:x2:tilde} as
\begin{align}
    {\bf x}_1 & = {\bf g}_1 m_1 +  {\bf F}_1 ({\bf y}_1 - {\bf F}_2 {\bf x}_1),
    \label{eq:x1}
    \\
    {\bf x}_2 & = {\bf g}_2 m_2 + {\bf F}_2 {\bf y}_2.
    \label{eq:x2}
\end{align}
Here, we introduce another set of parameters, $({\bf g}_1, {\bf F}_1, {\bf g}_2, {\bf F}_2)$, as functions of $(\tilde {\bf g}_1, \tilde {\bf F}_1, \tilde {\bf g}_2, \tilde {\bf F}_2)$ to represent \eqref{eq:x1}-\eqref{eq:x2}. The introduction of the new set of parameters makes the subsequent problem formulation and optimization easier to deal with.
The detailed derivation of the functional expressions for $({\bf g}_1, {\bf F}_1, {\bf g}_2, {\bf F}_2)$ from $(\tilde {\bf g}_1, \tilde {\bf F}_1, \tilde {\bf g}_2, \tilde {\bf F}_2)$ is presented in Appendix~\ref{app:system:equivalence}, \rev{whereas the derivation for the reverse case is provided in Appendix~\ref{app:system:equivalence2}.}
Given the functional expressions, both ${\bf F}_1$ and ${\bf F}_2$ are  strictly lower triangular.

Based on the equivalent conversion from \eqref{eq:x1:tilde}-\eqref{eq:x2:tilde} to \eqref{eq:x1}-\eqref{eq:x2},
we aim to design ${\bf g}_1$, ${\bf F}_1$, ${\bf g}_2$, and ${\bf F}_2$ and focus on the signal representation in \eqref{eq:x1}-\eqref{eq:x2}.
By putting the expressions of ${\bf x}_2$ and ${\bf x}_1$ (obtained in \eqref{eq:x1}-\eqref{eq:x2}) into \eqref{eq:y_vec}, we can 
rewrite \eqref{eq:y_vec} as
\begin{align}
    {\bf y}_1 & = {\bf x}_2 + {\bf n}_2
    \nonumber
    \\
    & =  {\bf g}_2 m_2 + {\bf F}_2 {\bf y}_2  + {\bf n}_2,
    \label{eq:y1:vec}
    \\
    {\bf y}_2 & = {\bf x}_1 + {\bf n}_1
    \nonumber
    \\ 
    & = {\bf g}_1 m_1 +  {\bf F}_1 ({\bf y}_1 - {\bf F}_2 {\bf x}_1) + {\bf n}_1
    \nonumber
    \\
    & = {\bf g}_1 m_1  + {\bf F}_1 {\bf g}_2 m_2 +
    ({\bf I} + {\bf F}_1{\bf F}_2){\bf n}_1 + {\bf F}_1{\bf n}_2.
    \label{eq:y2:vec}
\end{align}
To obtain the equality in the second line in \eqref{eq:y2:vec}, we put the expression of ${\bf y}_1$ (given in \eqref{eq:y1:vec}) into ${\bf y}_1$ in the fist line in \eqref{eq:y2:vec}.

We insert the expressions of ${\bf y}_1$ and ${\bf y}_2$ (obtained in \eqref{eq:y1:vec}-\eqref{eq:y2:vec}) back into \eqref{eq:x1}-\eqref{eq:x2}, and rewrite the transmit signals ${\bf x}_1$ and ${\bf x}_2$ in  \eqref{eq:x1}-\eqref{eq:x2} as
\begin{align}
    {\bf x}_1 
    & = {\bf g}_1 m_1  + {\bf F}_1 ({\bf g}_2 m_2 +
    {\bf F}_2{\bf n}_1 +{\bf n}_2),
    \label{eq:x1:mes+noise}
    \\
    {\bf x}_2
    & = {\bf g}_2 m_2 + {\bf F}_2 ({\bf g}_1 m_1  + {\bf F}_1 {\bf g}_2 m_2 +
    ({\bf I} + {\bf F}_1{\bf F}_2){\bf n}_1 + {\bf F}_1{\bf n}_2)
    \nonumber
    \\
    & = ({\bf I} + {\bf F}_2 {\bf F}_1 ) {\bf g}_2 m_2 +  {\bf F}_2{\bf g}_1 m_1  
    + {\bf F}_2({\bf I} + {\bf F}_1{\bf F}_2){\bf n}_1 + {\bf F}_2{\bf F}_1{\bf n}_2.
    \label{eq:x2:mes+noise}
\end{align}
Note that the transmit signals in \eqref{eq:x1:mes+noise}-\eqref{eq:x2:mes+noise} are represented as the sum of  messages and noises.
Using \eqref{eq:x1:mes+noise}-\eqref{eq:x2:mes+noise}, we formulate the transmit power of the users as
\begin{align}
    \mathbb{E}_{\rev{m_1,m_2,{\bf n}_1, {\bf n}_2}} \big[\|{\bf x}_1\|^2\big] 
    &= \|{\bf g}_1\|^2 +  \|{\bf F}_1 {\bf g}_2 \|^2 +  \|{\bf F}_1 {\bf F}_2 \|_F^2  \sigma_1^2
    + \| {\bf F}_1 \|_F^2  \sigma_2^2,
    \label{eq:power1}
    \\
    \mathbb{E}_{\rev{m_1,m_2,{\bf n}_1, {\bf n}_2}} \big[\|{\bf x}_2\|^2\big] 
    &= \| ({\bf I} + {\bf F}_2 {\bf F}_1 ) {\bf g}_2\|^2 + \|{\bf F}_2{\bf g}_1\|^2
    + \| {\bf F}_2( {\bf I} + {\bf F}_1{\bf F}_2 ) \|_F^2 \sigma_1^2
    + \| {\bf F}_2{\bf F}_1 \|_F^2 \sigma_2^2, 
    \label{eq:power2}
\end{align}
where the messages and the noises are assumed to be uncorrelated to each other. 



\subsection{Maximum Likelihood Decoding}
\label{ssec:linear:decoding}


Although encoding at the users is interrelated, decoding is an independent process at each user given the encoding schemes. Therefore, given the users' encoding schemes, i.e., ${\bf g}_1$, ${\bf F}_1$, ${\bf g}_2$, and ${\bf F}_2$, the optimal decoding strategy to minimize the sum-error is for each user to conduct its own optimal decoding. 
The optimal decoding for each user is to select the message that maximizes the a posteriori probability among all possible messages. Since we assume that all messages are equi-probable, the optimal decoding is the maximum likelihood decoding~\cite{butman1969general}.
In this subsection, we derive the maximum likelihood decoding for each user to estimate the message of the other user.

We first consider the decoding process at User 1. 
User 1 estimates $m_2$ given the receive signal ${\bf y}_1$ and its own message $m_1$ based on maximum likelihood, i.e., $\hat m_2 = \underset{m_2 \in \mathcal{M}_2} {\arg\max} \; \text{Pr}({\bf y}_1 | m_1, m_2)$. 
By using the expression of ${\bf x}_2$ in \eqref{eq:x2:mes+noise}, ${\bf y}_1$ is expressed as
\begin{align}
    {\bf y}_1 = {\bf x}_2 + {\bf n}_2 = {\bf F}_2 {\bf g}_1 m_1 + ({\bf I} + {\bf F}_2 {\bf F}_1){\bf g}_2 m_2 +  ({\bf I} + {\bf F}_2 {\bf F}_1){\bf F}_2 {\bf n}_1  + ({\bf I} + {\bf F}_2 {\bf F}_1){\bf n}_2.
    \label{eq:linear_decoding:y1}
\end{align}
We first define ${\bf r}_1$ with a linear transformation of ${\bf y}_1$ 
as
\begin{align}
      {\bf r}_1 & = ({\bf I} + {\bf F}_2 {\bf F}_1)^{-1} ({\bf y}_1 - {\bf F}_2{\bf g}_1m_1)
    \nonumber
    \\
    & = {\bf g}_2 m_2 + {\bf F}_2 {\bf n}_1 + {\bf n}_2,
    \label{eq:linear_decoding:r1}
\end{align}
where $({\bf I} + {\bf F}_2 {\bf F}_1)$ is always invertible for any strictly lower triangular matrices ${\bf F}_1$ and ${\bf F}_2$, since $({\bf I} + {\bf F}_2 {\bf F}_1)$ has all 1's on its diagonal entries.
We note that ${\bf r}_1$ is a sufficient statistic for estimating $m_2$ given $m_1$ since, by definition, the conditional distribution of ${\bf y}_1$ given ${\bf r}_1$ does not depend on $m_2$. In other words, ${\bf r}_1$ has all the information about $m_2$.
Then, we have $\hat m_2 = \underset{m_2 \in \mathcal{M}_2} {\arg\max} \;\; \text{Pr}({\bf y}_1 | m_1, m_2) = \underset{m_2 \in \mathcal{M}_2} {\arg\max} \;\; \text{Pr}({\bf r}_1 | m_1, m_2)$.
Note that ${\bf r}_1$ is conditionally Gaussian with conditional mean  and conditional covariance matrix, given~by
\begin{align}
    \mathbb{E}_{\rev{{\bf n}_1, {\bf n}_2}} \big[{\bf r}_1 \vert m_2 \big]  &=  {\bf g}_2 m_2
    \\
    \mathbb{E}_{\rev{{\bf n}_1, {\bf n}_2}} \big[({\bf r}_1 -{\bf g}_2 m_2)({\bf r}_1 -{\bf g}_2 m_2)^\top \vert m_2 \big] &=  {\bf F}_2 {\bf F}_2^\top \sigma_1^2 + \sigma_2^2 {\bf I} := {\bf Q}_2.
    \label{eq:Q2}
\end{align}
Thus, we obtain
\begin{align}
    \hat m_2 & = \underset{m_2 \in \mathcal{M}_2} {\arg\max} \;\; \text{Pr}({\bf r}_1 | m_1, m_2)
    \nonumber
    \\
    & = \underset{m_2 \in \mathcal{M}_2} {\arg\min} \;\; ({\bf r}_1 - {\bf g}_2m_2)^\top {\bf Q}_2^{-1} ({\bf r}_1 - {\bf g}_2 m_2).
    \label{eq:linear_decoding:r1_quad}
\end{align}
Note that ${\bf Q}_2$ defined in \eqref{eq:Q2} is invertible, since  
${\bf F}_2 {\bf F}_2^\top$ is positive semi-definite (PSD), and 
thus ${\bf F}_2 {\bf F}_2^\top \sigma_1^2 + \sigma_2^2 {\bf I}$ has all eigenvalues greater than or equal to $\sigma_2^2$, ensuring ${\bf Q}_2$ has full rank.
The quadratic form in \eqref{eq:linear_decoding:r1_quad} can be equivalently rewritten as 
\begin{equation}
    ({\bf r}_1 - {\bf g}_2 \tilde m_2)^\top {\bf Q}_2^{-1} ({\bf r}_1 - {\bf g}_2 \tilde m_2) + (m_2 - \tilde m_2)^2 {\bf g}_2^\top {\bf Q}_2^{-1} {\bf g}_2,
    \label{eq:linear_decoding:r1_quad:equi}
\end{equation}
by defining 
\begin{align}
    \tilde m_2 = \frac{{\bf g}_2^\top {\bf Q}_2^{-1}}{{\bf g}_2^\top {\bf Q}_2^{-1}{\bf g}_2} {\bf r}_1  := {\bf w}_2^\top {\bf r}_1 \in \mathbb{R}.
    \label{eq:linear_decoding:MVU:m2}
\end{align}
Note that $\tilde m_2$ in \eqref{eq:linear_decoding:MVU:m2} is the minimum variance unbiased (MVU) estimate of $m_2$, and
\begin{align}
    {\bf w}_2 = \frac{{\bf Q}_2^{-1} {\bf g}_2}{{\bf g}_2^\top {\bf Q}_2^{-1}{\bf g}_2}
    \label{eq:linear_coding:w2}
\end{align}
is the MVU estimator as a combining vector.
%
Using \eqref{eq:linear_decoding:r1_quad:equi}, we rewrite \eqref{eq:linear_decoding:r1_quad} as
\begin{align}
    \hat m_2 = \underset{m_2 \in \mathcal{M}_2} {\arg\min} \;\; (m_2 - \tilde m_2)^2.
\end{align}
Thus, the optimal decision for User $1$ is to select $m_2 \in \mathcal{M}_2$ that is closest to the MVU estimate $\tilde m_2 \in \mathbb{R}$.

Next, we formulate the receive signal-to-noise (SNR) in estimating $m_2$ at User 1. By putting ${\bf r}_1$ (in \eqref{eq:linear_decoding:r1}) into \eqref{eq:linear_decoding:MVU:m2}, we can rewrite \eqref{eq:linear_decoding:MVU:m2} as
\begin{align}
    \tilde m_2 = {\bf w}_2^\top {\bf g}_2 m_2 + {\bf w}_2^\top({\bf F}_2{\bf n}_1 + {\bf n}_2).
    \label{eq:linear_decoding:MVU:m2:reduced}
\end{align}
Since $\tilde m_2$ is also a sufficient statistic for estimating $m_2$, by using \eqref{eq:linear_decoding:MVU:m2:reduced}, we can define the receive SNR in estimating $m_2$  as
\begin{align}
    \text{SNR}_2 = \frac{\vert {\bf w}_{2}^\top {\bf g}_2 \vert^2} 
    {{\bf w}_{2}^\top {\bf Q}_2 {\bf w}_{2}}.
    \label{eq:SNR2}
\end{align}

Next, we consider the decoding process at User 2.  User 2 estimates $m_1$ given the receive signal ${\bf y}_2$ and its own message $m_2$ based on maximum likelihood. i.e., $\hat m_1 = \underset{m_1 \in \mathcal{M}_1} {\arg\max} \; \text{Pr}({\bf y}_2 | m_1, m_2)$. Recall  ${\bf y}_2 =  {\bf g}_1 m_1  + {\bf F}_1 {\bf g}_2 m_2 + ({\bf I} + {\bf F}_1{\bf F}_2){\bf n}_1 + {\bf F}_1{\bf n}_2$  in \eqref{eq:y2:vec}.
We first define ${\bf r}_2$ with a linear transformation of ${\bf y}_2$ 
as
\begin{align}
      {\bf r}_2 & = {\bf y}_2 - {\bf F}_1 {\bf g}_2 m_2.
      \nonumber
      \\
      & = {\bf g}_1 m_1  + ({\bf I} + {\bf F}_1{\bf F}_2){\bf n}_1 + {\bf F}_1{\bf n}_2.
    \label{eq:linear_decoding:r2}
\end{align}
We note that ${\bf r}_2$ is a sufficient statistic for estimating $m_1$ given $m_2$, since, by definition, the conditional distribution of ${\bf y}_2$ given ${\bf r}_2$ does not depend on $m_1$.
Then, we obtain $\hat m_1 = \underset{m_1 \in \mathcal{M}_1} {\arg\max} \;\; \text{Pr}({\bf y}_2 | m_1, m_2) = \underset{m_1 \in \mathcal{M}_1} {\arg\max} \;\; \text{Pr}({\bf r}_2 | m_1, m_2)$.
Note that ${\bf r}_2$ is conditionally Gaussian with conditional mean  and conditional covariance matrix, given by
\begin{align}
     \mathbb{E}_{\rev{{\bf n}_1, {\bf n}_2}} \big[{\bf r}_2 \vert m_1 \big]  &=  {\bf g}_1 m_1
    \\
    \mathbb{E}_{\rev{{\bf n}_1, {\bf n}_2}} \big[({\bf r}_2 -{\bf g}_1 m_1)({\bf r}_2 -{\bf g}_1 m_1)^\top \vert m_1 \big] &=  
    ({\bf I} + {\bf F}_1{\bf F}_2)({\bf I} + {\bf F}_1{\bf F}_2)^\top \sigma_1^2 + {\bf F}_1 {\bf F}_1^\top \sigma_2^2 := {\bf Q}_1
    \label{eq:Q1}
\end{align}
Thus, we obtain
\begin{align}
    \hat m_1 & = \underset{m_1 \in \mathcal{M}_1} {\arg\max} \;\; \text{Pr}({\bf r}_2 | m_1, m_2)
    \nonumber
    \\
    & = \underset{m_1 \in \mathcal{M}_1} {\arg\min} \;\; ({\bf r}_2 - {\bf g}_1m_1)^\top {\bf Q}_1^{-1} ({\bf r}_2 - {\bf g}_1 m_1).
    \label{eq:linear_decoding:r2_quad}
\end{align}
Note that ${\bf Q}_1$ defined in \eqref{eq:Q1} is invertible, since  
(i) $({\bf I} + {\bf F}_1{\bf F}_2)({\bf I} + {\bf F}_1{\bf F}_2)^\top \sigma_1^2$ has all eigenvalues greater than or equal to $\sigma_1^2$ for strictly lower triangular matrices ${\bf F}_1$ and ${\bf F}_2$, and (ii) ${\bf F}_1 {\bf F}_1^\top$ is PSD, and thus the summation of (i) and (ii) has all eigenvalues greater than or  equal to $\sigma_1^2$, ensuring 
${\bf Q}_1$ has full rank.
The quadratic form in \eqref{eq:linear_decoding:r2} is equivalently rewritten as 
\begin{equation}
    ({\bf r}_2 - {\bf g}_1 \tilde m_1)^\top {\bf Q}_1^{-1} ({\bf r}_2 - {\bf g}_1 \tilde m_1) + (m_1 - \tilde m_1)^2 {\bf g}_1^\top {\bf Q}_1^{-1} {\bf g}_1,
    \label{eq:linear_decoding:r2_quad:equi}
\end{equation}
by defining 
\begin{align}
    \tilde m_1 = \frac{{\bf g}_1^\top {\bf Q}_1^{-1}}{{\bf g}_1^\top {\bf Q}_1^{-1}{\bf g}_1} {\bf r}_2  := {\bf w}_1^\top {\bf r}_2 \in \mathbb{R}.
    \label{eq:linear_decoding:MVU:m1}
\end{align}
Note that $\tilde m_1$ in \eqref{eq:linear_decoding:MVU:m1} is the MVU estimate of $m_1$, and
\begin{align}
    {\bf w}_1 = \frac{{\bf Q}_1^{-1} {\bf g}_1}{{\bf g}_1^\top {\bf Q}_1^{-1}{\bf g}_1}
    \label{eq:linear_coding:w1}
\end{align}
is the MVU estimator as a combining vector.
%
Using \eqref{eq:linear_decoding:r2_quad:equi}, we rewrite \eqref{eq:linear_decoding:r2_quad} as
\begin{align}
    \hat m_1 = \underset{m_1 \in \mathcal{M}_1} {\arg\min} \;\; (m_1 - \tilde m_1)^2.
\end{align}
The optimal decision for User $2$ is to select $m_1 \in \mathcal{M}_1$ that is closest to the MVU estimate $\tilde m_1 \in \mathbb{R}$.

Next, we formulate the receive SNR in estimating $m_1$ at User 2. By putting ${\bf r}_2$ (in \eqref{eq:linear_decoding:r2}) into \eqref{eq:linear_decoding:MVU:m1}, we rewrite \eqref{eq:linear_decoding:MVU:m1} as
\begin{align}
    \tilde m_1 = {\bf w}_1^\top {\bf g}_1 m_1  + {\bf w}_1^\top  \big(({\bf I} + {\bf F}_1{\bf F}_2){\bf n}_1 + {\bf F}_1{\bf n}_2\big).
    \label{eq:linear_decoding:MVU:m1:reduced}
\end{align}
Since $\tilde m_1$ is also a sufficient statistic for estimating $m_1$, by using \eqref{eq:linear_decoding:MVU:m1:reduced}, we can define the receive SNR in estimating $m_1$  as
\begin{align}
    \text{SNR}_1 = \frac{\vert {\bf w}_{1}^\top {\bf g}_1 \vert^2} 
    {{\bf w}_{1}^\top {\bf Q}_1 {\bf w}_{1}}.
    \label{eq:SNR1}
\end{align}

We note that the combining vectors for decoding 
in \eqref{eq:linear_coding:w2} and \eqref{eq:linear_coding:w1} are represented as functions of the encoding schemes
of the users, i.e., ${\bf g}_1$, ${\bf F}_1$, ${\bf g}_2$, and ${\bf F}_2$.
By plugging \eqref{eq:linear_coding:w2} in \eqref{eq:SNR2} and \eqref{eq:linear_coding:w1} in \eqref{eq:SNR1}, we obtain the SNR expression that only depends on the encoding schemes, given by
\begin{align}
      \text{SNR}_i =  {\bf g}_i^\top {\bf Q}_i^{-1} {\bf g}_i, \quad i \in \{1,2\}.
      \label{eq:SNR_opt}
\end{align}
Thus, the joint design of the encoding and decoding schemes is equivalently reduced to the design of only the encoding schemes
with the derived form of SNRs in \eqref{eq:SNR_opt}.

\subsection{Conversion of Optimization Problem}
\label{ssec:linear:conversion}

For linear coding, we can represent the error probability metric $\mathcal{E}_i$ in \eqref{opt:obj} as a function of SNR, i.e., $\mathcal{E}_i(\text{SNR}_i)$, since the formula of $\text{SNR}_i$ given in \eqref{eq:SNR_opt} captures both the encoding and decoding schemes.\footnote{$\mathcal{E}_i$ is  a function of $P$, $\sigma_1^2$, $\sigma_2^2$, $K_1$, $K_2$, and $N$, in addition to the encoding and decoding schemes, as discussed in Sec.~\ref{ssec:opt}. In this work, we focus on the dependency of the encoding and decoding schemes on the error probability, since $P$, $\sigma_1^2$, $\sigma_2^2$, $K_1$, $K_2$, and $N$ are given.}
This modified input-output relationship in error probability makes it more convenient to analyze linear coding.
We note that $\mathcal{E}_i$ is either $\text{BLER}_i$ or $\text{BER}_i$; with $2^{K_i}$-ary PAM, $\text{BLER}_i$ is equivalent to symbol error rate (SER) of the message $m_i$. When satisfying $\mathbb{E}_{\rev{m_i}}[m_i]=0$ and $\mathbb{E}_{\rev{m_i}}[m^2_i]=1$, the BLER is then expressed as \cite{proakis2008digital}
\begin{equation}
    \text{BLER}_i = \frac{2^{K_i+1}-2}{2^{K_i}} Q\bigg( \sqrt{\frac{3\text{SNR}_i}{2^{2K_i}-1}} \bigg),
    \label{eq:SER:q_func}
\end{equation}
where $Q(x) = \frac{1}{\sqrt{2\pi}} \int_{x}^{\infty} (-\frac{u^2}{2}) du$ denotes the tail distribution function of the standard normal distribution.
We note that BER also depends on SNR. In the high SNR regime, $\text{BER}_i \approx \text{BLER}_i/K_i$ when Gray coding is used~\cite{tse2005fundamentals}, while in the low SNR regime, BER can be calculated from the relationship between SNR and the modulation order $2^{K_i}$~\cite{proakis2008digital}.
With the functional form $\mathcal{E}_i(\text{SNR}_i)$, we rewrite the optimization problem
\eqref{opt:obj}-\eqref{opt:const:power} as
\begin{align}
  & \underset{{\bf g}_1, {\bf F}_1, {\bf g}_2, {\bf F}_2}
    {\text{minimize}} & & 
 \mathcal{E}_1(\text{SNR}_1) + \mathcal{E}_2(\text{SNR}_2)
\nonumber
\\
& \text{subject to}
& &  
\mathbb{E}_{\rev{m_1,m_2,{\bf n}_1, {\bf n}_2}}\big[\|{\bf x}_1 \|^2\big] \le N P, \quad
\mathbb{E}_{\rev{m_1,m_2,{\bf n}_1, {\bf n}_2}}\big[\|{\bf x}_2 \|^2\big] \le N P,
\label{opt:linear:sum-error}
\end{align}
where $\text{SNR}_i$ and ${\bf x}_i$, $i \in \{1,2\}$, are functions of the variables, ${\bf g}_1$, ${\bf F}_1$, ${\bf g}_2$, and ${\bf F}_2$, and the expectation is taken over the distributions of \rev{the random variables,} ${m}_1$, ${m}_2$, ${\bf n}_1$, and ${\bf n}_2$. 
However, it is intractable to directly solve the optimization problem in \eqref{opt:linear:sum-error} 
due to (i) the coupling between the encoding schemes and the SNR expression 
in \eqref{eq:SNR_opt}
and (ii) the power constraints in \eqref{opt:linear:sum-error}.
We thus propose to transform it into a more tractable optimization problem through several steps as explained below. We start by stating the following remark. 

\begin{remark}
\label{lemma1}
The sum-error metric, $\mathcal{E}_1(\text{SNR}_1) + \mathcal{E}_2(\text{SNR}_2)$, is a decreasing function of $P$.
\end{remark}
%
We further discuss the validity of Remark~\ref{lemma1} within our linear coding framework in Appendix~\ref{app:linear:lemma1}.
%
Based on the relationship between sum-error and power in Remark~\ref{lemma1}, 
we introduce an alternative optimization problem to \eqref{opt:linear:sum-error}, which aims 
to minimize the maximum transmit power, given by 
\begin{align}
  & \underset{{\bf g}_1, {\bf F}_1, {\bf g}_2, {\bf F}_2, \bar P}
    {\text{minimize}} & & 
    \bar P
\nonumber
\\
& \text{subject to}
& &  
\mathcal{E}_1(\text{SNR}_1) + \mathcal{E}_2(\text{SNR}_2) = \omega
\nonumber
\\
& & & \mathbb{E}_{\rev{m_1,m_2,{\bf n}_1, {\bf n}_2}}\big[\|{\bf x}_1 \|^2\big] \le N \bar P, \quad
\mathbb{E}_{\rev{m_1,m_2,{\bf n}_1, {\bf n}_2}}\big[\|{\bf x}_2 \|^2\big] \le N \bar P,
\label{opt:linear:max-power}
\end{align}
where the auxiliary variable $\bar P$ is introduced.
Note that the optimal solutions for \eqref{opt:linear:sum-error} can be obtained by solving \eqref{opt:linear:max-power} through adjusting the value of $w$ based on Remark \ref{lemma1};
if the obtained objective value $\bar P$ in \eqref{opt:linear:max-power} is smaller than $P$, we keep decreasing the value of $\omega$  and solve the above problem until $\bar P$ reaches  $P$.
However, solving \eqref{opt:linear:max-power} under the sum-error equality constraint involves identifying the optimal pair of $\text{SNR}_1$ and $\text{SNR}_2$, since multiple pairs may satisfy the constraint.
To mitigate the optimization complexity, we consider individual constraints on  $\text{SNR}_1$ and $\text{SNR}_2$ for optimization rather than the sum-error constraint. We subsequently reformulate  \eqref{opt:linear:max-power} and obtain
\begin{align}
  & \underset{{\bf g}_1, {\bf F}_1, {\bf g}_2, {\bf F}_2,  \bar P}
    {\text{minimize}} & & 
    \bar P
\nonumber
\\
& \text{subject to}
& &  
\text{SNR}_1 = \eta_1, \quad 
\text{SNR}_2 = \eta_2, 
\nonumber
\\
& & & \mathbb{E}_{\rev{m_1,m_2,{\bf n}_1, {\bf n}_2}}\big[\|{\bf x}_1 \|^2\big] \le N \bar P, \quad
\mathbb{E}_{\rev{m_1,m_2,{\bf n}_1, {\bf n}_2}}\big[\|{\bf x}_2 \|^2\big] \le N \bar P,
\label{opt:linear:max-power:SNR}
\end{align}
where $\eta_i \in \mathbb{R}^+$ is the target SNR for the message $m_i$, $i\in \{1,2\}$.
In \eqref{opt:linear:max-power:SNR}, a feasible solution for $({\bf g}_1, {\bf F}_1, {\bf g}_2, {\bf F}_2)$ always exists
for any $\eta_1, \eta_2 \in \mathbb{R}^+$ because the SNR constraints can be satisfied by adjusting the magnitudes of ${\bf g}_1$ and ${\bf g}_2$ in \eqref{eq:SNR_opt}.

If $\bar P$ is an optimal value of the problem in \eqref{opt:linear:max-power:SNR}, at least one of the power constraints in \eqref{opt:linear:max-power:SNR} 
should be satisfied in equality, i.e., $\bar P = \max \{ \mathbb{E}_{\rev{m_1,m_2,{\bf n}_1, {\bf n}_2}} \big[\|{\bf x}_1 \|^2\big], \mathbb{E}_{\rev{m_1,m_2,{\bf n}_1, {\bf n}_2}} \big[\|{\bf x}_2 \|^2\big] \}/N$.
If $\bar P \neq \max \{ \mathbb{E}_{\rev{m_1,m_2,{\bf n}_1, {\bf n}_2}} \big[\|{\bf x}_1 \|^2\big],$ $\mathbb{E}_{\rev{m_1,m_2,{\bf n}_1, {\bf n}_2}} \big[\|{\bf x}_2 \|^2\big] \}/N$, 
$\bar P$ is not an optimal solution since $\max \{\mathbb{E}_{\rev{m_1,m_2,{\bf n}_1, {\bf n}_2}}$ $\big[\|{\bf x}_1 \|^2\big]$, $\mathbb{E}_{\rev{m_1,m_2,{\bf n}_1, {\bf n}_2}} \big[\|{\bf x}_2 \|^2\big] \}/N$ can be the smaller objective value in \eqref{opt:linear:max-power:SNR}.
This insight leads us to reformulate \eqref{opt:linear:max-power:SNR} equivalently~to
\begin{align}
  & \underset{{\bf g}_1, {\bf F}_1, {\bf g}_2, {\bf F}_2}
    {\text{minimize}} & & 
    {\max} \big\{ \mathbb{E}_{\rev{m_1,m_2,{\bf n}_1, {\bf n}_2}} \big[\|{\bf x}_1 \|^2\big], \mathbb{E}_{\rev{m_1,m_2,{\bf n}_1, {\bf n}_2}} \big[\|{\bf x}_2 \|^2\big] \big\}
\nonumber
\\
& \text{subject to}
& &  
\text{SNR}_1 = \eta_1, \quad 
\text{SNR}_2 = \eta_2.
\label{opt:linear:max-of-powers:SNR}
\end{align}
We show, in the following lemma, that the optimal solutions of the original sum-error minimization problem with the power constraints in \eqref{opt:linear:sum-error} can be obtained by solving the max-power minimization problem with the SNR constraints in \eqref{opt:linear:max-of-powers:SNR}.
%
\begin{lemma}
    The optimal solutions of \eqref{opt:linear:sum-error} can be obtained by solving the following problem, 
    defined as
\begin{align}
  & \underset{{\bf g}_1, {\bf F}_1, {\bf g}_2, {\bf F}_2}
    {\text{minimize}} & & 
 \mathcal{E}_1(\text{SNR}_1) + \mathcal{E}_2(\text{SNR}_2)
    \nonumber
    \\
    & \text{subject to}
    & &  
    ({\bf g}_1, {\bf F}_1, {\bf g}_2, {\bf F}_2) \in \mathcal{S},
    \label{opt:linear:SVEM}
\end{align}
where $\mathcal{S} = \bigcup_{\eta_1, \eta_2 \in \mathbb{R}^+} \mathcal{S}_{(\eta_1, \eta_2)}$.
Here, $ \mathcal{S}_{(\eta_1, \eta_2)}$ is the set of solutions $({\bf g}_1, {\bf F}_1, {\bf g}_2, {\bf F}_2)$ of \eqref{opt:linear:max-of-powers:SNR} given $\eta_1$ and $\eta_2$, while satisfying ${\max} \big\{ \mathbb{E}_{\rev{m_1,m_2,{\bf n}_1, {\bf n}_2}} \big[\|{\bf x}_1 \|^2\big]$, $\mathbb{E}_{\rev{m_1,m_2,{\bf n}_1, {\bf n}_2}} \big[\|{\bf x}_2 \|^2\big] \big\}$ $\le NP$.
    \label{lemma_optimization_conversion}
\end{lemma}




The proof of Lemma \ref{lemma_optimization_conversion} is provided in Appendix \ref{app:linear:lemma:optimality}.
Using Lemma \ref{lemma_optimization_conversion},  
instead of directly tackling \eqref{opt:linear:sum-error}, we solve \eqref{opt:linear:SVEM}. Specifically, we first determine the feasible solution set $\mathcal{S}$ by solving \eqref{opt:linear:max-of-powers:SNR} for all possible values of $\eta_1$ and $\eta_2$. After identifying $\mathcal{S}$, we find the solution that minimizes the sum-error. It is important to note that identifying $\mathcal{S}$ is the key to solve \eqref{opt:linear:SVEM}.
Next, we derive the following lemma, which is useful to find $\mathcal{S}$.
\begin{lemma}
    The optimal value of $ {\max} \big\{ \mathbb{E}_{\rev{m_1,m_2,{\bf n}_1, {\bf n}_2}} \big[\|{\bf x}_1 \|^2\big], \mathbb{E}_{\rev{m_1,m_2,{\bf n}_1, {\bf n}_2}} \big[\|{\bf x}_2 \|^2\big] \big\}$ in \eqref{opt:linear:max-of-powers:SNR} is an increasing function of $\eta_1$ when $\eta_2$ is fixed.
    \label{lemma:max_inc}
\end{lemma}
The proof of Lemma~\ref{lemma:max_inc} is provided in Appendix~\ref{app:linear:lemma2}.
Given that the search space for $\eta_1$ and $\eta_2$ is large in finding $\mathcal{S}$, we can use Lemma~\ref{lemma:max_inc} to reduce the search space of $\eta_1$ once $\eta_2$ is fixed.
That is,
Lemma~\ref{lemma:max_inc} allows us to develop an efficient search method along $\eta_1$, which will be discussed in detail in Sec.~\ref{ssec:linear:opt:sum-error:sol}.

To provide further tractability in solving \eqref{opt:linear:max-of-powers:SNR}, which is a step for finding $\mathcal{S}$ in \eqref{opt:linear:SVEM}, we treat the min-max problem in \eqref{opt:linear:max-of-powers:SNR} as a multi-objective  problem, where we aim to minimize both $\mathbb{E}_{\rev{m_1,m_2,{\bf n}_1, {\bf n}_2}} \big[\|{\bf x}_1 \|^2\big]$ and $\mathbb{E}_{\rev{m_1,m_2,{\bf n}_1, {\bf n}_2}} \big[\|{\bf x}_2 \|^2\big]$ and find the solution that minimizes ${\max} \big\{ \mathbb{E}_{\rev{m_1,m_2,{\bf n}_1, {\bf n}_2}} \big[\|{\bf x}_1 \|^2\big], \mathbb{E}_{\rev{m_1,m_2,{\bf n}_1, {\bf n}_2}} \big[\|{\bf x}_2 \|^2\big] \big\}$.
To solve the multi-objective problem, we use the \textit{weighted sum method} as an approximated solution technique,
which is a common choice in solving multi-objective problem~\cite{deb2011multi}. 
We note that, using the weighted sum method provides benefits in terms of tractability, in our case for developing linear coding methods.
We formulate the weighted sum-power minimization problem with a weighting coefficient $\alpha \in (0,1)$ as
%
\begin{align}
  & \underset{{\bf g}_1, {\bf F}_1, {\bf g}_2, {\bf F}_2}
    {\text{minimize}} & & 
 \alpha \mathbb{E}_{\rev{m_1,m_2,{\bf n}_1, {\bf n}_2}} \big[\|{\bf x}_1 \|^2\big] + (1-\alpha) \mathbb{E}_{\rev{m_1,m_2,{\bf n}_1, {\bf n}_2}} \big[\|{\bf x}_2 \|^2\big]
\nonumber
\\
& \text{subject to}
& &  
\text{SNR}_1 = \eta_1, \quad
\text{SNR}_2 = \eta_2.
\label{opt:linear:weighted-sum}
\end{align}


The solution obtained from solving \eqref{opt:linear:weighted-sum} for a specific weight $\alpha$ is feasible for \eqref{opt:linear:max-of-powers:SNR}, but it may not necessarily be optimal for \eqref{opt:linear:max-of-powers:SNR}. In general, solving weighted sum minimization problems does not guarantee to yield optimal solutions to non-convex multi-objective problems~\cite{marler2010weighted}. We note that choosing a proper weight in \eqref{opt:linear:weighted-sum} is crucial for finding the optimal or sub-optimal solution to \eqref{opt:linear:max-of-powers:SNR}.
However, it is challenging to determine the appropriate weight to achieve the optimal or sub-optimal solution of the min-max problem~\cite{marler2010weighted}.
A straightforward approach is to (i) solve \eqref{opt:linear:weighted-sum} given various values of weight $\alpha \in (0,1)$ to obtain feasible solutions for \eqref{opt:linear:max-of-powers:SNR} and (ii) determine the solution that minimizes the max-power among the feasible solutions. We develop an efficient search method along $\alpha$, which will be discussed in detail in Sec.~\ref{ssec:linear:opt:sum-error:sol}.

In summary, we transform the optimization problem from solving the sum-error minimization problem in \eqref{opt:linear:sum-error} to \eqref{opt:linear:SVEM}. The equivalence in optimization conversion is demonstrated through Lemma \ref{lemma_optimization_conversion}. Our focus then shifts to identifying the feasible solution set $\mathcal{S}$. Finding $\mathcal{S}$ involves solving the max-power minimization problem in \eqref{opt:linear:max-of-powers:SNR} by varying the variables $\eta_1$ and $\eta_2$. 
To make it more tractable to find $\mathcal{S}$,
we approximate solving \eqref{opt:linear:max-of-powers:SNR} to solving the weighted-sum power minimization problem in \eqref{opt:linear:weighted-sum} by varying the additional variable $\alpha$.
While the optimization transform from \eqref{opt:linear:sum-error} to \eqref{opt:linear:SVEM} preserves optimality, we may lose some optimality when finding $\mathcal{S}$ by solving \eqref{opt:linear:weighted-sum} instead of \eqref{opt:linear:max-of-powers:SNR}. The optimality is sacrificed for tractability during this step.

Solving \eqref{opt:linear:SVEM} involves the following steps. First, we identify the feasible solution set $\mathcal{S}$ by solving \eqref{opt:linear:weighted-sum} for various combinations of  $(\eta_1, \eta_2, \alpha)$. Then, we determine the solution that minimizes the sum-error within $\mathcal{S}$. 
As the initial step in developing solution methods, we focus on solving \eqref{opt:linear:weighted-sum} for a given $(\eta_1, \eta_2, \alpha)$ in Sec.~\ref{ssec:linear:opt:weighted-sum} and \ref{ssec:linear:opt:weighted-sum:sol}. Subsequently, 
in Sec.~\ref{ssec:linear:opt:sum-error:sol}, we provide an efficient algorithm to identify the feasible solution set $\mathcal{S}$ by varying $(\eta_1, \eta_2, \alpha)$, and ultimately, to find the solution for \eqref{opt:linear:SVEM}.

\subsection{Optimization Problem for Weighted Sum-Power Minimization}
\label{ssec:linear:opt:weighted-sum}

In general, at time $k \ge 2$, 
User $i$
feeds back a linear combination of the previously received symbols up to time $k-1$, i.e., $\{ {y}_i[\tau]\}_{\tau=1}^{k-1}$, where
$i \in \{1,2\}$.
This implies that the initially received symbols at the users are repetitively fed back to the other over a total of $N$ channel uses, e.g., the information of $y_2[1]$ at User 2 is fed back to User 1 over $N-1$ times from $k=2$ to $N$.
This repetitive feedback in both ways would make the design of the encoding schemes more complicated because the encoding schemes of the users are coupled.
To mitigate the complexity of designing the encoding schemes, 
we assume that 
User 2 only feeds back the recently received signal of ${\bf y}_2$ in \eqref{eq:x2}. 
 In other words, User 2 provides feedback only for $y[k-1]$ at time $k$. Consequently, ${\bf F}_2$ should be in the following form:
\begin{equation}
{\bf F}_2 = 
\begin{bmatrix}
0 & 0 & 0 & ... & 0\\
f_{2,2} & 0 & 0 & ... & 0 \\
0 & f_{2,3} & 0 & ... & 0  \\
\vdots & \vdots & \ddots & & 0 \\
0 & 0 & ... & f_{2,N} & 0
\end{bmatrix} 
\in \mathbb{R}^{N \times N}.
\label{eq:F2}
\end{equation} 
For convenience, we define a set $\mathcal{F}_2$ 
as a solution space for ${\bf F}_2$, i.e., ${\bf F}_2 \in \mathcal{F}_2$.
First, we will investigate the solution behavior for  feedback of User 2.
Specifically, we reveal in the following proposition
that it is optimal for User 2  not to utilize the last channel use for feedback 
to User 1, i.e., $f_{2,N}=0$, regardless of the encoding schemes.


\begin{proposition}
\label{pro:f2n}
In \eqref{opt:linear:weighted-sum}
with ${\bf F}_2 \in \mathcal{F}_2$,
it is optimal that $f_{2,N} = 0$.
\end{proposition}
%
The proof of Proposition~\ref{pro:f2n} is provided in Appendix~\ref{app:linear:pro:f2n}.
We next look into the solution behavior of the message encoding vector for User 2, ${\bf g}_2$. 
To this end, we first formulate the optimization problem \eqref{opt:linear:weighted-sum} only with respect to ${\bf g}_2$ by using the expression in \eqref{eq:power1}-\eqref{eq:power2} as follows:
\begin{align}
    & 
    \underset{{\bf g}_2}
    {\text{minimize}} & & 
 \alpha \|{\bf F}_1 {\bf g}_2 \|^2 
 + (1-\alpha) \| ({\bf I} + {\bf F}_2 {\bf F}_1 ) {\bf g}_2\|^2
 \nonumber
\\
& \text{subject to}
& &  
{\bf g}_2^\top {\bf Q}_2^{-1} {\bf g}_2 = \eta_2.
\label{opt:g2}
\end{align}
We note that ${\bf Q}_2$ is symmetric, so we can express it as ${\bf Q}_2 = {\bf U} {\Sigma} {\bf U}^\top $ by using the singular value decomposition.
By defining ${\bf Q}_{2,\text{sqrt}} = {\bf U} {\Sigma}_{\text{sqrt}} {\bf U}^\top$, where ${\Sigma}_{\text{sqrt}}$ is the diagonal matrix with entries that are the square roots of the corresponding diagonal entries in ${\Sigma}$, we rewrite  ${\bf Q}_2$ as ${\bf Q}_2 = {\bf Q}_{2,\text{sqrt}}^\top {\bf Q}_{2,\text{sqrt}}$. Note that ${\bf Q}_{2,\text{sqrt}} = {\bf Q}_{2,\text{sqrt}}^\top$.
Since ${\bf Q}_2$ is invertible, ${\bf Q}_{2,\text{sqrt}}$ is also invertible.
Defining ${\bf q}_2 = {\bf Q}_{2,\text{sqrt}}^{-1} {\bf g}_2$, we can obtain an equivalent optimization problem as
\begin{align}
    & 
    \underset{{\bf q}_2}
    {\text{minimize}} & & 
    {\bf q}_2^\top {\bf B} {\bf q}_2
 \nonumber
\\
& \text{subject to}
& &  
\|{\bf q}_2\|^2 = \eta_2,
\label{opt:q2}
\end{align}
where
\begin{align}
    {\bf B} 
    = \alpha {\bf Q}_{2,\text{sqrt}}  {\bf F}_1^\top  {\bf F}_1 {\bf Q}_{2,\text{sqrt}}
    + (1-\alpha) {\bf Q}_{2,\text{sqrt}}  ({\bf I} + {\bf F}_2 {\bf F}_1 )^\top ({\bf I} + {\bf F}_2 {\bf F}_1 )
     {\bf Q}_{2,\text{sqrt}}.
     \label{eq:linear:B}
\end{align}

We provide the following proposition to understand the solution behavior of ${\bf g}_2$ for any strictly lower triangular matrix ${\bf F}_1$ and any ${\bf F}_2 \in \mathcal{F}_2$.
\begin{proposition}
    \label{pro:B_lower_upper}
    For any strictly lower triangular matrix ${\bf F}_1$ and any ${\bf F}_2 \in \mathcal{F}_2$, 
    \begin{align}
    0
    \le \nu_{\min}[{\bf B}]  
    \le (1-\alpha) \sigma_2^2,
    \end{align}
where 
$\nu_{\min}[{\bf B}]$ denotes the smallest eigenvalue of ${\bf B}$ in \eqref{eq:linear:B}. 
\end{proposition}

The proof of Proposition~\ref{pro:B_lower_upper} is provided in Appendix~\ref{app:linear:pro:B_lower_upper}. 
We note that the value of $\nu_{\min}[{\bf B}] $ depends on the specific configurations of ${\bf F}_1$ and ${\bf F}_2$.
We find that the lower bound of 0 is loose in Proposition~\ref{pro:B_lower_upper}.
Intuitively, it is unlikely that $\nu_{\min}[{\bf B}] = 0 $, because this would imply that a certain choice of ${\bf q}_2$ (equivalently, a message encoding vector ${\bf g}_2$) could achieve zero transmit power (i.e., an objective value of 0) while still satisfying its message SNR constraint in \eqref{opt:q2}. 
We recognize that there may be a tighter lower bound. Therefore, we introduce the following conjecture on the tighter lower bound of $\nu_{\min}[{\bf B}]$.

\begin{conjecture}
    \label{conj:B}
    For any strictly lower triangular matrix ${\bf F}_1$ and any ${\bf F}_2 \in \mathcal{F}_2$, 
    \begin{align}
    \min \{\alpha \sigma_1^2, (1-\alpha) \sigma_2^2 \}
    \le \nu_{\min}[{\bf B}].
    \end{align}
\end{conjecture}

We have proved a special case of this conjecture when $N=3$ in Appendix~\ref{app:linear:conj:B}.
We note that, for any $N$, we have not found any example that violates the conjecture in our extensive numerical simulations, where ${\bf F}_1$ and ${\bf F}_2 \in \mathcal{F}_2$ are randomly generated.

\begin{proposition}
\label{pro:g:optimal}
If Conjecture~\ref{conj:B} is true, ${\bf g}_2 = [0, ...,  0, \sqrt{\eta_2}\sigma_2]^\top$ is optimal in  \eqref{opt:linear:weighted-sum} for any strictly lower triangular ${\bf F}_1$ and any ${\bf F}_2 \in \mathcal{F}_2$
 when $\alpha \ge  \frac{\sigma_2^2}{\sigma_1^2 + \sigma_2^2}$. This choice of ${\bf g}_2$ is optimal for any $\alpha \in [0,1]$ when $N=3$.
\end{proposition}
The proof of Proposition \ref{pro:g:optimal} is provided in
Appendix~\ref{app:linear:pro:g:optimal}. 
The result of Proposition \ref{pro:g:optimal}  shows
that it is optimal for User 2 to  transmit the message only over the last channel use when the weight coefficient in \eqref{opt:linear:weighted-sum} satisfies $\alpha \ge  \frac{\sigma_2^2}{\sigma_1^2 + \sigma_2^2}$.
%
%
\begin{corollary}
    \label{cor1}
    If Conjecture~\ref{conj:B} is true,
    when $\alpha <  \frac{\sigma_2^2}{\sigma_1^2 + \sigma_2^2}$, the  optimality loss of the objective value in  \eqref{opt:linear:weighted-sum} due to the choice of ${\bf g}_2 = [0, ...,  0, \sqrt{\eta_2}\sigma_2]^\top$ 
    in the worst case is $\eta_2 (\sigma_1^2 + \sigma_2^2) ( \frac{\sigma_2^2}{\sigma_1^2+\sigma_2^2} - \alpha )$.
\end{corollary}


The proof of Corollary~\ref{cor1} is provided in Appendix~\ref{app:linear:cor1}.\footnote{\rev{If Conjecture~\ref{conj:B} is not true, the optimality loss in  \eqref{opt:linear:weighted-sum} due to the choice of ${\bf g}_2 = [0, ...,  0, \sqrt{\eta_2}\sigma_2]^\top$ in the worst case is $(1-\alpha) \sigma_2^2 \eta_2$. 
The proof follows almost identically to that in Appendix~\ref{app:linear:cor1}, with the key difference being that the lower bound of $\nu_{\min}[{\bf B}]$ is zero from Proposition~\ref{pro:B_lower_upper}.}}
We note from Proposition \ref{pro:g:optimal} that there is no optimality loss when $\alpha \ge  \frac{\sigma_2^2}{\sigma_1^2 + \sigma_2^2}$.
We can extract further implications from Proposition \ref{pro:g:optimal}.
In the context of feedback utilization~\cite{schalkwijk1966coding, chance2011concatenated, butman1969general,palacio2021achievable}, one user transmits its message \textit{initially} and exchanges symbols containing the message and feedback information with the other user.
However, given that User 2 only transmits its message in the last channel use from Proposition \ref{pro:g:optimal}, User 2 does not utilize feedback. Consequently, User 1 does not function as a helper in providing feedback to User 2. 
Therefore, we can state the following remark.
%
\begin{remark}
    User 1 does not function as a helper when $\alpha \ge  \frac{\sigma_2^2}{\sigma_1^2 + \sigma_2^2}$ in \eqref{opt:linear:weighted-sum}. 
    \label{remark_helper}
\end{remark}

We further explore the implications of Remark~\ref{remark_helper}  in the context of the relationship between the weight coefficient $\alpha$ and the channel noise ratio ${\sigma_2^2}/({\sigma_1^2 + \sigma_2^2})$ within the framework  of weighted sum-power minimization in \eqref{opt:linear:weighted-sum}.
We consider a scenario where the inequality is completely fulfilled -- $\sigma_1^2$ is comparatively larger than  $\sigma_2^2$, and $\alpha$ is close to 1.
Under these conditions,
User 1 would not provide feedback because
delivering feedback across a noisy channel (characterized by the large $\sigma_1^2$) demands significant power, subsequently leading to a substantial increase in the weighted sum-power (due to the large $\alpha$) in \eqref{opt:linear:weighted-sum}.
In such a scenario, refraining from providing feedback would be advantageous in achieving a lower weighted sum-power.

The authors in  \cite{palacio2021achievable} introduced the concept of designating the user with lower channel noise as a helper and subsequently derived error exponents for GTWCs within this framework.
It's important to note that assigning the user with lower channel noise as a helper may not always be the optimal strategy for addressing the weighted sum-power minimization problem.
In this work, we investigate the relationship between the weight assigned in weighted sum-power minimization and the channel noise ratio. 
Our analysis reveals that even when a user (User 1) experiences lower channel noise (characterized by small $\sigma_1^2$), designating it as a helper might not be suitable in \eqref{opt:linear:weighted-sum} if the weight is larger than the channel noise ratio, i.e., $\alpha \ge  \frac{\sigma_2^2}{\sigma_1^2 + \sigma_2^2}$. 
Building upon the insights presented in \cite{palacio2021achievable}, our work extends the understanding of the helper concept to encompass weighted sum problems. 

Using Propositions~\ref{pro:f2n} and \ref{pro:g:optimal}, we next aim to simplify our optimization problem in \eqref{opt:linear:weighted-sum}.
In our optimization, we consider $\alpha \ge  \frac{\sigma_2^2}{\sigma_1^2 + \sigma_2^2}$.
From Proposition~\ref{pro:g:optimal}, we have ${\bf g}_2 = [0, ...,  0, \sqrt{\eta_2}\sigma_2]^\top$ as an optimal solution, which always satisfies $\text{SNR}_2  = \eta_2$ regardless of other variables.
Thus, we can remove the dependency of the constraint for $\text{SNR}_2$ in \eqref{opt:linear:weighted-sum}.
Further, to make \eqref{opt:linear:weighted-sum} more tractable, 
we first perform a matrix decomposition on ${\bf Q}_1$ and obtain ${\bf Q}_1 = {\bf Q}_{1,\text{sqrt}}^\top {\bf Q}_{1,\text{sqrt}}$ in the same way for the decomposition on ${\bf Q}_2$ as described below \eqref{opt:g2}.
Note that ${\bf Q}_{1,\text{sqrt}} = {\bf Q}_{1,\text{sqrt}}^\top$.
Since ${\bf Q}_1$ is invertible, ${\bf Q}_{1,\text{sqrt}}$ is also invertible.
Then,
we define ${\bf q}_1 = {\bf Q}_{1,\text{sqrt}}^{-1} {\bf g}_1$,
which implies 
    $\text{SNR}_1 = \| {\bf q}_1 \|^2
$ and 
$\| {\bf g}_1 \|^2 = {\bf q}^\top_1 {\bf Q}_1 {\bf q}_1$.
Consequently, we rewrite the transmit powers in \eqref{eq:power1} and \eqref{eq:power2} as
\begin{align}
    \mathbb{E}_{\rev{m_1,m_2,{\bf n}_1, {\bf n}_2}}\big[\|{\bf x}_1\|^2\big] 
    &= 
    {\bf q}^\top_1 {\bf Q}_1 {\bf q}_1 +
     \|{\bf F}_1 {\bf F}_2 \|_F^2 \sigma_1^2
    + \| {\bf F}_1 \|_F^2 \sigma_2^2
   \nonumber  \\
    &= 
    \|{\bf q}^\top_1 ({\bf I} + {\bf F}_1{\bf F}_2) \|^2 \sigma_1^2
    + 
    \| {\bf q}^\top_1 {\bf F}_1 \|^2 \sigma_2^2
    + \|{\bf F}_1 {\bf F}_2 \|_F^2 \sigma_1^2
    + \| {\bf F}_1 \|_F^2 \sigma_2^2,
    \label{eq:power1:q1}
    \\
    \mathbb{E}_{\rev{m_1,m_2,{\bf n}_1, {\bf n}_2}}\big[\|{\bf x}_2\|^2\big] 
    & =  
    \| {\bf g}_2 \|^2
    + \|{\bf F}_2 {\bf Q}_{1,\text{sqrt}} {\bf q}_1\|^2
    + 
    \| {\bf F}_2({\bf I} + {\bf F}_1{\bf F}_2 ) \|_F^2 \sigma_1^2
    + \| {\bf F}_2 {\bf F}_1
    \|_F^2 \sigma_2^2,
    \label{eq:power2:q1}
\end{align}
where the fact that ${\bf F}_1 {\bf g}_2 = {\bf 0}$ is used.
Finally, we simplify our optimization in \eqref{opt:linear:weighted-sum} as
\begin{align}
   & \underset{{\bf q}_1, {\bf F}_1, {\bf F}_2 \in \mathcal{F}_2}
    {\text{minimize}} & & 
 \alpha \mathbb{E}_{\rev{m_1,m_2,{\bf n}_1, {\bf n}_2}} \big[\|{\bf x}_1 \|^2\big] + (1-\alpha) \mathbb{E}_{\rev{m_1,m_2,{\bf n}_1, {\bf n}_2}} \big[\|{\bf x}_2 \|^2\big]
\nonumber
\\
& \text{subject to}
& &  
\| {\bf q}_1\|^2 = \eta_1.
\label{opt:linear:weighted-sum:3vars} 
\end{align}


\subsection{Optimization Solution for Weighted Sum-Power Minimization}
\label{ssec:linear:opt:weighted-sum:sol}


To solve the problem in \eqref{opt:linear:weighted-sum:3vars},
we divide it into two sub-problems and solve them alternately through a series of iterations.\footnote{Although this iterative optimization process does not guarantee convergence, it simplifies \eqref{opt:linear:weighted-sum:3vars} into two more tractable sub-problems. 
We demonstrate the convergence empirically through numerical experiments in Sec.~\ref{ssec:validation_linear_coding}.}
The first sub-problem is to solve for ${\bf q}_1$ and ${\bf F}_1$ given that ${\bf F}_2 \in \mathcal{F}_2$ is fixed, and the second sub-problem is to solve  for ${\bf F}_2 \in \mathcal{F}_2$ assuming ${\bf q}_1$ and ${\bf F}_1$ are fixed.

\subsubsection{First sub-problem for obtaining ${\bf q}_1$ and ${\bf F}_1$}
\label{sssec:problem1}


We assume a fixed value for ${\bf F}_2 \in \mathcal{F}_2$. 
%
We first show that $\mathbb{E}_{\rev{m_1,m_2,{\bf n}_1, {\bf n}_2}}\big[\|{\bf x}_2\|^2\big] $ is upper bounded by the sum of the scaled version of $\mathbb{E}_{\rev{m_1,m_2,{\bf n}_1, {\bf n}_2}}\big[\|{\bf x}_1\|^2\big] $ and some constant terms as follows:
\begin{align}
    & \mathbb{E}_{\rev{m_1,m_2,{\bf n}_1, {\bf n}_2}}[\|{\bf x}_2\|^2] 
    \overset{(a)}{=}  \| {\bf g}_2 \|^2 + \|{\bf F}_2 {\bf Q}_{1,\text{sqrt}} {\bf q}_1\|^2
    + 
    \| {\bf F}_2 \|_F^2 \sigma_1^2 
    + \| {\bf F}_2 {\bf F}_1 {\bf F}_2 \|_F^2 \sigma_1^2
    + \| {\bf F}_2 {\bf F}_1
    \|_F^2 \sigma_2^2
    \nonumber  \\
    & \hspace{.5cm}~~~~~~~  \le \| {\bf g}_2 \|^2 
    + \| {\bf F}_2 \|_F^2 \sigma_1^2 
   + f_{2,\max}^2 \big( \|{\bf Q}_{1,\text{sqrt}} {\bf q}_1\|^2 + \| {\bf F}_1 {\bf F}_2\|_F^2 \sigma_1^2 + \| {\bf F}_1 \|_F^2 \sigma_2^2 \big)
   \hspace{-2mm}  \nonumber \\
    & \hspace{.5cm} ~~~~~~~ = \| {\bf g}_2 \|^2 
    + 
    \| {\bf F}_2 \|_F^2 \sigma_1^2 
    + f_{2,\max}^2 \mathbb{E}_{\rev{m_1,m_2,{\bf n}_1, {\bf n}_2}}[\|{\bf x}_1\|^2],
    \label{eq:power2_org}
\end{align}
where $f_{2,{\rm max}}^2 = \underset{i=2,...,N-1}{\max} f_{2,i}^2$. To obtain the equality $(a)$ in \eqref{eq:power2_org}, we use the fact that 
$\textrm{tr}({\bf F}_2 {\bf F}_1 {\bf F}_2 {\bf F}_2^\top) = 0$ for ${\bf F}_2 \in \mathcal{F}_2$.
To derive the inequality in \eqref{eq:power2_org}, we exploit the property that $ \|{\bf F}_2 {\bf y}\|^2 \le f_{2,{\rm max}}^2 \|{\bf y}\|^2$ for any vector ${\bf y} \in \mathbb{R}^{N \times 1}$ and $\|{\bf F}_2 {\bf Y}\|_F^2 \le f_{2,{\rm max}}^2 \|{\bf Y}\|_F^2$ for any matrix ${\bf Y} \in \mathbb{R}^{N \times N}$ when ${\bf F}_2 \in \mathcal{F}_2$.
Accordingly, we obtain an upper bound on the objective function in \eqref{opt:linear:weighted-sum:3vars} as 
\begin{equation}
    { (1-\alpha) (\| {\bf g}_2 \|^2 
    + \| {\bf F}_2 \|_F^2 \sigma_1^2 )} + 
    (\alpha + f_{2,\max}^2 (1-\alpha)) \mathbb{E}_{\rev{m_1,m_2,{\bf n}_1, {\bf n}_2}} \big[\|{\bf x}_1 \|^2\big].
    \label{eq:Ex2_ineq}
\end{equation}

In the first sub-problem, instead of solving  \eqref{opt:linear:weighted-sum:3vars} directly, 
we aim to minimize the upper bound of the objective function  in \eqref{eq:Ex2_ineq}.\footnote{Although using the
upper bound as a surrogate objective may result in sub-optimal solutions, we employ it for tractability
in optimization.}
Since the other terms in \eqref{eq:Ex2_ineq} are constants except for $\mathbb{E}_{\rev{m_1,m_2,{\bf n}_1, {\bf n}_2}} \big[\|{\bf x}_1 \|^2\big]$, the first sub-problem is then reduced to
\begin{align}
   (\bm{\mathcal{{P}}_1}): \hspace{0.2cm} 
   & \underset{{\bf q}_1, {\bf F}_1}
    {\text{minimize}} & & 
 \mathbb{E}_{\rev{m_1,m_2,{\bf n}_1, {\bf n}_2}} \big[\|{\bf x}_1 \|^2\big]
\nonumber
\\
& \text{subject to}
& &  
\| {\bf q}_1\|^2 = \eta_1.
\label{opt:con:q1F1}
\end{align}



We will solve $\bm{\mathcal{{P}}_1}$
via (i) obtaining the optimal solution form of ${\bf F}_1$ in terms of ${\bf q}_1$, and then (ii) plugging the optimal solution form of ${\bf F}_1$ in $\mathbb{E}_{\rev{m_1,m_2,{\bf n}_1, {\bf n}_2}} \big[\|{\bf x}_1 \|^2\big]$ and solving the problem for ${\bf q}_1$.

\textbf{Solving for ${\bf F}_1$.} Note that ${\bf F}_1 \in \mathbb{R}^{N \times N}$ is a strictly lower triangular matrix given by
\begin{equation}
\begingroup 
\setlength\arraycolsep{2.5pt}
{\bf F}_1 = \begin{bmatrix}\setlength\arraycolsep{2pt}
0 &  0  & ... & 0\\
f_{1,2,1} & 0 & ... & 0 \\
\vdots & \ddots & \ddots & \vdots \\
f_{1,N,1}  & ... & f_{1,N,N-1} & 0
\end{bmatrix} = \begin{bmatrix}
0 & 0  & ... & 0\\
{\bf f}_{1,1} & 0 & ... & 0 \\
 & \ddots & \ddots  & \vdots \\
 &  &  {\bf f}_{1,N-1} & 0
\end{bmatrix},
\endgroup
\nonumber
\end{equation}
where ${\bf f}_{1,i} = [f_{1,i+1,i}, f_{1,i+2,i}, ..., f_{1,N,i}]^\top \in \mathbb{R}^{(N-i) \times 1}$, $i\in\{1,...,N-1\}$.
Considering ${\bf q}_{1} = [q_{1,1}, q_{1,2}, ..., q_{1,N}]^\top$, we define a vector that contains a portion of the entries of ${\bf q}_{1}$ as
\begin{equation}\label{eq:h_i}
    {\bf h}_{i} = [q_{1,i+1}, q_{1,i+2}, ..., q_{1,N}]^\top 
    \in \mathbb{R}^{(N-i) \times 1},
\end{equation}
where $i\in \{ 0,...,N-1\}$.
%
With the defined vectors $\{{\bf f}_{1,i}\}$ and $\{{\bf h}_{i}\}$, we can rewrite $\mathbb{E}_{\rev{m_1,m_2,{\bf n}_1, {\bf n}_2}} \big[\|{\bf x}_1 \|^2\big] $ in \eqref{eq:power1:q1} as
\begin{align}
    \mathbb{E}_{\rev{m_1,m_2,{\bf n}_1, {\bf n}_2}} \big[\|{\bf x}_1 \|^2\big] 
    =\sum_{i=1}^{N-1} \Phi_i({\bf f}_{1,i}) +  \sigma_1^2 \big( q_{1,N-1}^2 + q_{1,N}^2 \big),
    \label{eq:power1_hf}
\end{align}
where $\Phi_1({\bf f}_{1,1})\triangleq ({\bf h}_1^\top {\bf f}_{1,1} )^2 \sigma_2^2 
     +   {\bf f}_{1,1}^\top {\bf f}_{1,1} \sigma_2^2$ and 
     $\Phi_i({\bf f}_{1,i})\triangleq 
     \big( q_{1,i-1} + f_{2,i} {\bf h}_{i}^\top {\bf f}_{1,i} \big)^2 \sigma_1^2 
    + ( {\bf h}_{i}^\top {\bf f}_{1,i} )^2 \sigma_2^2  +{\bf f}_{1,i}^\top {\bf f}_{1,i}  (f_{2,i}^2 \sigma_1^2 + \sigma_2^2)$, 
    $i\in \{2,\cdots,N-1\}$.
The detailed derivation of $\Phi_i({\bf f}_{1,i})$, $i\in \{1,\cdots,N-1\}$, is presented in Appendix \ref{app:linear:phi}.

Using \eqref{eq:power1_hf}, our problem of interest (i.e., $\underset{{\bf F}_1}
    {\text{min}} ~ 
 \mathbb{E}_{\rev{m_1,m_2,{\bf n}_1, {\bf n}_2}} \big[\|{\bf x}_1 \|^2\big])$ can be
decomposed into $N-1$ independent sub-problems each in the form of $\underset{{\bf f}_{1,i}}
    {\text{min}} ~ 
 \Phi_i({\bf f}_{1,i})$,
 $i\in\{1,...,N-1\}$. 
Since each independent sub-problem is convex with respect to ${\bf f}_{1,i}$,
we find ${\bf f}_{1,i}$ optimally
by solving $\frac{\partial \Phi_i({\bf f}_{1,i})}{\partial {\bf f}_{1,i}} = {\bf 0}^\top$.
%
Obviously, we have ${\bf f}_{1,1} = {\bf 0}$. Also, for $i\in\{2,...,N-1\}$, we need to solve
\begin{align}
    & \frac{\partial \Phi_i({\bf f}_{1,i})}{\partial {\bf f}_{1,i}}  = (q_{1,i-1} + f_{2,i}{\bf h}_{i}^\top {\bf f}_{1,i})^\top {\bf h}_{i}^\top f_{2,i}\sigma_1^2
    + ({\bf h}_{i}^\top {\bf f}_{1,i})^\top {\bf h}_{i}^\top \sigma_2^2
    +  (  f_{2,i}^2 \sigma_1^2 + \sigma_2^2) {\bf f}_{1,i}^\top = {\bf 0}^\top.
    \label{eq:f1i_deriv}
\end{align}
We thus obtain 
\begin{align}
    {\bf f}_{1,i} = - \frac{ f_{2,i}\sigma_1^2}{f_{2,i}^2 \sigma_1^2 + \sigma_2^2}
    \frac{q_{1,i-1}}
    {1+ \| {\bf h}_{i}\|^2 }  {\bf h}_{i}, \quad i\in\{2,...,N-1\}.
    \label{eq:f1i_opt}
\end{align}
For a detailed derivation of \eqref{eq:f1i_opt} from \eqref{eq:f1i_deriv}, refer to Appendix \ref{app:linear:f1i}.

\textbf{Solving for ${\bf q}_1$.}
Putting the optimal solution of $\{{\bf f}_{1,i}\}_{i=2}^{N-1}$ obtained in \eqref{eq:f1i_opt} back into  \eqref{eq:power1_hf},    
we get
\begin{align}
    \mathbb{E}_{\rev{m_1,m_2,{\bf n}_1, {\bf n}_2}} \big[\|{\bf x}_1 \|^2\big] 
    & = 
    \sum_{i=2}^{N-1} \bigg[ 
    \bigg( q_{1,i-1} 
     - \frac{ f_{2,i}^2 \sigma_1^2}{f_{2,i}^2\sigma_1^2 + \sigma_2^2}
    \frac{q_{1,i-1} \|{\bf h}_{i}\|^2}
    {1+ \| {\bf h}_{i}\|^2 }  
    \bigg)^2 \sigma_1^2 
    + \bigg( \frac{ f_{2,i}\sigma_1^2}{f_{2,i}^2\sigma_1^2 + \sigma_2^2}
    \frac{q_{1,i-1} \|{\bf h}_{i}\|^2}
    {1+ \| {\bf h}_{i}\|^2 }  \bigg)^2 \sigma_2^2 
    \nonumber
    \\
    & \hspace{1cm}
     + \bigg( \frac{ f_{2,i}\sigma_1^2}{f_{2,i}^2 \sigma_1^2 + \sigma_2^2}
    \frac{q_{1,i-1} }
    {1+ \| {\bf h}_{i}\|^2 }  \bigg)^2 \|{\bf h}_{i}\|^2 (f_{2,i}^2\sigma_1^2 + \sigma_2^2) \bigg] 
    + \big( q_{1,N-1}^2 + q_{1,N}^2  \big) \sigma_1^2
       \nonumber \\
    & = \sum_{i=2}^{N-1}  \frac{\sigma_1^2 q_{1,i-1}^2 \big( f_{2,i}^2\sigma_1^2 + \sigma_2^2(1+ \|{\bf h}_{i}\|^2) \big)}
    {(f_{2,i}^2 \sigma_1^2 + \sigma_2^2)(1+ \|{\bf h}_{i}\|^2)} 
    + \big( q_{1,N-1}^2 + q_{1,N}^2  \big) \sigma_1^2.
\end{align}
Then, $\bm{\mathcal{{P}}_1}$ can be reduced to the following optimization problem:
\begin{align}
    & \underset{ {\bf q}_1 }{\text{minimize}} & & 
    \sum_{i=1}^{N-2}  \frac{\sigma_1^2 q_{1,i}^2 \big(  f_{2,i+1}^2\sigma_1^2 + \sigma_2^2(1+ \|{\bf h}_{i+1}\|^2) \big)}
    {(f_{2,i+1}^2\sigma_1^2 + \sigma_2^2)(1+ \|{\bf h}_{i+1}\|^2)} 
    + \big( q_{1,N-1}^2 + q_{1,N}^2  \big) \sigma_1^2
    \nonumber
\\
& \text{subject to}
& &  
\|{\bf q}_1\|^2  = \eta_1.
\label{eq:con:q1}
\end{align}
%
Defining $x_i = q_{1,i}^2 \ge 0$, we rewrite the objective function in \eqref{eq:con:q1} as
\begin{align}
    & \sum_{i=1}^{N-2}  
    \frac{f_{2,i+1}^2 \sigma_1^4 x_i }{{(f_{2,i+1}^2\sigma_1^2 + \sigma_2^2)(1+ x_{i+2} + ... + x_{N})}} 
    + \sum_{i=1}^{N-2} \frac{\sigma_1^2 \sigma_2^2 x_i}{f_{2,i+1}^2\sigma_1^2 + \sigma_2^2}
    + \sigma_1^2 (x_{N-1} + x_{N}),
    \label{eq:tmp1}
\end{align}
and the constraint in \eqref{eq:con:q1} as
    $\sum_{i=1}^{N} x_i  = \eta_1$.
%
Using the vector form of 
${\bf x} = [x_1, ..., x_N]^\top \in \mathbb{R}^{N \times 1}$, we can formulate an equivalent optimization problem to \eqref{eq:con:q1} as
\begin{align}
    & \underset{ {\bf x} }{\text{minimize}} & &  \sum_{i=1}^{N-1} \frac{{\bf u}_i^\top {\bf x} } { 1+ {\bf m}_i^\top {\bf x}}
    \nonumber
    \\
& \text{subject to}
& &  
{\bf 1}^\top {\bf x}  = \eta_1, \quad {\bf x} \ge {\bf 0},
\label{opt:con:x}
\end{align}
where ${\bf 1} = [1, ..., 1]^\top \in \mathbb{R}^{N \times 1}$ and ${\bf 0} = [0, ..., 0]^\top \in \mathbb{R}^{N \times 1}$. In~\eqref{opt:con:x}, ${\bf u}_i \in \mathbb{R}^{N \times 1}$ and ${\bf m}_i \in \mathbb{R}^{N \times 1}$, $i \in \{1,..., N-1\}$, are defined as
\begin{align}
    {\bf u}_i  &= \bigg[0, ..., 0, \underbrace{\frac{\vert f_{2,i+1} \vert^2 \sigma_1^4}{\vert f_{2,i+1} \vert^2\sigma_1^2 + \sigma_2^2}}_{ 
    i{\rm-th}}, 0, ..., 0 \bigg]^\top, \; i \in \{1,..., N-2\},
    \nonumber
    \\
    {\bf u}_{N-1} &= \bigg[ 
    \frac{\sigma_1^2 \sigma_2^2}{\vert f_{2,2} \vert^2\sigma_1^2 + \sigma_2^2}, ..., \frac{\sigma_1^2 \sigma_2^2}{\vert f_{2,N-1} \vert^2\sigma_1^2 + \sigma_2^2}, \sigma_1^2, \sigma_1^2
    \bigg]^\top,
    \nonumber
    \\
    {\bf m}_i &= [0, ..., 0, \underbrace{1}_{i{\rm-th}}, ..., 1]^\top, \; i \in \{1,..., N-2\}, \quad
    {\bf m}_{N-1} = [0, ..., 0]^\top,
    \nonumber
\end{align}
where ${\bf u}_i, {\bf m}_i \ge {\bf 0}$. Here, ${\bf u}_{N-1}$ captures the last two terms in \eqref{eq:tmp1}.
The problem in \eqref{opt:con:x} is a multi-objective linear fractional programming~\cite{freund2001solving}.
We thus can adopt commercial software~\cite{MatlabOTB} to solve this~problem.

\subsubsection{Second sub-problem for obtaining ${\bf F}_2$}
\label{sssec:problem2}

While fixing ${\bf q}_1$ and ${\bf F}_1$,
we formulate the second sub-problem as
\begin{align}
    & 
    (\bm{\mathcal{{P}}_2}): \hspace{0.2cm}  \underset{{\bf F}_2 \in \mathcal{F}_2}
    {\text{minimize}} & & 
 \alpha \mathbb{E}_{\rev{m_1,m_2,{\bf n}_1, {\bf n}_2}} \big[\|{\bf x}_1 \|^2\big] + (1-\alpha) \mathbb{E}_{\rev{m_1,m_2,{\bf n}_1, {\bf n}_2}} \big[\|{\bf x}_2 \|^2\big].
 \label{opt:obj:F2}
\end{align}
We aim to
minimize the objective of $\bm{\mathcal{{P}}_2}$ for each $f_{2,i}$, $i\in\{2,...,N-1\}$
by setting the derivative with respect to $f_{2,i}$ equal to zero.
Our methodology would yield a  sub-optimal solution given the non-convexity of the problem $\bm{\mathcal{{P}}_2}$.
We then obtain
\begin{align}
    & \alpha \frac{\partial \mathbb{E}_{\rev{m_1,m_2,{\bf n}_1, {\bf n}_2}} \big[\|{\bf x}_1 \|^2\big]}{\partial f_{2,i}}  + (1-\alpha) \frac{\partial \mathbb{E}_{\rev{m_1,m_2,{\bf n}_1, {\bf n}_2}} \big[\|{\bf x}_2 \|^2\big]}{\partial f_{2,i}} 
    = 
     2 \alpha \sigma_1^2 q_{1,i-1} {\bf h}_i^\top {\bf f}_{1,i}  + c_i f_{2,i},
     \label{eq:2ndsub:deriv}
\end{align}
where
\begin{align}
    & c_i  \triangleq 2 \alpha \sigma_1^2 \big( \vert {\bf h}_i^\top {\bf f}_{1,i} \vert^2    + \| {\bf f}_{1,i} \|^2  \big) +   2 (1-\alpha) p_{i-1}^2 + 2 (1-\alpha) \sigma_1^2 
    \nonumber
    \\
    & \hspace{.5cm} + 2 (1-\alpha) \sigma_1^2  \bigg( \sum_{j=i+1}^{N-1} f_{1,j,i}^2 f_{2,j+1}^2 + \sum_{k=2}^{i-2} f_{1,i-1,k}^2 f_{2,k}^2
    \bigg)
    + 2 \sigma_2^2 (1-\alpha) \sum_{j=1}^{i-2} f_{1,i-1,j}^2.
    \nonumber
\end{align}
The details of derivations used for  obtaining \eqref{eq:2ndsub:deriv} are presented in Appendix \ref{app:linear:deriv_f2}.
By setting the right-hand side of the equality in \eqref{eq:2ndsub:deriv} to zero, we obtain the solution for $f_{2,i}$ as
\begin{equation}
    f_{2,i} = -\frac{2 \alpha \sigma_1^2 q_{1,i-1} {\bf h}_i^\top {\bf f}_{1,i}}{c_i}, \quad i\in\{2,...,N-1\}.
    \label{eq:f2i}
\end{equation}

The pseudo-code of our iterative method to solve the overall optimization problem in \eqref{opt:linear:weighted-sum} is summarized in Algorithm \ref{al:linear:weighted_sum_power}.
We first set ${\bf g}_2 = [0, 0, ..., \sqrt{\eta_2} \sigma_2]^\top$ and ${f}_{2,N} = 0$, from Propositions~\ref{pro:f2n} and \ref{pro:g:optimal}, and initialize $\{ f_{2,i} \}_{i=2}^{N-1}$ randomly (in lines 3-4). 
Then, we solve the two sub-problems, discussed in Secs.~\ref{sssec:problem1} and ~\ref{sssec:problem2},
alternatively through a series of \textit{outer} iterations until the objective function value in \eqref{opt:linear:weighted-sum} converges (lines 5-18). 
For the first sub-problem (lines 6-9), we first solve the problem in \eqref{opt:con:x} with respect to ${\bf x} = [x_1, ..., x_N]^\top$ and obtain $q_{1,i} = \sqrt{x_i}$, $i \in \{1,...,N\}$. Then, we obtain the columns of ${\bf F}_1$, $\{ {\bf f}_{1,i} \}_{i=1}^{N-1}$, from \eqref{eq:f1i_opt}, using the values of $\{q_{1,i}\}_{i=1}^{N-1}$, and finally obtain ${\bf g}_1 = {\bf Q}_{1,\text{sqrt}} {\bf q}_1$ where ${\bf Q}_1$ is given in \eqref{eq:Q1}.
For the second sub-problem (lines 10-14), we obtain $f_{2,i}$ sequentially for $i \in \{2,...,N-1\}$ by \eqref{eq:f2i} and conduct several \textit{inner} iterations in obtaining $\{f_{2,i}\}_{i=2}^{N-1}$ until the objective function value in \eqref{opt:linear:weighted-sum} converges.

 \begin{algorithm}[t]
 \caption{Linear Encoding Schemes for Weighted Sum-Power Minimization in GTWC}
 \label{al:linear:weighted_sum_power}
 \begin{algorithmic}[1]
 \footnotesize
\State \textbf{Input.} 
Power constraint $NP$, noise variances $\sigma_1^2$ and $\sigma_2^2$, number of bits $K_1$ and $K_2$, number of channel uses $N$,  target SNRs $\eta_1$ and $\eta_2$, weight value $\alpha$, and stopping value $\epsilon$.
\State \textbf{Output.} 
Linear encoding schemes, ${\bf g}_1$, ${\bf F}_1$, ${\bf g}_2$, and ${\bf F}_2$.
\State Obtain the optimal solutions,
${\bf g}_2 = [0, 0, ..., \sqrt{\eta_2} \sigma_2]^\top$ and ${f}_{2,N} = 0$, from Propositions~\ref{pro:f2n} and \ref{pro:g:optimal}.
  \State \multiline{ Randomly generate $\{ f_{2,i} \}_{i=2}^{N-1}$. Calculate the objective function value $s_{\rm new}$ of \eqref{opt:linear:weighted-sum} with ${\bf g}_1$, ${\bf F}_1$, ${\bf g}_2$, and ${\bf F}_2$. Set $s_{\rm old} = 0$.}
  \While {$\vert s_{\rm new} - s_{\rm old} \vert > \epsilon$}
    \State \hspace{-3.8mm}  $\bullet$ \textit{\textbf{Sub-problem 1. Obtain ${\bf g}_1$ and ${\bf F}_1$}} 
    \State \multiline{Solve the problem in \eqref{opt:con:x} for ${\bf x} = [x_1, ..., x_N]^\top$ and obtain $q_{1,i} = \sqrt{x_i}$, $i \in \{1,...,N\}$.}
    \State Obtain the columns of ${\bf F}_1$, $\{ {\bf f}_{1,i} \}_{i=1}^{N-1}$, from \eqref{eq:f1i_opt}, using the values of $\{q_{1,i}\}_{i=1}^{N-1}$.
    \State Obtain ${\bf g}_1 = {\bf Q}_{1,\text{sqrt}} {\bf q}_1$ where ${\bf Q}_1$ is given in \eqref{eq:Q1}.
     \State  \hspace{-3.8mm} \textit{$\bullet$ \textbf{Sub-problem 2. Obtain ${\bf F}_2$}} 
     \State Calculate the objective function value $\nu_{\rm new}$ of \eqref{opt:linear:weighted-sum} with ${\bf g}_1$, ${\bf F}_1$, ${\bf g}_2$, and ${\bf F}_2$. Set $\nu_{\rm old} = 0$.
    \While {$\vert  \nu_{\rm new} - \nu_{\rm old} \vert > \epsilon$} 
        \State \multiline{ Obtain $f_{2,i}$ sequentially for $i \in \{2,...,N-1\}$ by \eqref{eq:f2i}.
        \\
        $\nu_{\rm old} \leftarrow \nu_{\rm new}$.
        \\
        Calculate the objective function value $\nu_{\rm new}$ of \eqref{opt:linear:weighted-sum} with the updated $\{ f_{2,i} \}_{i=2}^{N-1}$.}
    \EndWhile
    \State \hspace{-3.8mm} $\bullet$ \textit{\textbf{Update values for stopping criterion}} 
    \State $s_{\rm old} \leftarrow s_{\rm new}$.
    \State \multiline{ Calculate the objective function value $s_{\rm new}$ of \eqref{opt:linear:weighted-sum} with the updated ${\bf g}_1$, ${\bf F}_1$, and ${\bf F}_2$.}
     \EndWhile
 \end{algorithmic}
 \end{algorithm}

\subsection{Optimization Solution for Sum-Error Minimization}
\label{ssec:linear:opt:sum-error:sol}


We solved the weighted sum-power (WSP) minimization problem~\footnote{We solve it in a sub-optimal manner by trading off optimality for tractability. We note that optimality may be compromised due to the simplification of ${\bf F}_2$ in~\eqref{eq:F2}, the case when $\alpha < \frac{\sigma_2^2}{\sigma_1^2 + \sigma_2^2}$ (shown in Corollary~\ref{cor1}), and the iterative optimization process. The performance loss in WSP minimization is further analyzed in Sec.~\ref{ssec:validation_linear_coding}} (in \eqref{opt:linear:weighted-sum}) in Sec.~\ref{ssec:linear:opt:weighted-sum}, and presented the solution algorithm, Algorithm~\ref{al:linear:weighted_sum_power}.
We can express the input-output relationship of  Algorithm~\ref{al:linear:weighted_sum_power}  as
\begin{equation}
    ( {\bf g}_1, {\bf F}_1, {\bf g}_2, {\bf F}_2) = \text{Min-WSP} (\sigma_1^2, \sigma_2^2, N, \eta_1, \eta_2, \alpha),
\end{equation}
where $\eta_1$, $\eta_2$, and $\alpha$ are given. 
We note that the objective value in \eqref{opt:linear:weighted-sum} is the sum of the transmit powers, where the transmit powers, $\mathbb{E}_{\rev{m_1,m_2,{\bf n}_1, {\bf n}_2}} \big[\|{\bf x}_1 \|^2\big]$ and $\mathbb{E}_{\rev{m_1,m_2,{\bf n}_1, {\bf n}_2}} \big[\|{\bf x}_2 \|^2\big]$, can be calculated from the obtained solutions, ${\bf g}_1$, ${\bf F}_1$, ${\bf g}_2$, and ${\bf F}_2$, by \eqref{eq:power1}-\eqref{eq:power2}.
To explicitly indicate the dependency of the powers,  $\mathbb{E}_{\rev{m_1,m_2,{\bf n}_1, {\bf n}_2}} \big[\|{\bf x}_1 \|^2\big]$ and $\mathbb{E}_{\rev{m_1,m_2,{\bf n}_1, {\bf n}_2}} \big[\|{\bf x}_2 \|^2\big]$, on the values of $\eta_1$, $\eta_2$, and $\alpha$, we denote 
$ \mathbb{E}_{\rev{m_1,m_2,{\bf n}_1, {\bf n}_2}} \big[\|{\bf x}_1 \|^2\big]$ and $\mathbb{E}_{\rev{m_1,m_2,{\bf n}_1, {\bf n}_2}} \big[\|{\bf x}_2 \|^2\big]$ as
$P_1(\eta_1, \eta_2, \alpha)$ and $P_2(\eta_1, \eta_2, \alpha)$, respectively.



Our next goal is to solve the sum-error minimization problem in \eqref{opt:linear:SVEM} (equivalently, \eqref{opt:linear:sum-error}) by using the solution algorithm for the WSP minimization problem in \eqref{opt:linear:weighted-sum}, i.e., Algorithm~\ref{al:linear:weighted_sum_power}.
A naive approach is to first run  Algorithm~\ref{al:linear:weighted_sum_power} for every possible combinations of $(\eta_1, \eta_2, \alpha)$, and then find the feasible solution set $\mathcal{S}$ in \eqref{opt:linear:SVEM}. Then, we select the best solution of $( {\bf g}_1, {\bf F}_1, {\bf g}_2, {\bf F}_2)$ that minimizes the sum-error, 
$\mathcal{E}_1(\eta_1) + \mathcal{E}_2(\eta_2)$. 
However, the search space for $(\eta_1, \eta_2, \alpha)$ would be prohibitively large since the search space is continuous and three-dimensional.

To mitigate such complexities, we propose 
a method that reduces
 the three-dimensional search to a one-dimensional search only over $\eta_2$ through  two distinct one-dimensional search methods: (i) the bisection method for finding $\eta_1$ and (ii) the golden-section search method for searching over~$\alpha$.

\subsubsection{Reducing the search space for $\alpha$ via the golden-section search method~\cite{chong2013introduction}}
\label{sssec:alpha}


Given $\eta_1$ and $\eta_2$, we consider  solving  
\eqref{opt:linear:weighted-sum} for various $\alpha$ and  obtaining the solution of \eqref{opt:linear:max-of-powers:SNR} by selecting the best $\alpha$, denoted by $\alpha^\star$, and its corresponding solution, $( {\bf g}_1, {\bf F}_1, {\bf g}_2, {\bf F}_2)$.
However, searching over $\alpha$ 
requires lots of computations since $\alpha$ is continuous over the region $(0,1)$.
Therefore,  we develop an approximate searching method by assuming that the objective function  in \eqref{opt:linear:max-of-powers:SNR}, $\max \big\{ P_1(\eta_1, \eta_2, \alpha), P_2(\eta_1, \eta_2, \alpha)  \big\}$, is unimodal along $\alpha \in (0,1)$. 
This assumption is motivated by our observation of unimodality through numerical experiments, which is demonstrated in the simulation section, Sec.~\ref{ssec:validation_linear_coding}.
Unimodal functions have a unique maximal or minimal point, for which the golden-section search method can be employed to find the point~\cite{chong2013introduction}.
Thus,
we adopt the golden-section search method and obtain   
\begin{equation}
    \alpha^\star (\eta_1, \eta_2)  = \underset{\alpha \in (0,1)}{\min} \max \big\{ P_1(\eta_1, \eta_2, \alpha), P_2(\eta_1, \eta_2, \alpha)  \big\}.
    \label{eq:alpha_star}
\end{equation}
where $\alpha^\star (\eta_1, \eta_2)$ is obtained depending on the values of $\eta_1$ and $\eta_2$.
In summary, we obtained the solution of \eqref{opt:linear:max-of-powers:SNR} given $\eta_1$ and $\eta_2$ by applying the golden-section search method over $\alpha$ to the problem  \eqref{opt:linear:weighted-sum} without requiring exhaustive search over $\alpha$.


\subsubsection{Reducing the search space for $\eta_1$ via the bisection method~\cite{ferziger1981numerical}}
We recall that the key in solving \eqref{opt:linear:SVEM} is to find the feasible solution set $\mathcal{S}$, which requires 
solving \eqref{opt:linear:max-of-powers:SNR} over all possible values of ($\eta_1$, $\eta_2$) and finding the best solution.
However, the computational complexity for searching over continuous two-dimensional space is prohibitive. 
Therefore, we use the fact that 
the objective function of \eqref{opt:linear:max-of-powers:SNR} is increasing over $\eta_1$ from Lemma \ref{lemma:max_inc}, and utilize a one-dimensional search method for finding $\eta_1$ given $\eta_2$.
Furthermore, we are interested in finding $\eta_1$  such that the max-power reaches the power constraint $NP$. If the max-power is less than $NP$, we can always find a higher value of $\eta_1$ by increasing the max-power to $NP$ given $\eta_2$ by Lemma~\ref{lemma:max_inc}. The higher value of $\eta_1$ (i.e., $\text{SNR}_1$) with the same value of $\eta_2$ will result in a lower sum-error value in \eqref{opt:linear:SVEM}.
When seeking a particular functional value and its corresponding point in an increasing function, we can utilize the bisection method as an efficient one-dimensional search approach~\cite{ferziger1981numerical}.
Hence, 
we apply the bisection method to find $\eta^\star_1(\eta_2)$, which depends on the value of $\eta_2$, such that $u(\eta^\star_1(\eta_2))=0$, where 
\begin{equation}
    u(\eta_1)= \max \{ P_1(\eta_1, \eta_2, \alpha^\star(\eta_1, \eta_2)), P_2(\eta_1, \eta_2, \alpha^\star(\eta_1, \eta_2))  \} - N P.
    \label{eq:bisection_func}
\end{equation}
Note that $u(\eta_1)$ is an increasing function with respect to $\eta_1$ from Lemma \ref{lemma:max_inc}.
We then obtain $\eta_1^\star(\eta_2)$ such that $ \max \{ P_1(\eta_1^\star(\eta_2), \eta_2, \alpha^\star(\eta_1^\star, \eta_2)), P_2(\eta_1^\star(\eta_2), \eta_2, \alpha^\star(\eta_1^\star, \eta_2))  \}= N P$. 

For each value of $\eta_2$ in the search space, we obtain the 
tuple $(\eta_1^\star(\eta_2), \eta_2, \alpha^\star(\eta_1^\star, \eta_2))$ and find the best case that yields the minimum sum-error.
%
%
%
%
Overall, by adopting the bisection method and golden-section search method in solving \eqref{opt:linear:SVEM}, we reduced the three-dimensional search space to one-dimensional search only over $\eta_2$ and mitigated the complexity.
The overall algorithm is provided in Algorithm~\ref{al:linear:sum-error}.






 \begin{algorithm}[t]
 \caption{Linear Encoding and Decoding Schemes for Sum-Error Minimization in GTWC}
 \label{al:linear:sum-error}
 \begin{algorithmic}[1]
 \footnotesize
\State \textbf{Input.} 
Power constraint $NP$, noise variances $\sigma_1^2$ and $\sigma_2^2$, number of bits $K_1$ and $K_2$, and number of channel uses $N$.
\State \textbf{Output.} 
Linear encoding schemes ${\bf g}_1$, ${\bf F}_1$, ${\bf g}_2$, and ${\bf F}_2$, linear decoding schemes ${\bf w}_1$ and ${\bf w}_2$, and SNR values $\eta_1$ and $\eta_2$.
\State Set a finite search space for $\eta_2$ and pick a value for $\eta_2$ from the set. Set a Boolean variable $\text{SEARCH}=\text{True}$.
  \While {\text{SEARCH}}
    \State \multiline{Apply the bisection method to find $\eta_1^\star(\eta_2)$ such that $u(\eta_1^\star(\eta_2)) = 0$, 
    where $u(\eta_1) = \max \{ P_1(\eta_1, \eta_2, \alpha^\star(\eta_1,\eta_2)),$ $P_2(\eta_1, \eta_2, \alpha^\star(\eta_1,\eta_2))  \} - NP$ in \eqref{eq:bisection_func}.
    Here, $\alpha^\star(\eta_1,\eta_2)$ is obtained by \eqref{eq:alpha_star} through running Algorithm \ref{al:linear:weighted_sum_power} over $\alpha$, i.e., $( {\bf g}_1, {\bf F}_1, {\bf g}_2, {\bf F}_2)$ = \text{Min-WSP}($\sigma_1^2$, $\sigma_2^2$, $N$, $\eta_1$, $\eta_2$, $\alpha$), and calculating 
    $P_1(\eta_1, \eta_2, \alpha)$ and $P_2(\eta_1, \eta_2, \alpha)$ from \eqref{eq:power1}-\eqref{eq:power2} with the obtained solution $( {\bf g}_1, {\bf F}_1, {\bf g}_2, {\bf F}_2)$. 
    }
    \State Set an unexplored point in the search space of $\eta_2$. Once all points are searched, set $\text{SEARCH}=\text{False}$.
    \EndWhile 
    \State Find the tuple of $( {\bf g}_1, {\bf F}_1, {\bf g}_2, {\bf F}_2, \eta_1, \eta_2)$ that minimizes  $\mathcal{E}_1(\eta_1) + \mathcal{E}_2(\eta_2)$.
    \State Obtain the decoding schemes ${\bf w}_1$ and ${\bf w}_2$ by using ${\bf g}_1$, ${\bf F}_1$, ${\bf g}_2$, and ${\bf F}_2$ from \eqref{eq:linear_coding:w1} and \eqref{eq:linear_coding:w2}. 
 \end{algorithmic}
 \end{algorithm}

\subsection{Medium/Long Block-Lengths and Modulation Orders}
\label{ssec:linear:long}
We have focused on exchanging a message pair, $m_1$ and $m_2$, between the two users, where the encoding/decoding schemes 
are obtained 
by 
Algorithm~\ref{al:linear:sum-error}.
We recall that the message $m_i$ contains $K_i$ bits via $2^{K_i}$-ary PAM, $i \in \{1,2\}$.
Let us define a long block-length as $L_i$ to distinguish from a short block-length $K_i$, where $K_i$ denotes the number of \textit{processing bits} contained per each message $m_i$.
Given a total of $L_i$ bits, User $i$ has a total of $M_i = L_i/K_i$ message symbols, each containing $K_i$ bits and modulated with $2^{K_i}$-ary PAM.
We define the total number of channel uses for exchanging $L_1$ and $L_2$ bits between the two users as $N_\text{L}$ and the rate of User $i$ as  $r_i=L_i/N_\text{L}$. 
We consider that the values of $L_i$ and  $N_\text{L}$ are fixed, while the value of $K_i$ can be adjusted by User $i$.
We assume the same number of messages at the users, i.e., $M=M_1=M_2$, by properly choosing $K_i$ such that $L_1/K_1 = L_2/K_2$.
The number of channel uses for exchanging a message pair is then $N=N_\text{L}/M$. Through this \textit{successive transmission}, the users can exchange a total of $M$ message pairs over $N_\text{L}$ channel uses.


We provide new insights into how the value of $K_i$ affects  sum-error rates in two-way channels, given that  the rate $r_i=L_i/N_\text{L}$ is fixed, i.e., $L_i$ and $N_\text{L}$ are fixed.
Specifically, depending on the value of $K_i$, there is a trade-off between the minimum distance of the constellation and the feedback benefit in obtaining better sum-error rates.
First, as $K_i$ increases, the modulation order $2^{K_i}$ increases. Each messages is one of $2^{K_i}$ possible symbols, representing $K_i$ bits. For larger $K_i$, the minimum distance between the $2^{K_i}$ symbols decreases in signal space, making it more vulnerable to symbol errors.
Second, as $K_i$ increases, the number of messages $M_i=L_i/K_i$ decreases, and the number of channel uses per message symbol $N=N_\text{L}/M_i = K_i/r_i$ increases. The increasing number of channel uses for exchanging a pair of messages results in greater benefit from feedback, leading to an improvement in error performance for the recipient of the help. This would result in better sum-error rates.
In summary, as $K_i$ increases, the higher modulation order $2^{K_i}$ may lead to a loss in error performance by decreased minimum constellation distance, but the increased number of channel uses $N=K_i/r_i$ can improve sum-error performance due to greater feedback benefits. Therefore, sum-error performance varies depending on the values of~$K_i$.
The corresponding simulations are presented in Sec.~\ref{ssec:sim:short}.

Our linear coding schemes can be straightforwardly extended to incorporating concatenated coding by using our schemes for inner coding in a similar way  in feedback-enabled GOWCs proposed by~\cite{chance2011concatenated}.
In concatenated coding, outer coding can be implemented with any error correction codes, such as turbo codes or LDPC, where the $L_i$ source bits are encoded to $L^{\text{(c)}}_i$ bits with outer coding rate $r_\text{out} = L_i/L^{\text{(c)}}_i$.
For inner coding, we consider modulating $L^{\text{(c)}}_i$ bits to $M_i = L^{\text{(c)}}_i/K_i$ messages each with $2^{K_i}$-ary PAM, and each message pair is exchanged with our linear coding schemes.
Since this work focuses on proposing new frameworks for two-way coding, we evaluate our coding methodology with a simple coding setup without adopting concatenated coding.

\subsection{Enhancement of the Linear Coding Scheme through Alternate Channel Uses}
\label{ssec:linear:alternate}






In two-way channels, it is important to utilize feedback at one user to minimize sum-errors.
When the channel from User 2 to 1 is better than that of the reverse under certain conditions, User 2  acts as a helper to provide feedback to User 1~\cite{palacio2021achievable} and sends its message in the last channel use (by Proposition~\ref{pro:g:optimal}). In contrast, User 1 does not provide feedback to User 2 in this asymmetric channel (by Remark~\ref{remark_helper}). Therefore, it can be understood that User 1 functions similarly to a transmitter as a recipient of help in feedback-enabled one-way channels~\cite{schalkwijk1966coding, chance2011concatenated, butman1969general, kim2020deepcode, kim2023feedback}, while User 2 acts as a receiver providing feedback to User 1.
That is, to utilize feedback, User 1 waits for the transmit symbol to return, much like a transmitter in the feedback-enabled one-way channels. 
However, the number of channel uses for receiving feedback differs between feedback-enabled one-way and two-way channels.
In feedback-enabled one-way channels, it takes only a single channel use for the transmitter to send the transmit symbols and receive feedback about the transmit symbols from the receiver.
In contrast, in two-way channels, it takes two channel uses for each user to receive feedback about its transmit symbols. As a result, User 1 in two-way channels waits for the transmit symbol to return in two channel uses, which can lead to alternating channel uses. 

We observe this pattern of alternate channel uses at the users in two-way channels from the solutions generated by Algorithm~\ref{al:linear:sum-error}.
Specifically, when exchanging a message pair, we observe that Users 1 and 2 only utilize about half of the channel uses (refer to the case of $N=3$ in Fig.~\ref{fig:power_linear_no_alt}(\subref{fig:power_linear_no_alt(N3)}) and $N=5$ in Fig.~\ref{fig:power_linear_alt}(\subref{fig:power_linear(K=1,N=5)})).
More generally, when $N$ is odd, it has been shown that User 1 uses only odd-numbered channel uses, while User 2  uses only even-numbered channel uses and the last channel use. It is interesting to see that although we do not impose any constraints on the separation of the channel usages between the two users, the separation of the channel usages is caused by solving the optimization problem. Specifically, the solution ${\bf x}$ of the sub-problem \eqref{opt:con:x} has non-zero and zero values in alternating positions.

It is important to note that the separation of the channel uses results in a waste of channel uses. To address this inefficiency,
we introduce a strategy of using alternate channel uses for exchanging \textit{multiple} pairs of messages, so that the two users fully utilize the available channel uses. For ease of explanation, we consider the example of exchanging $M=2$ message pairs, $(m_1,m_2)$ and $(m'_1,m'_2)$, over $N_\text{L}=6$ channel uses under power constraint  $N_\text{L}P=6P$.
The naive approach for exchanging the two message pairs is the successive 
transmission, as discussed in Sec.~\ref{ssec:linear:long}; the first message pair is exchanged over the first $N=3$ channel uses under power constraint $3P$, and the second message pair is then exchanged over the next $N=3$ channel uses under power constraint $3P$. However, this approach results in a waste of channel uses. Therefore, we propose an alternate channel use strategy with the following steps.

\textbf{Step 1.} 
We first consider exchanging a pair of messages $(m_1,m_2)$ over $N=2N_\text{L}/M-1=5$ channel uses and power constraint $(N_\text{L}/M)P=3P$. Then, we obtain the transmit signals as
\begin{equation}
    {\bf x}_1 = [a_1, 0, b_1, 0, c_1], \quad {\bf x}_2 = [0, a_2, 0, b_2, c_2], 
    \label{eq:first_message}
\end{equation}
where $a_i$, $b_i$, and $c_i$, $i \in \{1,2\}$, are non-zero values.
To exchange another pair of messages $(m_1',m_2')$ over $N=5$ channel uses under power constraint $3P$, we have the transmit signals as
\begin{equation}
    {\bf x}_1' = [a_1', 0, b_1', 0, c_1'], \quad 
    {\bf x}_2' = [0, a_2', 0, b_2', c_2'],
    \label{eq:second_message}
\end{equation}
where $a'_i$, $b'_i$, and $c'_i$, $i \in \{1,2\}$, are non-zero values. Note that there are zero-valued entries in  \eqref{eq:first_message} and \eqref{eq:second_message}, implying that the corresponding channel uses are not utilized.

\textbf{Step 2.} 
To fully utilize the channel uses, we consider that 
the users exchange the two message pairs via an alternate transmission of the transmit signals in \eqref{eq:first_message} and \eqref{eq:second_message}. To this end, each User $i$, $i \in \{1,2\}$, transmits ${\bf x}_i$ in \eqref{eq:first_message}  and ${\bf x}_i'$ in \eqref{eq:second_message} in an alternate manner over $N_\text{L}=6$ channel uses.
Formally, User $i$ generates $\bar {\bf x}_i \in \mathbb{R}^{N+1}$ by putting the non-zero values of ${\bf x}_i$  and ${\bf x}_i'$ alternately in $\bar {\bf x}_i$.
From the above example, we obtain
\begin{equation}
    \bar {\bf x}_1 = [a_1, a_1', b_1, b_1', c_1, c_1'], \quad 
    \bar {\bf x}_2 = [c_2', a_2, a_2', b_2, b_2', c_2].
    \label{eq:alternate_channel}
\end{equation}
Note that the transmit signals in \eqref{eq:alternate_channel} satisfy the power constraint $N_\text{L}P=6P$ while utilizing $N_\text{L}=6$ channel uses.
This strategy also satisfies  causality of the encoding process.
We note that the transmit signal  $c_2'$ in $\bar {\bf x}_2$ of \eqref{eq:alternate_channel} is not restricted to follow a causal processing since it is only composed of a scaled version of the message $m_2'$ from Proposition~\ref{pro:g:optimal}, and thus it
can be transmitted at any channel use. The further simulations are provided in Sec.~\ref{ssec:sim:alternate}

We have so far explored the linear approach for two-way coding, which offer low computational complexity (further elaborated in Sec.~\ref{sec:comp}). However, the simplified linear coding approach may limit the flexibility in code design. To address this, we propose a non-linear coding scheme based on deep learning in the next section. 

\section{Learning-Based Coding via RNNs}
\label{sec:RNN}


In this section, we propose a non-linear coding methodology based on deep  learning frameworks which allow higher degrees of freedom in coding.
First, we adopt a state propagation-based encoding (Sec.~\ref{ssec:RNN:state_enc}), and then, we discuss the composition of our learning-based coding structure in two parts: encoding  (Sec.~\ref{ssec:RNN:encoding}) and decoding (Sec.~\ref{ssec:RNN:decoding}).
Afterwards, we discuss how to train our coding architecture and how to make an inference 
(Sec.~\ref{ssec:RNN:training}), and finally in Sec.~\ref{ssec:RNN:modulo}, we discuss a modulo-based approach to process long block-lengths of bits with the proposed RNN-based coding architecture. 

\subsection{State Propagation-Based Encoding}
\label{ssec:RNN:state_enc}

Recall the optimization problem \eqref{opt:obj}-\eqref{opt:const:power}, where the design variables are $2N$ different encoding functions: $\{f_{1,k}\}_{k=1}^N$ and  $\{f_{2,k}\}_{k=1}^N$, and two decoding functions, $g_1$ and $g_2$.
It is expected that the $N$ encoding functions at each user are correlated with one another, since the inputs used for encoding at each timestep in \eqref{eq:encoding} overlap.
%
%
%
%
Based on the correlation of encoding processes across time, we adopt a state propagation-based encoding technique~\cite{kim2023feedback},
where only two functions are used for encoding at each user:
(i) signal-generation and (ii) state-propagation.
By designing only these two functions instead of the $N$ encoding functions, the design complexity for encoding could be significantly reduced.

We first define the signal-generation function of User $i$ as $f_i: \mathbb{R}^{K_i+N_s+1} \to \mathbb{R} $.
We then re-write the encoding process in \eqref{eq:encoding} as
\vspace{-.5mm}
\begin{align}
    x_i[k] = f_i({\bf b}_i, y_i[k-1], {\bf s}_i[k]),
    \quad k \in \{1,...,N\},
    \label{eq:func_f}
\end{align}
where we assume $y_i[0]=0$. Here, ${\bf s}_i[k] \in \mathbb{R}^{N_s} $ is the state vector of size $N_s$, which propagates over time through the state-propagation function $h_i:\mathbb{R}^{K_i+N_s+1} \to \mathbb{R}^{N_s}$, which is given by
\begin{align}
    {\bf s}_i[k] = h_i({\bf b}_i, y_i[k-1], {\bf s}_i[k-1]), \quad k \in \{1,...,N\}.
    \label{eq:func_h}
\end{align}
In \eqref{eq:func_h}, the current state ${\bf s}_i[k]$ is updated from the prior state ${\bf s}_i[k-1]$ by incorporating ${\bf b}_i$ and $y_i[k-1]$. For the initial condition, we assume ${\bf s}_i[0]={\bf 0}$.
We note that the functional form  
$x_i[k] = f_{i,k}({\bf b}_i, y_i[1], ..., y_i[k-1])$ in \eqref{eq:encoding} is represented by the the two equations \eqref{eq:func_f} and \eqref{eq:func_h}; The current state vector ${\bf s}_i[k]$ as an input in \eqref{eq:func_f} is a function of the previous state ${\bf s}_i[k-1]$ in \eqref{eq:func_h}, where ${\bf s}_i[k-1]$ contains $y_i[k-2]$ as an input. 
By the recursive nature of the
function $h_i$ in \eqref{eq:func_h}, all previous receive signals, $y_i[1], ..., y_i[k-1]$ are captured for encoding as in \eqref{eq:encoding}.
This encoding model in \eqref{eq:func_f}-\eqref{eq:func_h} can be seen as a general and non-linear extension of the state-space model used for linear encoding in feedback-enabled systems~\cite{elia2004bode}.

By using this technique, we only need to construct two encoding functions for each User $i$, $f_i$ and $h_i$ -- instead of $N$ distinct functions $\{f_{i,k}\}_{k=1}^N$ in \eqref{eq:encoding} -- and the decoding function $g_i$. 
We thus re-write the optimization problem in \eqref{opt:obj}-\eqref{opt:const:power} as
\begin{align}
&  \underset{ f_1, f_2,  h_1,  h_2,  g_1,  g_2} {\text{minimize}} & & 
\mathcal{E}_1(f_1, f_2,  h_1,  h_2,  g_2) + \mathcal{E}_2(f_1, f_2,  h_1,  h_2,  g_1)
\label{opt:obj:state-based}
\\
& \hspace{3mm} \text{subject to}
& &  
\mathbb{E}_{ {\bf b}_1, {\bf b}_2, {\bf n}_1, {\bf n}_2 } \bigg[ \sum_{k=1}^N  x_i^2[k]   \bigg] \le N P, \quad i \in \{1,2\},
\label{opt:const:power:state-based}
\end{align}
where the dependency on the error probability $\mathcal{E}_i$, $i \in \{1,2\}$, has been modified from \eqref{opt:obj} to \eqref{opt:obj:state-based} for state propagation-based encoding.

The problem 
\eqref{opt:obj:state-based}-\eqref{opt:const:power:state-based} is still non-trivial since the encoding/decoding functions, $f_i$, $h_i$, and  $g_i$, can take arbitrary forms. 
A proper design of such non-linear functions can potentially provide
higher degrees of flexibility in encoders' and decoders' behavior, as compared to the linear coding discussed in the previous section.
To design such non-linear functions, in this section, we develop a learning-based coding architecture, which will be discussed in two parts: encoding (Sec.~\ref{ssec:RNN:encoding}) and decoding (Sec.~\ref{ssec:RNN:decoding}).



\begin{figure}[t]
    \centering
    \includegraphics[width=.9\linewidth]{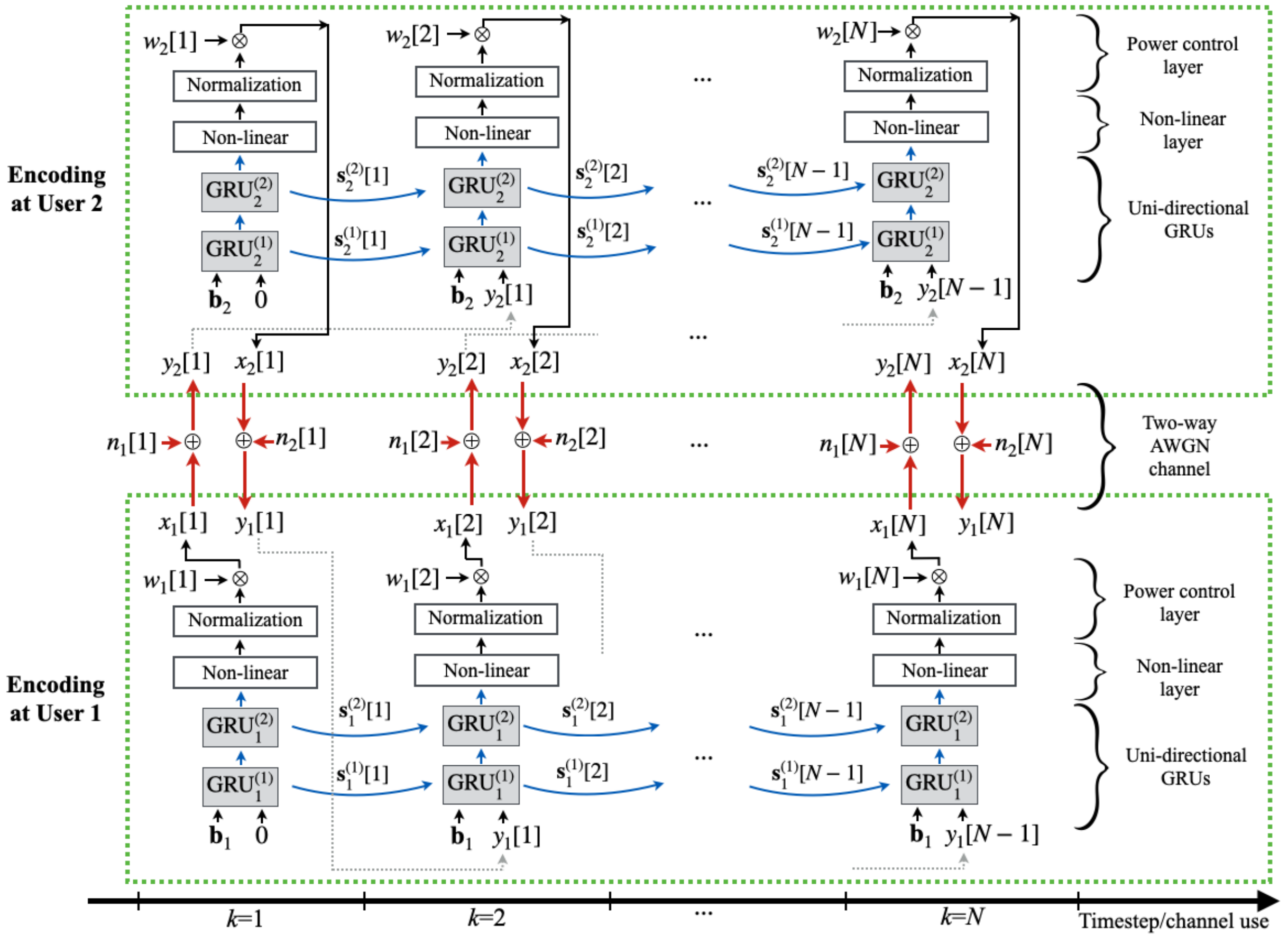}
    \caption{Our proposed RNN-based encoding architecture for  two-way channels. A pair of interactive RNNs captures the coupling effect caused by the encoding processes of the two users.}
    \label{fig:RNN_encoders}
\end{figure}


\subsection{Encoding}
\label{ssec:RNN:encoding}

We follow the state propagation-based encoding  approach discussed in Sec.~\ref{ssec:RNN:state_enc}. 
The concept of state propagation over time motivates us to employ RNNs as components of our learning architecture, similar to \cite{kim2023feedback}. 
Specifically, we utilize gated recurrent units (GRUs), a type of RNNs, to effectively capture the long-term dependency of time-series information~\cite{goodfellow2016deep}.
For User $i$, $i\in \{1,2\}$, the state-propagation function $h_i$ in \eqref{eq:func_h} consists of two layers of GRUs, while the signal-generation function $f_i$ in \eqref{eq:func_f} consists of a non-linear layer and a power control layer in  sequence. 
The overall encoding architecture of the two users and their interaction through the two-way channel are depicted in Fig.~\ref{fig:RNN_encoders}.

\subsubsection{GRUs for state propagation}
We adopt two layers of uni-directional GRUs at each user to capture the time correlation of the receive signals in a causal manner. 
Formally, we represent the input-output relationship at each layer at time $k$ at User $i$ as
\begin{align}
    {\bf s}_i^{(1)}[k] & = \text{GRU}_i^{(1)}({\bf b}_i, y_i[k-1], {\bf s}_i^{(1)}[k-1]),
    \label{eq:s1}
    \\
    {\bf s}_i^{(2)}[k] & = \text{GRU}_i^{(2)}({\bf s}_i^{(1)}[k], {\bf s}_i^{(2)}[k-1]),
    \label{eq:s2}
\end{align}
where $\text{GRU}_i^{(\ell)}$  represents a functional form of GRU processing at layer $\ell$ of User $i$ and ${\bf s}_i^{(\ell)}[k] \in \mathbb{R}^{N_{i,\ell}^{(\text{enc})}}$ is the state vector obtained by $\text{GRU}_i^{(\ell)}$ at time $k$, where $i \in \{1,2\}$, $\ell \in \{1,2\}$, and $k \in \{1,...,N\}$. 
For the initial conditions, we assume ${\bf s}_i^{(\ell)}[0]={\bf 0}$,  $i \in \{1,2\}$, $\ell \in \{1,2\}$.

Equations \eqref{eq:s1}-\eqref{eq:s2} can be represented as a functional form of the state propagation-based encoding in \eqref{eq:func_h}. By defining the overall state vector as ${\bf s}_i[k] = [{\bf s}_i^{(1)}[k], {\bf s}_i^{(2)}[k]]$, we obtain ${\bf s}_i[k] = h_i({\bf b}_i, y_i[k-1], {\bf s}_i[k-1])$, where $h_i$ denotes the process of two layers of GRUs in \eqref{eq:s1}-\eqref{eq:s2}.
Note that ${\bf s}_i[k]$ propagates over time through the GRUs by incorporating the current input information into its state.
Because the bit vector ${\bf b}_i$ with length $K_i$ is handled as a \textit{block} to generate transmit signals with any length $N$, 
our method is flexible enough to support any rate $r_i = K_i/N$.
Furthermore, the GRUs are interactive between the users since each user's GRU incorporates the previously received symbol as input, as shown in Fig.~\ref{fig:RNN_encoders}. The pair of interactive GRUs effectively captures the interplay of the users' encoding processes.

\subsubsection{Non-linear layer}
We adopt an additional non-linear layer at the output of the GRUs. The state vector at the last layer of the GRUs, i.e., ${\bf s}_i^{(2)}[k]$, is taken as an input to this additional non-linear layer.
Formally, we can represent the process of the non-linear layer at User $i$ as
\begin{align}
    \tilde x_i[k] = \phi( {\bf w}_{\text{enc},i}^T {\bf s}_i^{(2)}[k] + {b}_{\text{enc},i}  ), \quad k \in \{1,...,N\},
    \label{eq:enc:non-linear}
\end{align}
where ${\bf w}_{\text{enc},i} \in \mathbb{R}^{N_{i,2}^{(\text{enc})}}$ and ${b}_{\text{enc},i} \in \mathbb{R}$ are the trainable weights and biases, respectively, and $\phi:
\mathbb{R} \to \mathbb{R}$ is an activation function. In this work, we employ the hyperbolic tangent for $\phi$~\cite{goodfellow2016deep}.

Since $\tilde x[k]$ (as the output of the hyperbolic tangent) ranges in $(-1,1)$, it is possible to
use a scaled version of $\tilde x[k]$, i.e., $\sqrt{P} \tilde x[k]$, as a transmit symbol that satisfies the power constraint $\sum_{k=1}^N  ( \sqrt{P} \tilde x[k] )^2 \le N P$.
However, this does not ensure maximum or efficient utilization of the transmit power budget. Power control over the sequence of transmit signals is essential in the design of encoding schemes using feedback signals in order to achieve robust error performance, e.g., in feedback-enabled communications~\cite{schalkwijk1966coding, chance2011concatenated,kim2020deepcode, kim2023feedback}.

\subsubsection{Power control layer} We introduce a power control layer to optimize for the power distribution, while  satisfying the power constraint in \eqref{opt:const:power:state-based}. This layer consists of two sequential modules: normalization and power-weight multiplication.
The transmit symbol of User $i$ at time $k$
is then generated by
\begin{align}
    {x}_i[k] = w_i[k]  \gamma^{(J)}_{i,k}(\tilde x_i[k]), \quad k \in \{1,...,N\},
    \label{eq:norm_power}
\end{align}
where 
$w_i[k]$ is a trainable power weight satisfying $\sum_{k=1}^N w^2_i[k] = N P$, $i \in \{1,2\}$, and
$\gamma^{(J)}_{i,k}: \mathbb{R} \rightarrow \mathbb{R}$ is a normalization function applied to $\tilde x_i[k]$ in a form of 
$\gamma_{i,k}^{(J)}(x) = (x-m_{i,k}(J))/d_{i,k}(J)$. Here, $m_{i,k}(J)$ and $d^2_{i,k}(J)$ are the sample mean and sample variance of $x$ at User $i$ at time $k$ calculated from the data with size $J$.


Through the power control layer, 
the power weights $\{w_i[k]\}_{i,k}$ are optimized via training in a way that minimizes the sum of errors in \eqref{opt:obj:state-based}.
At the same time, the power constraint in \eqref{opt:const:power:state-based} should be satisfied.
However, satisfying the power constraint in 
an (ensemble) average sense 
is non-trivial because the distributions of $\{x_i[k]\}_{k=1}^N$ are unknown.
Therefore, we approach it in an empirical sense as adopted in \cite{kim2023feedback}:
(i) During training, we use standard batch normalization; we normalize ${\tilde x}_i[k]$ with the sample mean and sample variance calculated from each batch of data with size $N_\text{batch}$ at each $k$. 
Note that $N_\text{batch}$ denotes the size of batch used in a single iteration during training, and $J$ represents the total available training data.
(ii) After training, we calculate and save the sample mean $m_{i,k}(J)$ and sample variance $d^2_{i,k}(J)$ from the entire training data with size $J$.
(iii) For inference, we use the saved mean and variance for normalization.

In the following lemma, we show that the above procedure guarantees satisfaction of the equality power constraint 
in an asymptotic sense
with a large number of training data used for normalization.

\begin{lemma}
\label{lemma:RNN:power}
Given the power control layer at User $i$ in \eqref{eq:norm_power}, the power constraint in \eqref{opt:const:power:state-based}  converges to $N P$ almost surely, i.e., $\mathbb{E}_{ {\bf b}_1, {\bf b}_2, {\bf n}_1, {\bf n}_2 } \big[ \sum_{k=1}^N  x_i^2[k]   \big]  \xrightarrow{a.s.} N P$, as the number of training data $J$ used for the normalization in \eqref{eq:norm_power} tends to infinity.
\end{lemma}
The proof of Lemma~\ref{lemma:RNN:power} is provided in Appendix~\ref{app:RNN:lemma3}.

\subsection{Decoding}
\label{ssec:RNN:decoding}

\begin{figure}[t]
    \centering
    \includegraphics[width=.9\linewidth]{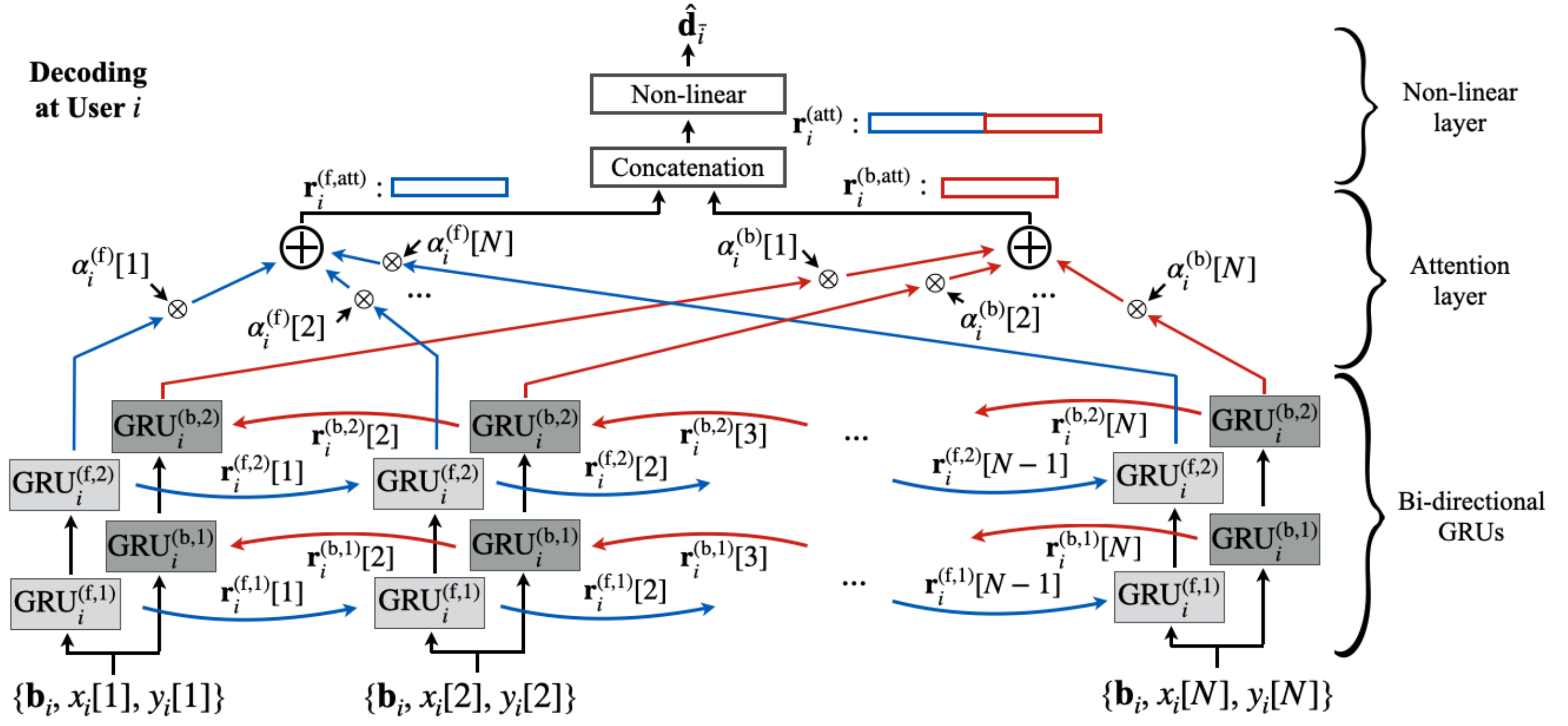}
    \caption{Our proposed RNN-based decoding architecture at User $i$, $i \in \{1,2\}$. Bi-directional RNNs with the attention mechanism are introduced to exploit correlations among receive symbols  in which the encoders’ coupling behavior is captured. 
    Here, $\bar i$ denotes the index of the counterpart of User $i$, i.e., $\bar i=2$ if $i=1$ while $\bar i=1$ if $i=2$.}
    \label{fig:RNN_decoder}
\end{figure}

The overall decoding architecture at User $i$, $i \in \{1,2\}$, is shown in Fig.~\ref{fig:RNN_decoder}.
The decoding function $g_i$ for User $i$ in \eqref{eq:dec_func_g} consists of  bi-directional GRUs, an attention layer, and a non-linear layer in sequence. We discuss each of them in detail below.

\subsubsection{Bi-directional GRUs}
At each user, we utilize two layers of bi-directional GRUs to capture the time correlation of the receive symbols both in the forward and backward directions over the sequence of the receive symbols. 
We represent the input-output relationship of the forward directional GRUs at time $k$ of User $i$ as
\begin{align}
    {\bf r}_i^{(\text{f},1)}[k] & = \text{GRU}_i^{(\text{f},1)} ({\bf b}_i, x_i[k], y_i[k], {\bf r}_i^{(\text{f},1)}[k-1] ),
    \nonumber
    \\
\textrm{and} \;\;\;    {\bf r}_i^{(\text{f},2)}[k] & = \text{GRU}_i^{(\text{f},2)} ({\bf r}_i^{(\text{f},1)}[k], {\bf r}_i^{(\text{f},2)}[k-1] ),
    \label{eq:dec:rf1}
\end{align}
and that of the backward directional GRUs as
\begin{align}
    {\bf r}_i^{(\text{b},1)}[k] & = \text{GRU}_i^{(\text{b},1)} ({\bf b}_i, x_i[k], y_i[k], {\bf r}_i^{(\text{b},1)}[k+1] ),
    \nonumber
    \\
\textrm{and} \;\;\;    {\bf r}_i^{(\text{b},2)}[k] & = \text{GRU}_i^{(\text{b},2)} ({\bf r}_i^{(\text{b},1)}[k], {\bf r}_i^{(\text{b},2)}[k+1] ),
\label{eq:dec:rf2}
\end{align}
where $\text{GRU}_i^{(\text{f},\ell)}$ and $\text{GRU}_i^{(\text{b},\ell)}$ represent functional forms of GRU processing at layer $\ell$ of User $i$ in the forward and backward direction, respectively.
Here, ${\bf r}_i^{(\text{f},\ell)}[k] \in \mathbb{R}^{N_{i,\ell}^{\text{(dec)}}}$ and ${\bf r}_i^{(\text{b},\ell)}[k] \in \mathbb{R}^{N_{i,\ell}^{\text{(dec)}}}$ are the state vectors obtained by $\text{GRU}_i^{(\text{f}, \ell)}$ and $\text{GRU}_i^{(\text{b}, \ell)}$, respectively, at time $k$, where $i \in \{1,2\}$, $\ell \in \{1,2\}$, and $k \in \{1,...,N\}$. 
For the initial conditions,
${\bf r}_i^{(\text{f},\ell)}[0]= {\bf 0}$ and ${\bf r}_i^{(\text{b},\ell)}[N+1]= {\bf 0}$, where $i \in \{1,2\}$ and $\ell \in \{1,2\}$.

\subsubsection{Attention layer}
We consider the forward/backward state vectors at the last layer, i.e., $\{{\bf r}_i^{(\text{f},2)}[k]\}_{k=1}^N$ and $\{{\bf r}_i^{(\text{b},2)}[k]\}_{k=1}^N$, as inputs to the attention layer.
Each state vector contains different feature information depending on both its direction and timestep $k$: the forward state vector ${\bf r}_i^{(\text{f},2)}[k]$ captures the implicit correlation information of the input tuples of the previous timesteps, i.e., $\{{\bf b}_i, x_i[1], y_i[1]\}$, ..., $\{{\bf b}_i, x_i[k], y_i[k]\}$, 
while the backward state vector ${\bf r}_i^{(\text{b},2)}[k]$ captures that of the later timesteps, i.e., $\{{\bf b}_i, x_i[k], y_i[k]\}$, ..., $\{{\bf b}_i, x_i[N], y_i[N]\}$.
Although the state vectors at each end, i.e., ${\bf r}_i^{(\text{f},2)}[N]$ and ${\bf r}_i^{(\text{b},2)}[1]$, contain the information of all input data tuples, the long-term dependency cannot be fully captured~\cite{bengio1993problem}. Therefore, we adopt the attention layer~\cite{bahdanau2014neural}, which merges the state vectors in the form of a summation.
Formally,
\begin{align}
    {\bf r}_i^{(\text{f,att})} &=  \sum_{k=1}^N \alpha_i^{(\text{f})}[k] {\bf r}_i^{(\text{f},2)}[k] \in \mathbb{R}^{N_{i,2}^{\text{(dec)}}},
    \quad 
\textrm{and} \;\;\;
    {\bf r}_i^{(\text{b,att})} =  \sum_{k=1}^N \alpha_i^{(\text{b})}[k] {\bf r}_i^{(\text{b},2)}[k] \in \mathbb{R}^{N_{i,2}^{\text{(dec)}}},
    \label{eq:dec:att_weight}
\end{align}
where $\alpha_i^{(\text{f})}[k] \in \mathbb{R}$ and $\alpha_i^{(\text{b})}[k] \in \mathbb{R}$ are the trainable \textit{attention weights} applied to the forward and backward state vectors, respectively, $k \in \{1,...,N\}$.
We capture the forward and backward directional information separately by stacking the two vectors as
\begin{align}
    {\bf r}_i^{(\text{att})} = [ {\bf r}_i^{(\text{f,att})}; {\bf r}_i^{(\text{b,att})} ] \in \mathbb{R}^{2N_{i,2}^{\text{(dec)}}}.
    \label{eq:dec:att_vector}
\end{align}
%
The attention mechanism enables the decoder at User $i$ to fully capture the 
interaction between the two users over noisy channels
by exploiting all timesteps' data tuples, $\{{\bf b}_i, x_i[k], y_i[k]\}_{k=1}^N$.

\subsubsection{Non-linear layer}
\label{sssec:non-linear}
We utilize a non-linear layer to finally obtain the estimate of the other user's bit vector $\hat {\bf b}_{\bar i}$, 
 by using the feature vector ${\bf r}_i^{(\text{att})}$ in \eqref{eq:dec:att_vector}, where  $\bar i$ denotes the index of the counterpart of User $i$, i.e., $\bar i=2$ if $i=1$ while $\bar i=1$ if $i=2$.
The input-output relationship at the non-linear layer is given by
\begin{align}
    \hat {\bf d}_{\bar i} =  \theta ({\bf W}_{\text{dec},i} {\bf r}_i^{(\text{att})}+ {\bf v}_{\text{dec},i})   \in (0,1)^{M_{\bar i}}
    \label{eq:decoder:softmax}
\end{align}
where 
${\bf W}_{\text{dec},i} \in \mathbb{R}^{ M_{\bar i} \times 2N_{i,2}^{(\text{dec})}}$ and ${\bf v}_{\text{dec},i} \in \mathbb{R}^{M_{\bar i}}$ are the trainable weights and biases, respectively, and
$\theta: \mathbb{R}^{M_{\bar i}} \rightarrow \mathbb{R}^{M_{\bar i}}$ is an activation function.

The dimension $M_{\bar i}$ and the activation function $\theta$ are chosen differently depending on the performance metric of interest.
When BLER is considered for a metric, we utilize the softmax activation function and set $M_{\bar i}=2^{K_{\bar i}}$. Then, $\hat {\bf d}_{\bar i}$ in \eqref{eq:decoder:softmax} denotes the probability distribution of $2^{K_{\bar i}}$ possible outcomes of $\hat {\bf b}_{\bar i}$. Since the softmax function allows for classification with multiple classes, we can minimize the block error of the bit vectors, i.e., $\text{Pr}[ \{{\bf b}_i \neq \hat {\bf b}_i \} ]$, by treating each possible outcome of ${\bf b}_i$ as a class. For example, if ${\bf b}_i \in \{0,1\}^{2}$, there are four possible classes, $[0,0]$, $[0,1]$, $[1,0]$, and $[1,1]$.
On the other hand, 
when BER is considered for a metric,  we consider the sigmoid activation function and set $M_{\bar i}=K_{\bar i}$, where each entry of $\hat {\bf d}_{\bar i}$ denotes the probability distribution of each entry of $\hat {\bf b}_{\bar i}$.
Since the sigmoid function allows classification for binary classes, we can minimize the error of each bit, i.e., $\text{Pr}[ \{{ b}_i[\ell] \neq \hat {b}_i[\ell] \} ]$, by conducting binary classification for each bit.


\subsection{Training and Inference}
\label{ssec:RNN:training}

\begin{figure}[t]
    \centering
    \includegraphics[width=.5\linewidth]{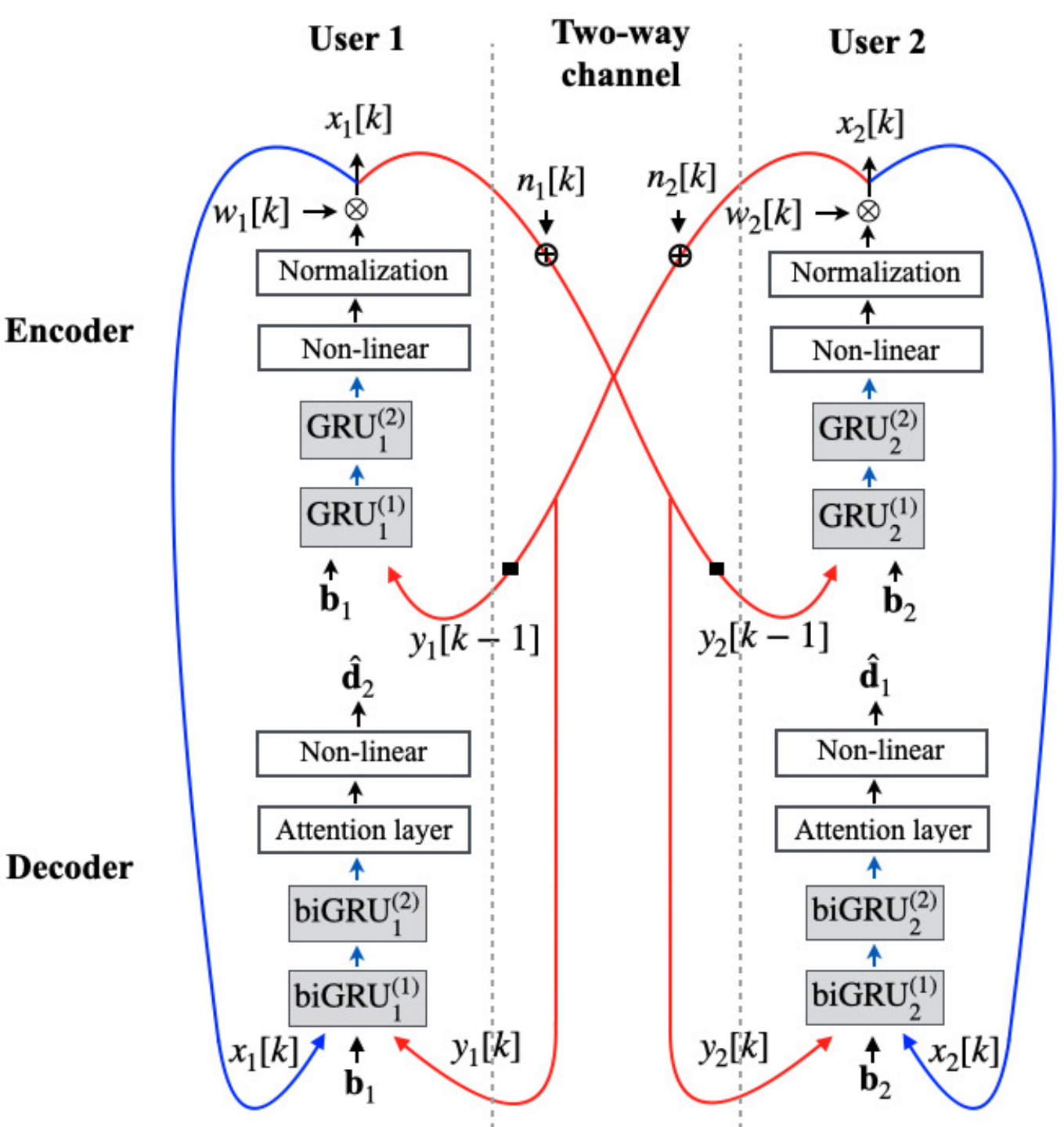}
    \caption{Our proposed encoding/decoding architecture  in a compact form for two-way channels.}
    \label{fig:RNN_compact}
\end{figure}

\subsubsection{Model training} 
We consider different loss functions depending on the performance metric of interest. 
When BLER is considered for a metric, we have adopted the softmax function for multi-class classification, discussed in Sec.~\ref{sssec:non-linear}. 
The commonly used loss function for training with the softmax function is cross-entropy (CE) loss. This loss function has been demonstrated to be effective in terms of the performance of multi-class classification tasks~\cite{goodfellow2016deep}.
Therefore, for sum-BLER minimization, we consider the sum of CE loss defined by
\begin{equation}
\mathcal{L}_{\text{sum-BLER}} = \sum_{i=1}^2 \text{CE}({\bf d}_i, \hat {\bf d}_i)  =
- \sum_{i=1}^2 \sum_{\ell=1}^{2^{K_i}} {d}_{i}[\ell] \log \hat {d}_{i}[\ell],
\end{equation}
where ${d}_{i}[\ell]$ and $\hat {d}_{i}[\ell]$ are the $\ell$-th entry of ${\bf d}_{i} \in \{0,1\}^{2^{K_i}}$ and $\hat {\bf d}_{i} \in (0,1)^{2^{K_i}}$, respectively. Here, ${\bf d}_i$ is the target vector, which is a one-hot representation of ${\bf b}_i \in \{0,1\}^{K_i}$. That is, only one entry of ${\bf d}_i$ has a value of 1, while the rest entries are zero.
Note that  $\hat {\bf d}_{i}$ is the inference output from the softmax function in \eqref{eq:decoder:softmax}.
By treating the entire bit vector as a block through the use of one-hot vectors,
we transform our problem of minimizing sum-BLER, i.e., $\sum_{i=1}^2 \text{Pr}[ \{{\bf b}_i \neq \hat {\bf b}_i \} ]$, into a \textit{block-level} classification problem.

On the other hand,  when BER is considered for a metric, we have adopted the sigmoid function for binary classification, discussed in Sec.~\ref{sssec:non-linear}. 
Binary cross-entropy (BCE) loss is a commonly used loss function for training with sigmoid function and has been shown to be effective for binary classification.~\cite{goodfellow2016deep}.
Therefore, for sum-BER minimization, we consider the sum of binary cross entropy (BCE) loss defined by
\begin{equation}
\mathcal{L}_{\text{sum-BER}} = \sum_{i=1}^2 \text{BCE}({\bf b}_i, \hat {\bf d}_i)  =
- \sum_{i=1}^2 \sum_{\ell=1}^{K_i} 
\bigg( {b}_i[\ell] \log(\hat {d}_{i}[\ell]) + (1-{b}_i[\ell])\log(1-\hat {d}_{i}[\ell]) \bigg),
\end{equation}
 where ${b}_{i}[\ell] \in \{0,1\}$ is the $\ell$-th entry of ${\bf b}_{i} \in \{0,1\}^{K_i}$. Note that 
$\hat {\bf d}_{i} \in (0,1)^{K_i}$ denotes the sigmoid output from \eqref{eq:decoder:softmax}.
By treating each bit for training, we convert our problem of minimizing sum-BER, i.e.,
$ \sum_{i=1}^2 (1/K_i) \sum_{\ell=1}^{K_i}  \text{Pr}[(\{{b}_i[\ell] \neq \hat { b}_i[\ell]\})]$,
to a \textit{bit-level} classification problem.

To minimize either the sum-BLER or sum-BER, we jointly train the encoders and decoders of both users within the structure of an autoencoder, as illustrated in  Fig.~\ref{fig:RNN_compact}. 
Specifically, the autoencoder framework allows us to optimize the encoder and decoder neural networks together within a single training process to achieve the common objective of minimizing the sum-BLER or sum-BER.
The overall algorithm for training our RNN-based coding architecture is provided in Algorithm~\ref{alg:training}.



\subsubsection{Inference}
Once the coding architecture is trained, the goal of the decoder at each User $i$ is to recover the other user's bit vector $\hat {\bf b}_{\bar i} \in \{0,1\}^{K_{\bar i}}$ from the decoding output $\hat {\bf d}_{\bar i}$ in \eqref{eq:decoder:softmax}. We note that the recovery process is different  depending on the performance metric of interest. 
First, for sum-BLER minimization, User $i$ 
forces the entry with the largest value of $\hat {\bf d}_{\bar i} \in (0,1)^{2^{K_{\bar i}}}$ to 1, while setting the rest of the entries to $0$, 
and then maps the obtained one-hot vector to a bit  vector, which is then $\hat {\bf b}_{\bar i}$.
%
For sum-BER minimization, User $i$ obtains $\hat {\bf b}_{\bar i}$ by rounding each entry of $\hat {\bf d}_{\bar i} \in (0,1)^{K_{\bar i}}$ to be either 0 or 1. That is, $\hat {b}_{\bar i}[\ell]$ is a rounded version of $\hat {d}_{\bar i}[\ell]$.

The RNN-based coding architecture is trained to solve \eqref{opt:obj:state-based}-\eqref{opt:const:power:state-based}, which in turn produces distinct encoding/decoding modules 
at specific channel SNRs of the users. 
We consider the scenario where the users store multiple encoding/decoding architectures, each trained at distinct channel SNRs. During the inference stage, the users select and use a specific encoding/decoding architecture corresponding to the channel SNRs that the system configures. In this manner, the users adaptively replace their coding architectures depending on the measured channel SNRs.



\begin{algorithm}[t]
   \caption{Training for the proposed two-way coding architecture based on RNN autoencoder}
   \label{alg:training}
   \footnotesize
\begin{algorithmic}
   \State {\bfseries Input:} Training data
   $\{{\bf b}_1^{(j)}, {\bf b}_2^{(j)}, {\bf n}_1^{(j)},  {\bf n}_2^{(j)}  \}_{j=1}^{{J}}$, number of epochs $N_\text{epoch}$, and batch size $N_\text{batch}$.
   \State {\bfseries Output:} Model parameters.
   \State Initialize the model parameters.
   \For{$e=1$ {\bfseries to} $N_\text{epoch}$}
   \For{$t=1$ {\bfseries to} $J/N_\text{batch}$}
    \State\multiline{Extract $N_{\text{batch}}$ data tuples $\{{\bf b}_1^{(j)}, {\bf b}_2^{(j)}, {\bf n}_1^{(j)},  {\bf n}_2^{(j)}  \}_{j \in \mathcal{I}_{\text{batch},t}}$ from the entire training data, where  $\mathcal{I}_{\text{batch},t}$ denotes the indices of the extracted data tuples with $\vert \mathcal{I}_{\text{batch},t} \vert = N_\text{batch}$.}\\
    \State \multiline{Update the model parameters using the gradient decent on the defined loss function;}
    \State \multiline{(i) For sum-BLER minization, we define the loss function as
    \begin{equation}
        \sum_{j \in \mathcal{I}_{\text{batch},t}} \mathcal{L}_{\text{sum-BLER},j} 
        = - \sum_{j \in \mathcal{I}_{\text{batch},t}} \sum_{i=1}^2 \sum_{\ell=1}^{2^{K_i}} {d}^{(j)}_{i}[\ell] \log \hat {d}^{(j)}_{i}[\ell],
    \end{equation}
    where ${d}_{i}^{(j)}[\ell]$ and $\hat {d}_{i}^{(j)}[\ell]$ are the $\ell$-th entry of ${\bf d}^{(j)}_{i}  \in \{0,1\}^{2^{K_i}}$ and $\hat {\bf d}^{(j)}_{i}  \in (0,1)^{2^{K_i}}$, respectively. Here,    
    ${\bf d}_i^{(j)} = \text{one-hot}({\bf b}_i^{(j)})$ is the target vector, and $\hat{\bf d}_i^{(j)}$ is the block-level classification output obtained by the $j$-th data tuple passing through the overall encoder-decoder architecture in 
    \eqref{eq:s1}-\eqref{eq:decoder:softmax}.}
    \State \multiline{(ii) For sum-BER minization, we define the loss function as
    \begin{equation}
        \sum_{j \in \mathcal{I}_{\text{batch},t}} \mathcal{L}_{\text{sum-BER},j} 
        = - \sum_{j \in \mathcal{I}_{\text{batch},t}} \sum_{i=1}^2 \sum_{\ell=1}^{K_i} 
        \bigg( {b}_i^{(j)}[\ell] \log(\hat {d}^{(j)}_{i}[\ell]) + (1-{b}_i^{(j)}[\ell])\log(1-\hat {d}^{(j)}_{i}[\ell]) \bigg),
    \end{equation}
    where ${b}^{(j)}_{i}[\ell]$ and $\hat {d}_{i}^{(j)}[\ell]$ are the $\ell$-th entry of ${\bf b}^{(j)}_{i}  \in \{0,1\}^{K_i}$ and $\hat {\bf d}^{(j)}_{i} \in (0,1)^{K_i}$, respectively.
    Here, ${\bf b}^{(j)}_{i}$ is the target vector, and $\hat{\bf d}_i^{(j)}$ is the bit-level classification output obtained by the $j$-th data tuple passing through 
    \eqref{eq:s1}-\eqref{eq:decoder:softmax}.}
   \EndFor
   \EndFor
\end{algorithmic}
\end{algorithm}

\subsection{Modulo-Based Approach for Long Block-Lengths}
\label{ssec:RNN:modulo}


A direct application of the proposed coding architecture for long block-lengths is not feasible, since the larger input/output sizes of the encoders and decoders result in a substantial increase in complexity. 
In particular, the larger input/output sizes lead to an increased number of computations at the non-linear layer in the decoder (in \eqref{eq:decoder:softmax}) and potentially larger  state vector sizes at the encoder (in \eqref{eq:s1}-\eqref{eq:s2}) and the decoder (in \eqref{eq:dec:rf1}-\eqref{eq:dec:rf2}), aiming to capture  hidden features from the input data.
%
%
To address this issue, we consider a modulo approach for processing long block-lengths of bits by successively applying our coding architecture built for short block-lengths~\cite{kim2023feedback}.

In the long block-length regime,  we define the entire block-length at User $i$ as $L_i > K_i$, while $K_i$ denotes the number of processing bits input to our coding architecture at a time, $i \in \{1,2\}$.
We consider that our two-way coding architecture has been trained for block-lengths $K_1$ and $K_2$, i.e., for conveying $K_1$ bits from User 1 to 2 and $K_2$ bits from User 2 to 1 over $N$ channel uses.
To process the entire block-length $L_1$ at User 1 and $L_2$ at User 2, each User $i$ first divides the $L_i$ bits into $\lceil L_i/K_i \rceil$ chunks each with length $K_i$. Note that when $L_i$ is not a multiple of $K_i$, we can simply pad zeros following the residual bits in the last chunk. Then, the two users exchange their signals to convey their chunks of $K_1$ and $K_2$ bits by using our coding architecture in a time-division~manner. 
 

This modulo-based approach provides two benefits. First, it reduces the complexity of the network architecture by simplifying the encoding and decoding processes through successive applications of the neural networks trained for shorter block-lengths. Second, it allows generalization across various block-lengths (multiple of $K_i$ bits) without necessitating re-training. 

\section{Computational Complexity of Linear and RNN-Based Coding}
\label{sec:comp}

In this section, we analyze the computational complexity of our proposed linear and RNN-based coding schemes.
We first look into the case of linear coding.
%
%
In Algorithm \ref{al:linear:sum-error}, we solve for the encoding schemes, i.e., ${\bf g}_1$, ${\bf F}_1$, ${\bf g}_2$, and ${\bf F}_2$,  given the fixed number of channel uses, $N$, and the noise variances, $\sigma_1^2$ and $\sigma_2^2$. Subsequently, we  obtain the decoding (combining) schemes, ${\bf w}_1$ and ${\bf w}_2$, by \eqref{eq:linear_coding:w1} and
\eqref{eq:linear_coding:w2}
as functions of the encoding schemes. Once the encoding/decoding schemes are determined, they are used for  encoding via \eqref{eq:x1}-\eqref{eq:x2} and decoding described in Sec.~\ref{ssec:linear:decoding} 
 without being recalculated. 
Specifically, the complexity of encoding at User 1 is 
$\mathcal{O}(N^2)$ from \eqref{eq:x1}, where ${\bf F}_1{\bf F}_2$ can be pre-calculated, while
User 2 has the complexity of $\mathcal{O}(N)$ in \eqref{eq:x2}, where the computation of ${\bf F}_2 {\bf y}_2$ has $\mathcal{O}(N)$ complexity since ${\bf F}_2$ is in the form of \eqref{eq:F2}.
For decoding, User 1 has complexity of $\mathcal{O}(N^2)$, where the decoding process at User 1 consists of (i) $\tilde {\bf y}_1 = ({\bf I} + {\bf F}_2 {\bf F}_1)^{-1} ({\bf y}_1 - {\bf F}_2 {\bf g}_1 m_1) $  and (ii) $ {\hat m}_2 = {\bf w}_{2}^\top \tilde {\bf y}_1$, described above \eqref{eq:SNR2}. Here, $({\bf I} + {\bf F}_2 {\bf F}_1)^{-1}$ and $ {\bf F}_2 {\bf g}_1$ can be pre-computed.
On the other hand, User 2 has $\mathcal{O}(N)$ complexity, where the decoding process at User 2 is composed of (i) $\tilde {\bf y}_2 =  {\bf y}_2 - {\bf F}_1 {\bf g}_2 m_2$ and (ii) $ {\hat m}_1 = {\bf w}_{1}^\top \tilde {\bf y}_2$, described above \eqref{eq:SNR1}. Here, ${\bf F}_1 {\bf g}_2$ can be pre-computed.



We next investigate the computational complexity for RNN-based coding.
In Algorithm \ref{alg:training}, we obtain the model parameters for the non-linear coding architecture depicted in Fig.~\ref{fig:RNN_compact}. The constructed architecture is then used for encoding and decoding with the fixed model parameters without further training. 
Although we implemented two layers of GRUs at the encoder (in Fig.~\ref{fig:RNN_encoders}) and decoder (in Fig.~\ref{fig:RNN_decoder}), we consider a general numbers of layers for complexity analysis, and define the number of GRU layers at the encoder and decoder of User $i$ as 
$N_{i}^{\text{(enc,lay)}}$ and $N_{i}^{\text{(dec,lay)}}$, respectively.
Accordingly, we denote the number of neurons at each GRU layer at the encoder and decoder of User $i$ as $N_{i}^{(\text{enc})}$ and $N_{i}^{(\text{dec})}$, respectively, assuming the same number of neurons is applied to each GRU layer.
For encoding, User $i$ has the complexity of $\mathcal{O}( N N_{i}^{(\text{enc})}(N_{i}^{(\text{enc})}N_{i}^{\text{(enc,lay)}}+K_i) )$.
For decoding, User $i$ has the complexity of 
$\mathcal{O}(N_{i}^{\text{(dec,lay)}} N (N_{i}^{(\text{dec})})^2 + K_{\bar i} N N_{i}^{(\text{dec})})$ when the sigmoid function is used, while it has 
$\mathcal{O}(N_{i}^{\text{(dec,lay)}} N (N_{i}^{\text{(dec)}})^2 + K_{\bar i} N N_{i}^{\text{(dec)}} + 2^{K_{\bar i}} N_{i}^{\text{(dec)}})$ when the softmax function is used, where   $\bar i$ denotes the index of the counterpart of User $i$, i.e., $\bar i=2$ if $i=1$ while $\bar i=1$ if $i=2$.

The encoding and decoding complexity at the users for linear coding and RNN-based coding are summarized in Tables~\ref{table:comp:linear} and \ref{table:comp:RNN}. 
Linear coding has a low coding complexity due to the simplified system model adopting linear operation for encoding and decoding in \eqref{eq:x1}-\eqref{eq:x2}, \eqref{eq:linear_coding:w2}, and \eqref{eq:linear_coding:w1}.
While the RNN-based coding causes higher encoding/decoding complexity, it benefits from higher degrees of flexibility thanks to the non-linearity provided by deep neural networks, leading to better error performance under many practical noise scenarios (shown in Sec.~\ref{sec:sim}).


\begin{table}[t]
\parbox{.35\linewidth}{
\centering
\caption{Linear coding complexity.}
\label{table:comp:linear}
{
\begin{tabular}{
|c||c|c| }
 \hline
 & User 1 & User 2  \\
 \hline
 Encoder  & $\mathcal{O}(N^2)$ & $\mathcal{O}(N)$ \\
 \hline
 Decoder & $\mathcal{O}(N^2)$ & $\mathcal{O}(N)$ \\
 \hline
\end{tabular}}
}
\hfill
\parbox{.65\linewidth}{
\centering
\caption{RNN-based coding complexity.}
\label{table:comp:RNN}
{
\begin{tabular}{
|c||c| }
 \hline
 & User $i$, $i \in \{1,2\}$  \\
 \hline
 Encoder  & $\mathcal{O}( N N_{i}^{(\text{enc})}(N_{i}^{(\text{enc})}N_{i}^{\text{(enc,lay)}}+K_i) )$ \\
 \hline
 Decoder with sigmoid   & $\mathcal{O}(N_{i}^{\text{(dec,lay)}} N (N_{i}^{(\text{dec})})^2 + K_{\bar i} N N_{i}^{(\text{dec})})$ \\
 \hline
 Decoder with softmax   & $\mathcal{O}(N_{i}^{\text{(dec,lay)}} N (N_{i}^{\text{(dec)}})^2 + K_{\bar i} N N_{i}^{\text{(dec)}} + 2^{K_{\bar i}} N_{i}^{\text{(dec)}})$ \\
 \hline
\end{tabular}}
}
\end{table}








\section{Experimental Results}
\label{sec:sim}


In this section, we investigate both our linear and learning-based coding methodologies through numerical experiments.
We first describe the simulation setup (Sec.~\ref{ssec:sim:parameter}) and validate our coding methodologies in a short block-length regime (Sec.~\ref{ssec:sim:short}).
In addition, we discuss the enhancement of our linear coding scheme by thoroughly investigating its solution behavior (Sec.~\ref{ssec:sim:alternate}).
We also show the effectiveness of our coding methodologies in a medium/long block-length regime (Sec.~\ref{ssec:sim:medium/long}). 
Then, we analyze the benefits of two-way coding (Sec.~\ref{ssec:sim:threshold}), and show that the best coding strategy is different in SNR regimes (Sec.~\ref{ssec:sim:best}).
Moreover, we investigate the error performance by varying the rates (Sec.~\ref{ssec:sim:coding_rate}) and provide  information theoretic insights on power distribution of both linear and RNN-based coding (Sec.~\ref{ssec:sim:power}). Finally, we analyze the block-length gain of our RNN-based coding (Sec.~\ref{ssec:sim:blockgain}).

\subsection{Simulation Parameters}
\label{ssec:sim:parameter}

We assume $P = 1$, i.e., the average power constraint per channel use is unity. 
We denote the average SNR for the channel of User $i$ as $\text{SNR}^{\text{ch}}_i = NP/(N\sigma_i^2) = 1/\sigma_i^2$. 
We note that the channel SNR,  $\text{SNR}_i^{\text{ch}}$, is different from the message SNR described in Sec.~\ref{sec:linear}, which is the post-processed SNR for message transmissions after the two-way linear encoding/decoding schemes are applied.
In simulations, we consider $\text{SNR}_1^{\text{ch}} \le \text{SNR}_2^{\text{ch}}$ without loss of generality.
For linear coding, we consider 30 different
initializations of $\{f_{2,i}\}_{i=2}^{N-1}$ with $f_{2,i} \sim \mathcal{U}(0,1)$, and select the best solution for Algorithm~\ref{al:linear:weighted_sum_power}.
The threshold for the stopping criterion in lines 5 and 12 of Algorithm~\ref{al:linear:weighted_sum_power} is $\epsilon = 10^{-3}$.
For Algorithm~\ref{al:linear:sum-error}, we consider a finite search space for $\eta_2$ over the region $(0,\eta_{2,\max})$ where $\eta_{2,\max} = NP/\sigma_2^2$.
The bisection method is carried out for $\eta_1$ over  $(0,\eta_{1,\max})$ where $\eta_{1,\max}$ is  set to a scaled value of the maximum SNR of the feedback scheme as $\eta_{1,\max} = 2NP(\sigma_1^2 + \sigma_2^2)/\sigma_2^2$~\cite{chance2011concatenated}, while the golden-section search method is performed for $\alpha$ over the range $[ \sigma_2^2/(\sigma_1^2+\sigma_2^2),1)$ to comply with the range of $\alpha$ in Proposition~\ref{pro:g:optimal}.

For RNN-based coding, the number of training data samples that we used is $J=10^7$, the batch size is $N_\text{batch} = 2.5 \times 10^4$, and the number of epochs is $N_\text{epoch} = 100$. We use the ADAM optimizer~\cite{goodfellow2016deep} and a decaying learning rate, where the initial rate is 0.01 and the decaying ratio is  $\gamma = 0.95$ applied for every epoch.
We also use gradient clipping for
training, where the gradients are clipped when the norm of gradients is larger than 1.
We adopt two layers of uni-directional
GRUs at the encoder and two layers of bi-directional GRUs at the decoder for each user, with $N_\text{neurons}=N_{i}^{(\text{enc})}=N_{i}^{(\text{dec})}=50$ neurons at each GRU.
We initialize each neuron in a GRU with the uniform distribution $U(-\zeta, \zeta)$, where $\zeta = 1/\sqrt{N_\text{neurons}}$, and
all the power weights and attention weights to 1.
We train our neural network model under particular noise scenarios, i.e., $\text{SNR}_1^{\text{ch}}$ and $\text{SNR}_2^{\text{ch}}$, and conduct inference under the same noise scenario. For inference, we consider $10^{10}$ data tuples to calculate BER and BLER.

\subsection{Short Block-Length}
\label{ssec:sim:short}

We first investigate the performance of our coding strategies in a short block-length regime.
We consider the same block-length of $L_1=L_2=6$ bits at the users and $N_\text{L}=18$ channel uses for exchanging $L_1$ and $L_2$ bits between the users. The rate is then $r_1=r_2=1/3$.
For baselines, we first consider repetition coding, where User $i$ modulates each bit of ${\bf b}_i \in \{0,1\}^{L_i}$ with binary phase-shift keying (BPSK) and transmits each modulated symbol repetitively over $N_\text{L}/L_i=3$ channel uses. 
We consider tail-biting convolutional coding (TBCC)~\cite{ma1986tail} as another baseline, adopted in LTE standards~\cite{LTEstandard} for short block-length codes. We consider the trellis with $(7, [133,171,165])$ of a constraint length of $7$ and the generator $[133,171,165]$ in octal representation, and BPSK modulation.
As a benchmark, we adopt a lower bound (LB) of the BLER for open-loop (OL) one-way communications, denoted by OL-LB $i$ for User $i$, by using the results of \cite{polyanskiy2010channel} and Lemma 39 therein for our case with an average power constraint.\footnote{\rev{A tighter lower bound on the BLER for the one-way channel might be achievable for short block-lengths, depending on the specific values of the coding rate, block-length, and channel SNR, as discussed in~\cite{yavas2024third}.}}
We calculate OL-LB $i$, given the coding rate, block-length, and channel SNR of User $i$. 
We consider the sum of the lower bounds, i.e., $\text{Sum-OL-LB} = \text{OL-LB 1} + \text{OL-LB 2}$, as a benchmark for two-way communications.
This represents a lower bound for ``open-loop" two-way communications, i.e., when cooperation is not allowed. Comparing with this benchmark, we can measure how much cooperation improves the BLER performance through our two-way coding methods.\footnote{\rev{We find that it would be interesting to develop a lower bound on the sum-BLERs or an upper bound on coding rates for \textit{cooperation}-enabled two-way communication systems. However, deriving such bounds in the finite block length regime for two-way systems is inherently challenging due to the cooperative nature of the users.
Since our work focuses on developing practical two-way coding schemes, we leave this interesting avenue for future work.}}
%
For our linear and RNN-based coding, there are  several ways to exchange $L_1=L_2=6$ bits between the users over $N_\text{L}=18$ channel uses by varying the value of $K_i$.\footnote{For linear coding, User $i$ modulates $L_i$ bits to $M_i=L_i/K_i$ messages with $2^{K_i}$-ary PAM, and $M_i$ messages are exchanged between the users.
For RNN-based coding, we construct a coding architecture that supports $K_1$ and $K_2$ bits at the users.}
We consider the case of $K=K_1=K_2$ for both linear and RNN-based coding and investigate our coding schemes under different values of $K$. For RNN-based coding, we use two different coding architectures: (i) $K_1=K_2=6$ processing bits with the softmax function in \eqref{eq:decoder:softmax}  and (ii) $K_1=K_2=3$ bits with the sigmoid function.

\begin{figure}[t]
  \centering
\begin{subfigure}{.49\linewidth}
  \centering
  \includegraphics[width=\linewidth]{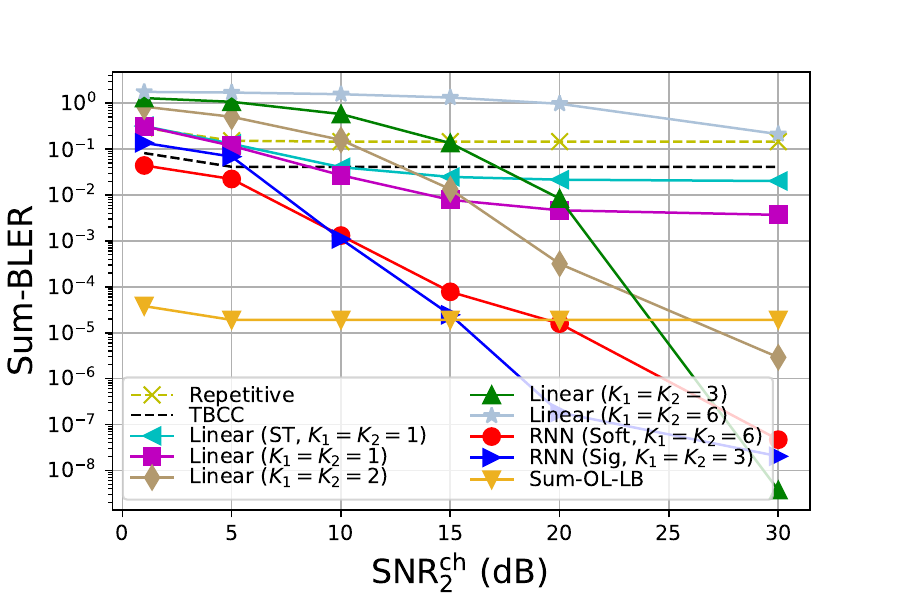}
  \caption{ $\text{SNR}^{\text{ch}}_1 = 1$dB.
  }
  \label{fig:short:SNR1=1}
\end{subfigure}
\begin{subfigure}{.49\linewidth}
  \centering
  \includegraphics[width=\linewidth]{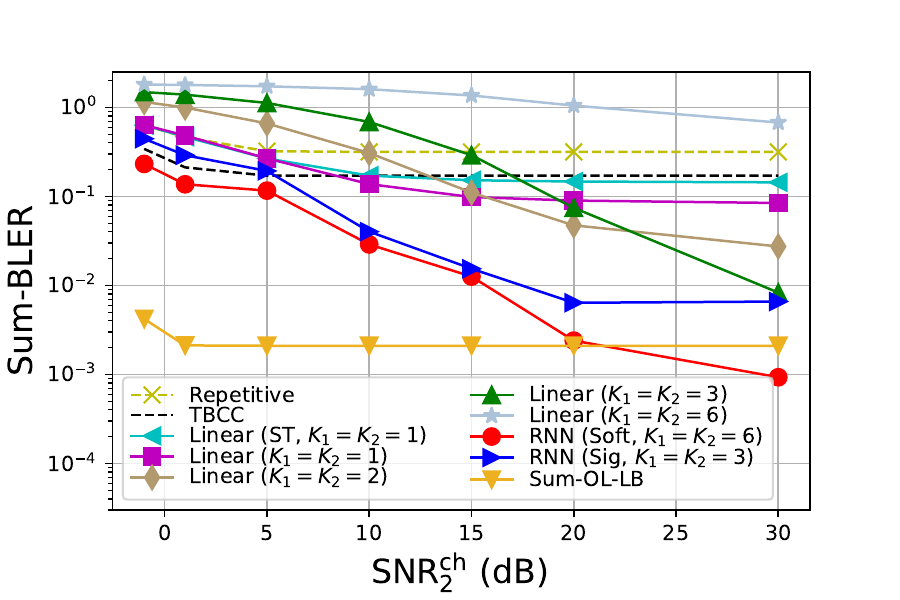}
  \caption{ $\text{SNR}^{\text{ch}}_1 = -1$dB.
  }
  \label{fig:short:SNR1=-1}
\end{subfigure}
  \caption{Sum-BLER with short block-lengths of $L_1=L_2=6$ bits and $N_\text{L}=18$ channel uses with $r_1=r_2=1/3$. RNN-based coding shows robustness to high channel noises, while linear coding performs well in very low noise~scenarios. Here, $\text{sum-BLER}$ of the channel coding schemes remains almost the same over $\text{SNR}^{\text{ch}}_2$, since 
$\text{sum-BLER}$ is dominated by $\text{BLER}_1$ and $\text{BLER}_1$ is constant due to a constant value of $\text{SNR}^{\text{ch}}_1$. The proposed linear and RNN-based schemes benefit from interactive two-way coding and yield a significant performance improvement.}
  \label{fig:short}
\end{figure} 


Fig.~\ref{fig:short} shows the sum-BLER performance under varying $\text{SNR}^{\text{ch}}_2$, 
where $\text{SNR}^{\text{ch}}_1=1$dB in Fig.~\ref{fig:short}(\subref{fig:short:SNR1=1}) and $\text{SNR}^{\text{ch}}_1=-1$dB in Fig.~\ref{fig:short}(\subref{fig:short:SNR1=-1}).
The channel coding schemes, i.e., repetitive and TBCC, and the benchmark Sum-OL-LB yield almost constant value in sum-BLER when $\text{SNR}^{\text{ch}}_1 < \text{SNR}^{\text{ch}}_2$, since 
$\text{sum-BLER}$ is dominated by  $\text{BLER}_1$. Note that $\text{BLER}_1$ is constant due to a constant value of $\text{SNR}^{\text{ch}}_1$. 
On the other hand, our two-way coding schemes, i.e., linear and RNN, yield better sum-BLER performance across different values of  $\text{SNR}^{\text{ch}}_2$. The interactive exchange of information between the users allows
User 2, which has  a better channel SNR than that of User 1, to help User 1, which results in a performance improvement of $\text{BLER}_1$, ultimately improving sum-BLER.
It is interesting that cooperation enables users to achieve a lower sum-BLER than Sum-OL-LB. This is because User 1's error performance is improved by the feedback from User 2, resulting in an error rate even lower than the open-loop lower bound. The behavior of individual BLER curves is further analyzed in Sec.~\ref{ssec:sim:threshold} and Fig.~\ref{fig:threshold}.
This phenomenon is also observed in feedback-enabled one-way communications; as shown in Fig.~2 in \cite{kim2020deepcode}, using feedback in coding improves the BLERs, achieving performance better than the open-loop lower bound.

For linear coding, we adopt the successive transmission for exchanging multiple messages between the users discussed in Sec.~\ref{ssec:linear:long}, denoted by ``Linear (ST)". We also consider adopting the alternate channel use strategy for linear coding, denoted by ``Linear", discussed in Sec.~\ref{ssec:linear:alternate}.
Note that this strategy improves the error performance as shown in the plots of Fig.~\ref{fig:short}, and thus we adopt this strategy throughout simulations.
%
%
As discussed in Sec.~\ref{ssec:linear:long}, the value of $K_i$ for linear coding affects sum-error performance in asymmetric channels due to the trade-off between the minimum distance of the constellation and the feedback benefit depending on $K_i$.
In Figs.~\ref{fig:short}(\subref{fig:short:SNR1=1}) and (\subref{fig:short:SNR1=-1}),
at high channel SNR at User 2, e.g., at $\text{SNR}^{\text{ch}}_2=30$dB, high modulation up to $K_i=3$  gives better performance since the feedback benefit from using a larger number of channel uses $N=K_i/r_i$ outweighs the performance loss due to the high modulation order $2^{K_i}$ and the subsequently decreased constellation distance.
Under lower channel SNRs, 
$K_i=2$ performs the best at $\text{SNR}^{\text{ch}}_2=20$dB, while $K_i=1$ yields the best performance
when $\text{SNR}^{\text{ch}}_2 \le 15$dB.
This indicates that, at lower channel SNRs, smaller values of $K_i$ perform better than larger values.
This is because, at low channel SNRs, higher modulation order results in a significant loss in error performance  due to the decreased constellation distance, which outweighs the benefits of feedback.
Note that with $K_i=6$, the sum-error performance increases drastically across all User 2's channel SNR values since the performance loss due to the high modulation order dominates over the feedback benefit.
%
%
In Fig.~\ref{fig:short}(\subref{fig:short:SNR1=1}),
we observe that linear coding with $K_1=K_2=3$ outperforms the other schemes when $\text{SNR}^{\text{ch}}_2$ is high.
For the rest of the $\text{SNR}^{\text{ch}}_2$ regions, RNN-based coding yields the best performance.
In Fig.~\ref{fig:short}(\subref{fig:short:SNR1=1}) with  $\text{SNR}^{\text{ch}}_1=1$dB,
RNN-based coding with the softmax function and $K_1=K_2=6$ bits performs better 
when $\text{SNR}^{\text{ch}}_2 <10$dB, 
while RNN-based coding with the sigmoid function and $K_1=K_2=3$ bits performs better when $\text{SNR}^{\text{ch}}_2 >10$dB.
This implies that using the softmax function for decoding 
provides robustness to channel noises due to the block-level classification of the bit vector, which is aligned with the result of feedback-enabled GOWCs~\cite{kim2023feedback}.
In Fig.~\ref{fig:short}(\subref{fig:short:SNR1=-1}), where we consider a more noisy scenario with $\text{SNR}^{\text{ch}}_1=-1$dB, 
RNN-based coding with the softmax function yields the best performance  over all the $\text{SNR}^{\text{ch}}_2$ regions, showing its robustness to the high channel noises.



\subsection{Linear Coding Enhanced by Alternate Channel Uses}
\label{ssec:sim:alternate}





\begin{figure}[t]
  \centering
\begin{subfigure}{.49\linewidth}
  \centering
  \includegraphics[width=\linewidth]{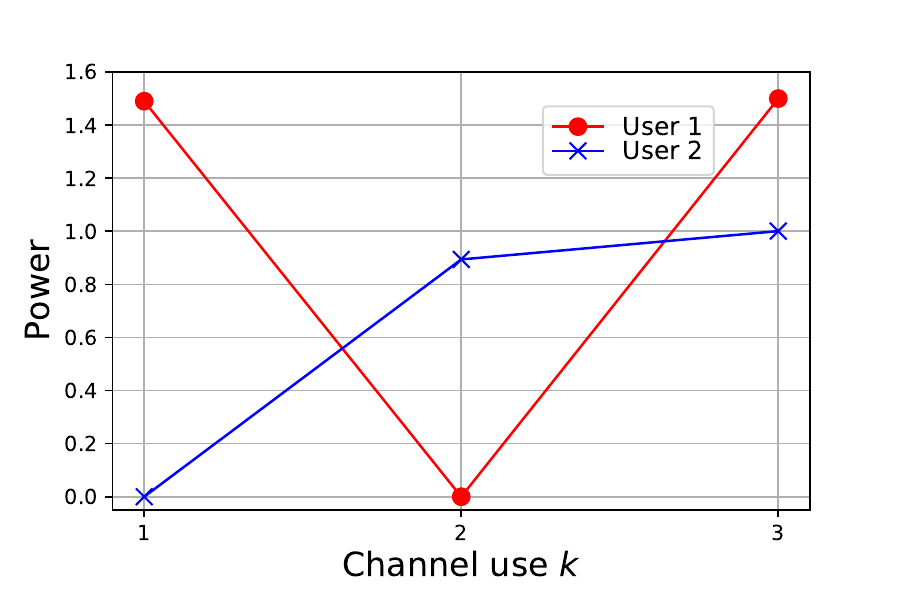}
  \caption{ $M=1$ message pair over $N=3$ channel uses.
  }
  \label{fig:power_linear_no_alt(N3)}
\end{subfigure}
\begin{subfigure}{.49\linewidth}
  \centering
  \includegraphics[width=\linewidth]{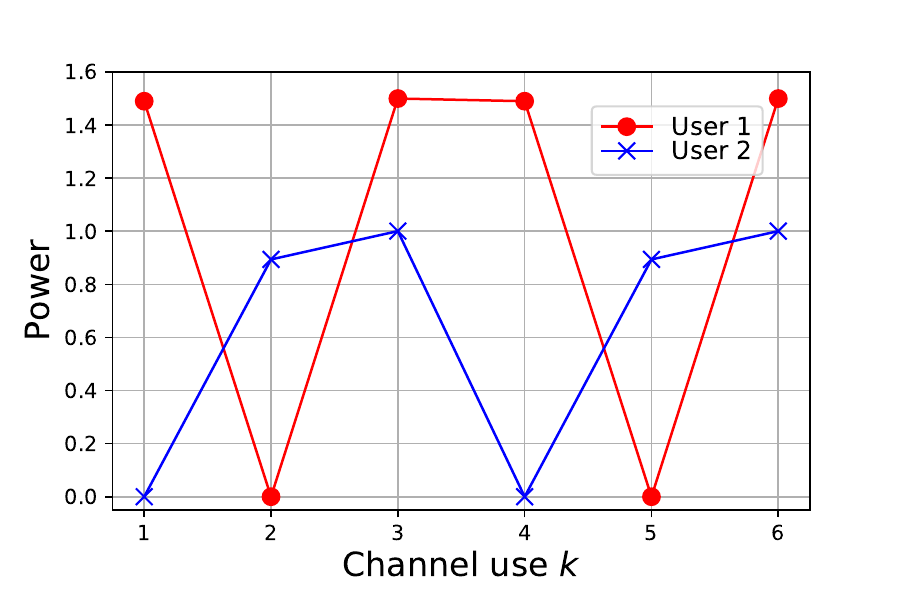}
  \caption{ $M=2$ message pairs over $N_\text{L}=6$ channel uses.
  }
  \label{fig:power_linear_no_alt(N6)}
\end{subfigure}
  \caption{Power distribution from linear coding along the channel uses.
  In (a), a pair of messages is exchanged over $N=3$ channel uses with power constraint $NP=3$. In (b), two pairs of messages are exchanged with power $N_\text{L}P=6$ over $N_\text{L}=6$ channel uses via the successive transmission of the two message pairs.
  Some of the channel uses are not utilized by the users, which causes a waste of channel uses.}
  \label{fig:power_linear_no_alt}
\end{figure}
\begin{figure}[t]
  \centering
\begin{subfigure}{.49\linewidth}
  \centering
  \includegraphics[width=\linewidth]{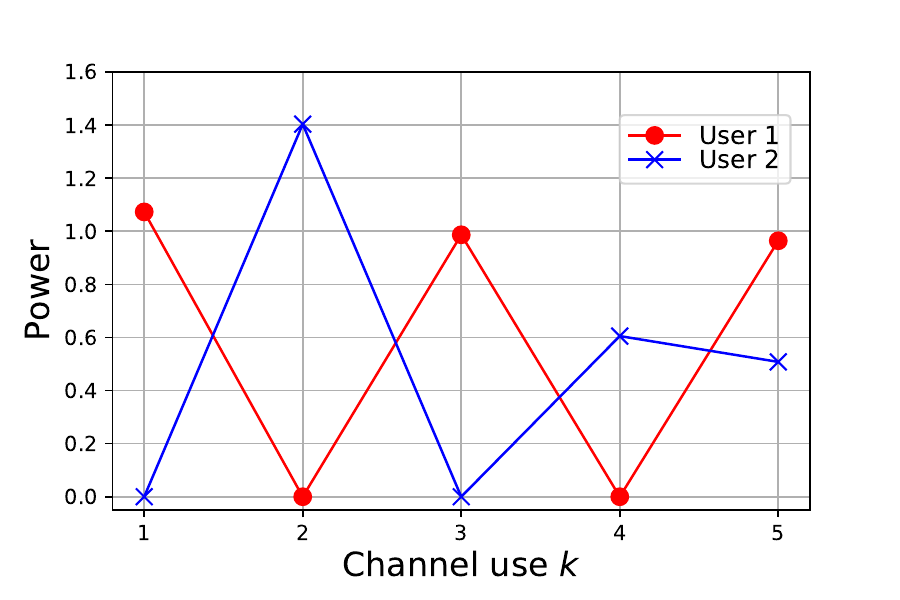}
  \caption{ $M=1$ message pair over $N=5$ channel uses.
  }
  \label{fig:power_linear(K=1,N=5)}
\end{subfigure}
\begin{subfigure}{.49\linewidth}
  \centering
  \includegraphics[width=\linewidth]{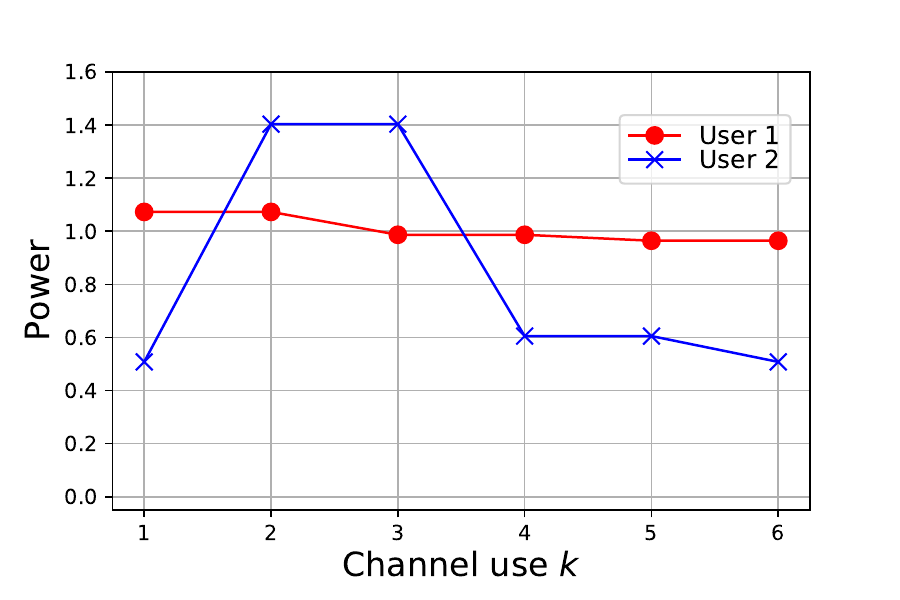}
  \caption{  $M=2$ message pairs over $N_\text{L}=6$ channel uses.
  }
  \label{fig:power_linear(K=2,N=6)}
\end{subfigure}
  \caption{Power distribution from linear coding along the channel uses.
  In (a), a pair of messages is exchanged over $N=5$ channel uses with power constraint $3P$. In (b), two pairs of messages are exchanged  over $N_\text{L}=6$ channel uses with power constraint $N_\text{L}P=6$ where we adopt the strategy of alternate channel uses. Through this strategy, the users can fully utilize the channel uses.}
  \label{fig:power_linear_alt}
\end{figure}

We consider that the users exchange $M=2$ message pairs, $(m_1,m_2)$ and $(m'_1,m'_2)$, over $N_\text{L}=6$ channel uses under power constraint  $N_\text{L}P=6P$, where $\text{SNR}^{\text{ch}}_1 = 1$dB and $\text{SNR}^{\text{ch}}_2 = 20$dB.
We first consider the successive transmission discussed in Sec.~\ref{ssec:linear:long}. 
Fig.~\ref{fig:power_linear_no_alt}(\subref{fig:power_linear_no_alt(N3)}) shows the power distribution for exchanging a message pair over $N=3$ channel uses with the proposed linear coding scheme.
The two message pairs are then exchanged via the successive transmission with the power distribution in Fig.~\ref{fig:power_linear_no_alt}(\subref{fig:power_linear_no_alt(N6)}).
From the power distribution in Fig.~\ref{fig:power_linear_no_alt}(\subref{fig:power_linear_no_alt(N3)}) and Fig.~\ref{fig:power_linear_no_alt}(\subref{fig:power_linear_no_alt(N6)}), 
Users 1 and 2 only utilize about half of the channel uses, resulting in a waste of channel uses.
Next, we consider the same setup for the strategy of alternate channel uses.
%
Based on the alternate channel use strategy, we plot the power distribution in Fig.~\ref{fig:power_linear_alt}.
Fig.~\ref{fig:power_linear_alt}(\subref{fig:power_linear(K=1,N=5)}) shows the power distribution for exchanging a pair of messages over $N=5$ channel uses discussed in \textbf{Step 1} in Sec.~\ref{ssec:linear:alternate}.
Fig.~\ref{fig:power_linear_alt}(\subref{fig:power_linear(K=2,N=6)}) shows the power distribution for exchanging two pairs of messages over $N_\text{L}=6$ channel uses when adopting the alternate channel use strategy discussed in \textbf{Step 2} in Sec.~\ref{ssec:linear:alternate}. 
This alternate channel use strategy allows the users to fully utilize the channel uses within the same power constraint compared to the successive transmission in Fig.~\ref{fig:power_linear_no_alt}(\subref{fig:power_linear_no_alt(N6)}), and thus leads  to the error performance improvement as demonstrated in Fig.~\ref{fig:short}.
Throughout simulations, we adopt this strategy for  linear coding.

\subsection{Medium/Long Block-Lengths}
\label{ssec:sim:medium/long}


\begin{figure}[t]
  \centering
\begin{subfigure}{.49\linewidth}
  \centering
  \includegraphics[width=\linewidth]{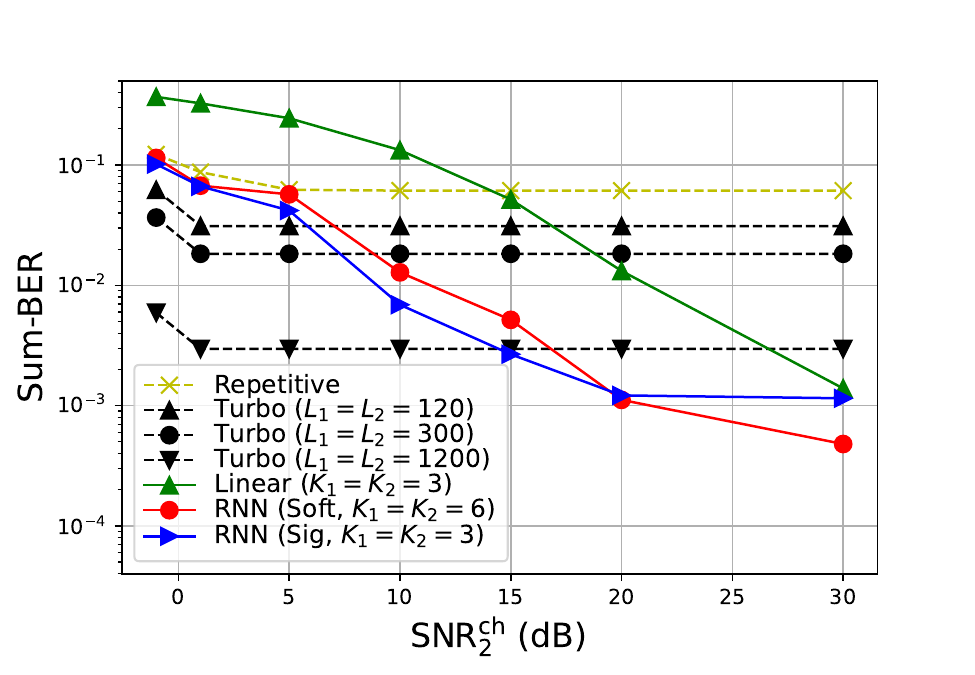}
  \caption{ Sum-BER with medium/long block-lengths.
  }
  \label{fig:ber_long}
\end{subfigure}
\begin{subfigure}{.49\linewidth}
  \centering
  \includegraphics[width=\linewidth]{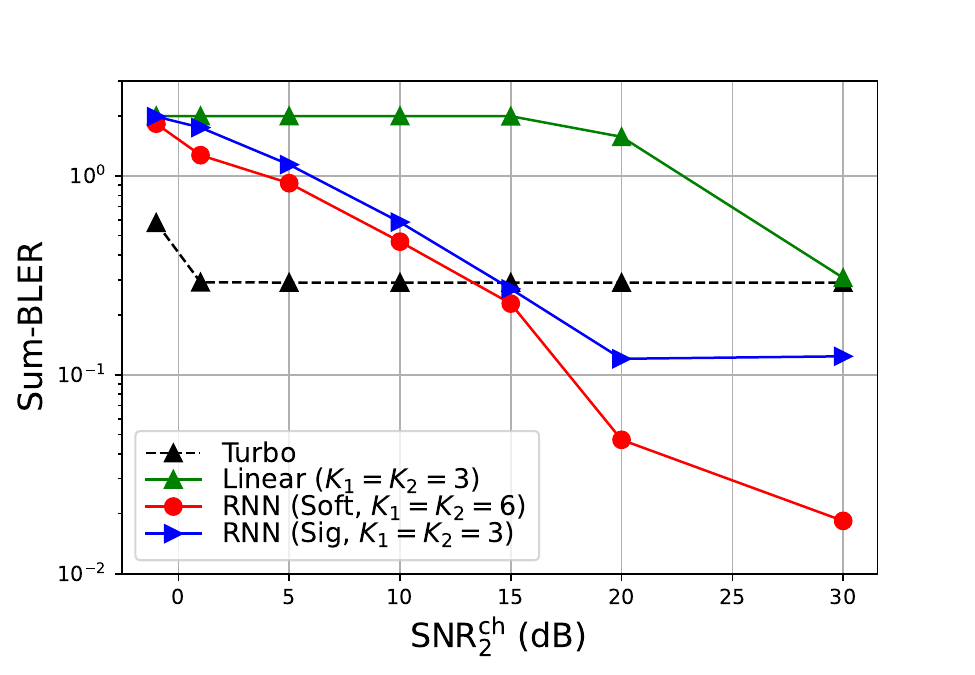}
  \caption{ Sum-BLER with medium block-length ($L_1=L_2=120$).
  }
  \label{fig:bler_medium}
\end{subfigure}
  \caption{Sum-error with medium/long block-lengths for $L_1,L_2$ bits where $\text{SNR}^{\text{ch}}_1=-1$dB and rate $r_1=r_2=1/3$. Under the high noise scenario with $\text{SNR}^{\text{ch}}_1=-1$dB, RNN-based coding performs well in terms of balancing the communication reliability when the channel noises become more asymmetric.}
  \label{fig:medlong}
\end{figure} 


We next analyze a medium and long block-length regime, where we consider turbo coding~\cite{berrou1996near} adopted in LTE standards as another baseline for medium/long block-length codes. 
We consider the trellis with $(4,[13, 15])$, BPSK  modulation, and $10$ decoding iterations. 
Fig.~\ref{fig:medlong}(\subref{fig:ber_long}) shows sum-BER curves under varying $\text{SNR}^{\text{ch}}_2$ where $\text{SNR}^{\text{ch}}_1=-1$dB and $r_1=r_2=1/3$. 
Since turbo coding exploits the block-length gain, a longer block of bits leads to a better sum-BER performance. However, for given $L_1$ and $L_2$, the sum-BER performances are almost constant over $\text{SNR}^{\text{ch}}_2$ since sum-BER is dominated by a fixed value of $\text{BER}_1$.
On the other hand, our proposed two-way coding schemes, i.e., linear and RNN-based coding, significantly improve $\text{BER}_1$ as $\text{SNR}^{\text{ch}}_2$ increases, due to the two-way interactions between the users,  which leads to substantial improvements in the sum-BER.

Fig.~\ref{fig:medlong}(\subref{fig:bler_medium}) shows the sum-BLER performance for medium block-length with $L_1=L_2=120$, where $\text{SNR}^{\text{ch}}_1=-1$dB and rate $r_1=r_2=1/3$.
Since the SNR condition is poor, i.e., $\text{SNR}^{\text{ch}}_1=-1$dB,  turbo coding does not perform well, resulting in a performance degradation in terms of sum-BLER.
However, the two-way coding schemes, in particular RNN-based coding, improve the error performance due to their robustness to channel noises, and ultimately balance the communication reliability, i.e., obtain much smaller sum-BLER values.

\subsection{Threshold for Two-Way Coding Benefits}
\label{ssec:sim:threshold}

\begin{figure}[t]
  \centering
\begin{subfigure}{.49\linewidth}
  \centering
  \includegraphics[width=\linewidth]{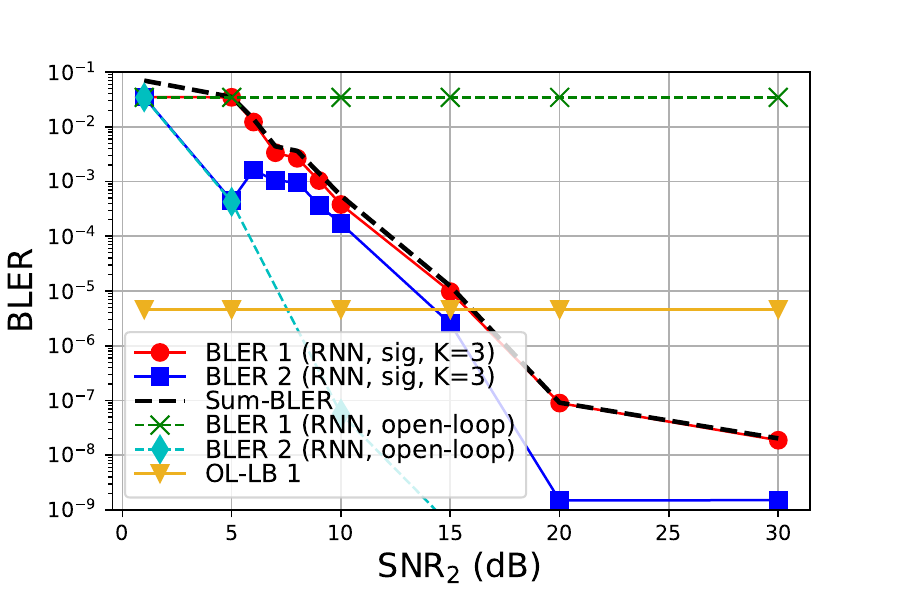}
  \caption{ $\text{SNR}^{\text{ch}}_1 = 1$dB. 
  }
  \label{fig:threshold:SNR1=1dB}
\end{subfigure}
\begin{subfigure}{.49\linewidth}
  \centering
  \includegraphics[width=\linewidth]{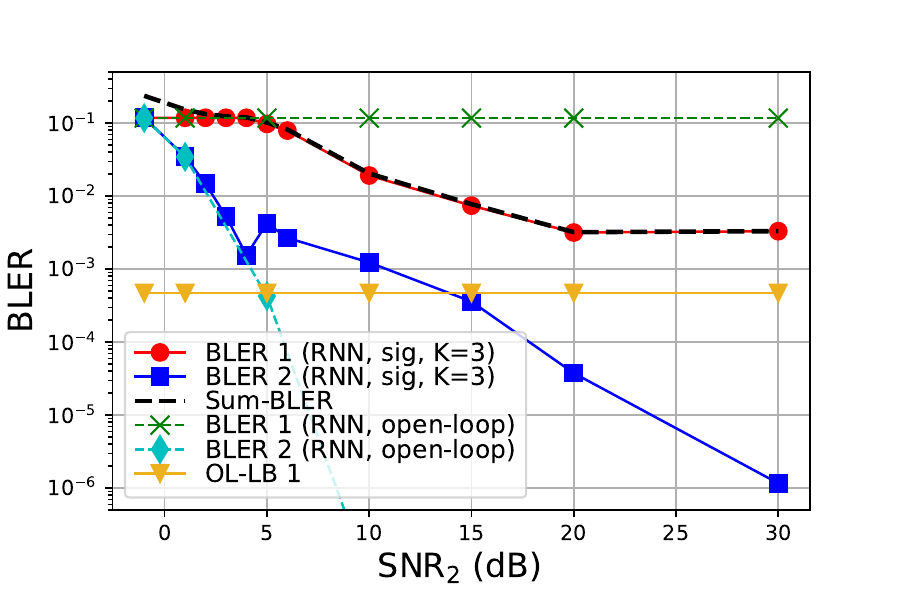}
  \caption{ $\text{SNR}^{\text{ch}}_1 = -1$dB.
  }
  \label{fig:threshold:SNR1=-1dB}
\end{subfigure}
  \caption{Each of BLER curves obtained by RNN-based coding with the sigmoid function and $K_1=K_2=3$ bits. The users start to benefit from two-way coding when $\text{SNR}^{\text{ch}}_2$ is larger than a threshold, which is 6dB in (a) and 5dB in (b), respectively.}
  \label{fig:threshold}
\end{figure}


Next, we investigate each error curve to analyze how communication reliability can be balanced by our two-way coding scheme under asymmetric channels. 
Fig.~\ref{fig:threshold} shows the BLER curves obtained by our RNN-based coding scheme with the sigmoid function and $K_1=K_2=3$ bits, where  BLER is measured over the  $K_1=K_2=3$ bits.
We consider $\text{SNR}^{\text{ch}}_1=1$dB in Fig.~\ref{fig:threshold}(\subref{fig:threshold:SNR1=1dB}) and $\text{SNR}^{\text{ch}}_1=-1$dB in Fig.~\ref{fig:threshold}(\subref{fig:threshold:SNR1=-1dB}).
To understand the interrelation between 
the two users through the proposed two-way coding, we introduce a baseline by modifying our RNN-based coding to generate  two independent open-loop codes. Specifically,  in Fig.~\ref{fig:RNN_compact}, we provide $x_i[k-1]$ as an input to the encoder of User $i$ rather than $y_i[k-1]$ to remove the dependency between the encoders of the users, which is denoted by ``RNN, open-loop" in the plots. 
Under the asymmetric channels with $\text{SNR}^{\text{ch}}_1 < \text{SNR}^{\text{ch}}_2$, $\text{BLER}_1$ and $\text{BLER}_2$ obtained by ``RNN, open-loop" is the upper and lower bound, respectively, of those obtained by our two-way coding scheme.
In the figures, $\text{BLER}_1$ with open-loop coding yields a constant value, since $\text{BLER}_1$ is only determined by the fixed value of $\text{SNR}^{\text{ch}}_1$, while $\text{BLER}_2$ with open-loop coding decreases  along $\text{SNR}^{\text{ch}}_2$.
We note that the two independent open-loop codes do not balance the communication reliability when the channels are asymmetric.

%
%
%

From the figures, we observe that our two-way coding scheme balances the communication reliability under asymmetric channels due to the interactive exchange of information between the users.
$\text{BLER}_1$ is improved relative to that of the open-loop codes, while $\text{BLER}_2$ is sacrificed as compared to the open-loop coding. 
This is because User 2 (with higher SNR) helps User 1 (with lower SNR)  through the interactive exchange of information to improve $\text{BLER}_1$ by implicitly providing feedback information to User 1.
This behavior is in line with the insight discussed in \cite{palacio2021achievable} that a user with lower channel noise can act as a helper.
Specifically, we can identify a \textit{threshold} of $\text{SNR}^{\text{ch}}_2$, above which the users start to benefit from the two-way interaction in minimizing sum-errors. 
When $\text{SNR}^{\text{ch}}_1=1$dB, the threshold is $\text{SNR}^{\text{ch}}_2=6$dB as indicated in Fig.~\ref{fig:threshold}(\subref{fig:threshold:SNR1=1dB}), 
whereas, the threshold is $\text{SNR}^{\text{ch}}_2=5$dB in Fig.~\ref{fig:threshold}(\subref{fig:threshold:SNR1=-1dB}) when $\text{SNR}^{\text{ch}}_1=-1$dB. Below the threshold, the BLERs of the users match the open-loop BLERs, which implies that User 2 does not provide feedback to User 1. 
%


We also observe that $\text{BLER}_2$ increases slightly at the threshold and then decreases afterwards, while $\text{BLER}_1$ starts to decrease at the threshold. Since our objective is to minimize the ``sum" of the BLERs, this behavior in the asymmetric channel indicates that User 2 begins to sacrifice its performance to improve User 1's performance.
This may lead to a slight increase in $\text{BLER}_2$ even with a higher SNR; for example, $\text{BLER}_2$ at $\text{SNR}^{\text{ch}}_2=6$dB is larger than at $\text{SNR}^{\text{ch}}_2=5$dB as shown in Fig.~\ref{fig:threshold}(\subref{fig:threshold:SNR1=1dB}).
Overall, it is important to note that sum-BLER (the dashed black line) starts to substantially decrease at the threshold.
In Fig.~\ref{fig:threshold}(\subref{fig:threshold:SNR1=1dB}), $\text{BLER}_1$ is even lower than OL-LB~1.\footnote{\rev{While we acknowledge that the lower bound, OL-LB~1, is  intended for the non-cooperative case, it still provides valuable insights into the extent to which error performance can be improved through cooperation.}}
This demonstrates that cooperation between the users enables User 1 to achieve a BLER lower than the lower bound achieved when cooperation is not allowed.


\subsection{The Superior Scheme over Different SNR Regions}
\label{ssec:sim:best}

\begin{figure}[t]
    \centering    \includegraphics[width=.5\linewidth]{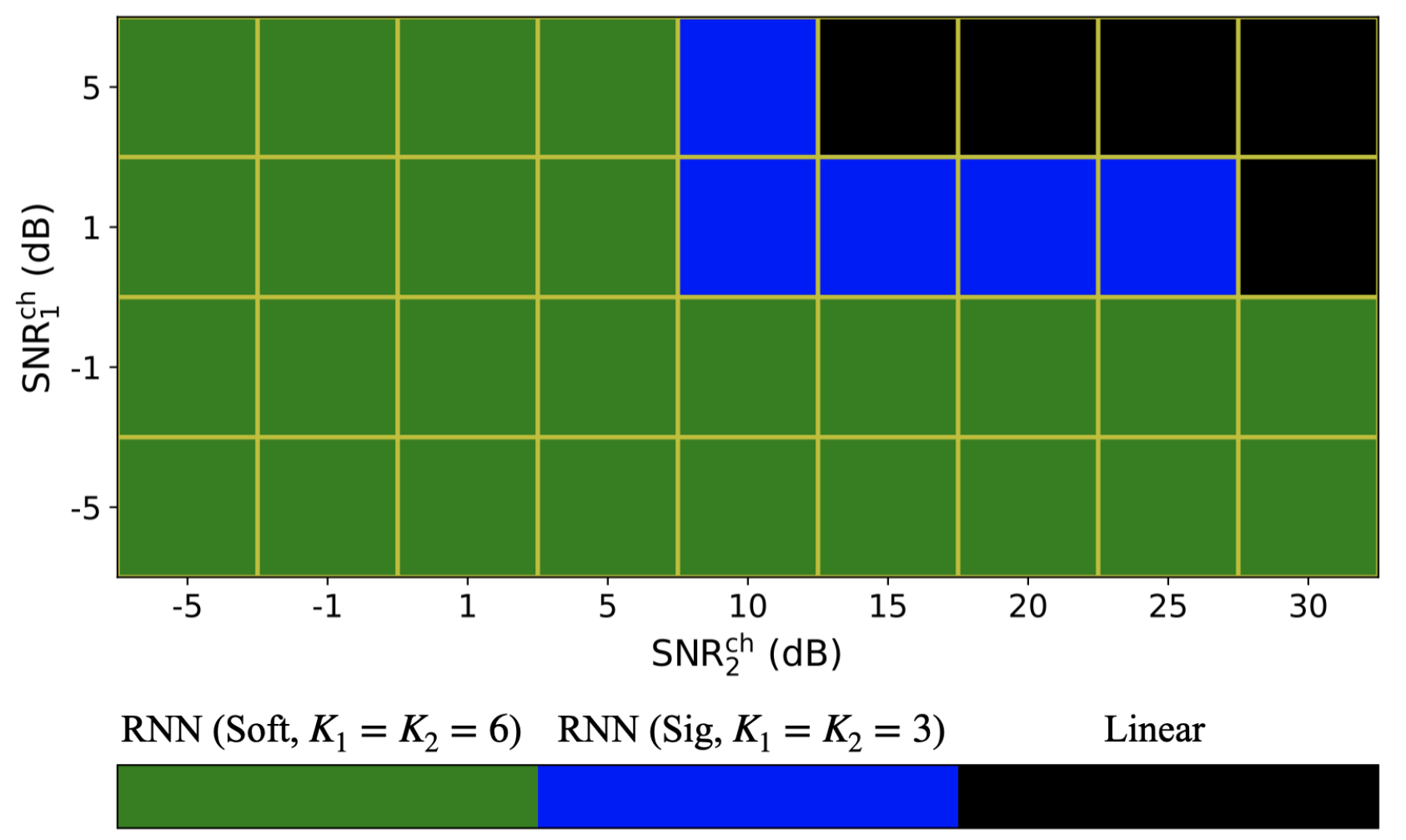}
    \caption{The superior scheme in sum-BLER for $L_1=L_2=6$ bits under various SNR scenarios. When SNRs are high and asymmetric, linear coding performs best, while RNN-based coding excels in low SNRs.}
    \label{fig:2D_best}
\end{figure}

Next, we explore different coding scheme among repetitive coding, TBCC, linear coding, and RNN-based coding in terms of sum-BLER for $L_1=L_2=6$ under various SNR scenarios. The  results are demonstrated in Fig.~\ref{fig:2D_best}.
In high and asymmetric SNR regions, we observe that linear coding  outperforms the other schemes. 
For moderate and asymmetric SNRs, RNN-based coding with the sigmoid function performs the best among the schemes.
In low SNR regions, RNN-based coding with the softmax function and $K_1=K_2=6$ bits yields the best performance, demonstrating its robustness to  channel noises.
At low/medium SNR, the noise component is relatively large, making the channel environment highly stochastic. Learning-based coding helps to extract features from this high-stochastic environment, while linear coding may not adapt well to high noise levels due to the limited degrees of freedom in coding caused by its simplified linear assumptions.
At high SNR, the noise component is relatively small, making the channel  environment more deterministic. 
Learning-based coding employing deep neural networks, such as in our RNN-based coding, may introduce unnecessary complexity in such straightforward conditions. In contrast, linear coding techniques often offer  analytical solutions, leading to  potentially better performance in low-randomness channel environments.
%
Under symmetric SNR scenarios ($\text{SNR}^{\text{ch}}_1=\text{SNR}^{\text{ch}}_2=-5,-1,1,5$dB), RNN-based coding behaves like two independent open-loop coding  when the channels are symmetric, as depicted in Fig.~\ref{fig:threshold}, which outperforms TBCC. This result suggests that our learning architecture can be also utilized to develop error correction codes. 
While our study focuses on two-way coding behavior, the construction of error correcting codes is a potential area for future research.

\subsection{Varying Rates}
\label{ssec:sim:coding_rate}

\begin{figure}[t]
    \centering
    \includegraphics[width=.5\linewidth]{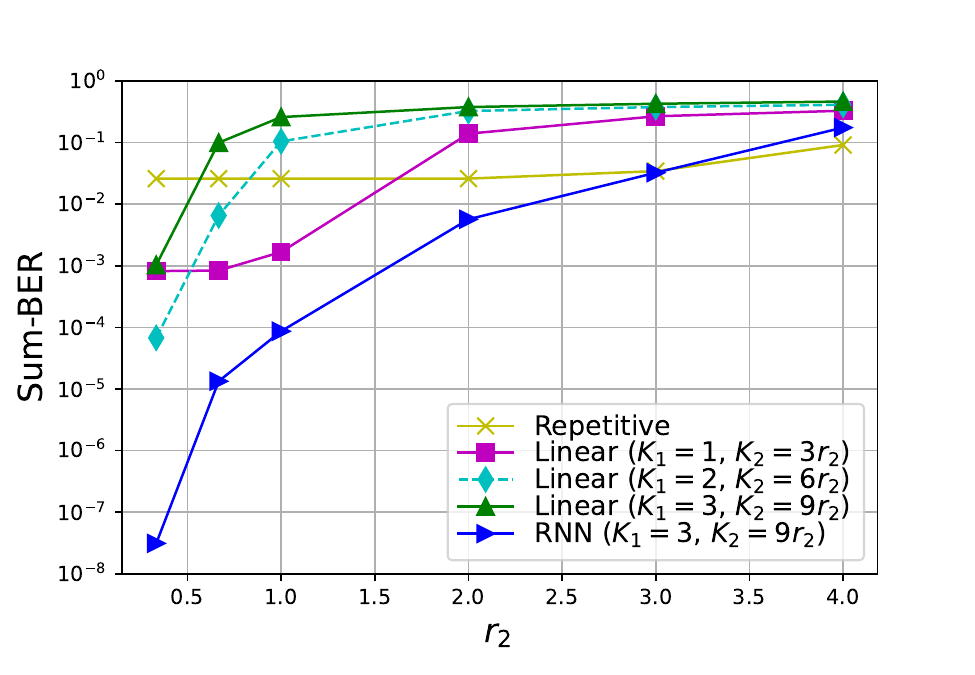}
    \caption{Sum-BER with various rates of $r_2$, where $\text{SNR}^{\text{ch}}_1 = 1$dB, $\text{SNR}^{\text{ch}}_2 = 20$dB, and $r_1=1/3$. RNN-based coding outperforms the counterparts in sum-BER even while supporting high rates.}
    \label{fig:coding_rate}
\end{figure}


We so far looked into the benefit of two-way coding in terms of improving sum-error performance when rates $r_i$ are fixed.
However, when $\text{SNR}^{\text{ch}}_2$ is high enough, while helping User 1 to improve sum-error, User 2 can increase its rate $r_2$. 
To investigate the relationship between the rates $r_i$ and sum-error, we show the sum-BER performances with varying rates of $r_2=K_2/N$ in Fig.~\ref{fig:coding_rate}, where $r_1=1/3$, $\text{SNR}^{\text{ch}}_1 = 1$dB, and $\text{SNR}^{\text{ch}}_2 = 20$dB.
For linear coding, we consider three different modulation bits at User 1, $K_1 = 1,2,3$. 
Since $N=K_1/r_1$ with $r_1=1/3$, we have $N=3,6,9$ for $K_1 = 1,2,3$, respectively.
Given $N$, we set the modulation bits at User 2 as $K_2=Nr_2$.
We consider varying $r_2$ from $1/3$ to $4$ in the simulation.\footnote{The rate $r_i$ in this work is defined as the number of bits per channel use. Our coding framework aligns better with coded modulation frameworks~\cite{ungerboeck1982channel} because it integrates both modulation and channel coding, particularly by incorporating cooperation into coding.}
We consider the strategy of alternate channel uses discussed in Sec.~\ref{ssec:linear:alternate} for linear coding, where two message pairs are exchanged in an alternate manner. 
Specifically, $2K_1$ and $2K_2$ 
bits are modulated to two messages at User $1$ denoted by $m_1, m_1'$ and at User $2$ denoted by $m_2, m_2'$, respectively, and the two pairs of  messages, $(m_1,m_2)$ and $(m_1',m_2')$, are exchanged over 
$2N$
channel uses in an alternate manner.
For RNN-based coding, the number of bits input to the encoders of Users 1 and 2 are $K_1=3$ and $K_2= N r_2= 9r_2$, respectively.
From the figure, we observe that RNN-based coding outperforms the counterparts even under very high rates, $r_2 \le 3$, while linear coding outperforms the repetitive coding when the rate is quite low, $r_2 \le 1.5$.








\subsection{Power Distribution in Linear and RNN-Based Coding}
\label{ssec:sim:power}

Next, we investigate the power distribution of linear coding and RNN-based coding. We consider that two users convey $L_1=L_2=6$ bits to one another over $N_\text{L}=18$ channel uses, where $\text{SNR}^{\text{ch}}_1 = 1$dB and $\text{SNR}^{\text{ch}}_2 = 20$dB. We consider $K_1=K_2$.
We note that the power distribution of linear coding and RNN-based coding differs based on the number of processed bits at a time, i.e., $K_1$ and $K_2$. 
In linear coding, $K_i$ bits are modulated into a message, and through the alternate channel use strategy, two message pairs are exchanged over $2N = 2N_\text{L}/L_i = 6K_i$ channel uses.  
In RNN-based coding, $K_i$ bits are input to the RNN-based coding architecture, which generates $N = N_\text{L}/L_i = 3K_i$ transmit symbols. This means that a message pair each containing $K_i$ bits is exchanged over  $N = 3K_i$ channel uses.
Therefore, depending on the processed bits at a time ($K_i$), the utilization of the channel uses will be different, which is reflected in the power distribution over the channel uses.

\begin{figure}[t]
  \centering
\begin{subfigure}{.49\linewidth}
  \centering
  \includegraphics[width=\linewidth]{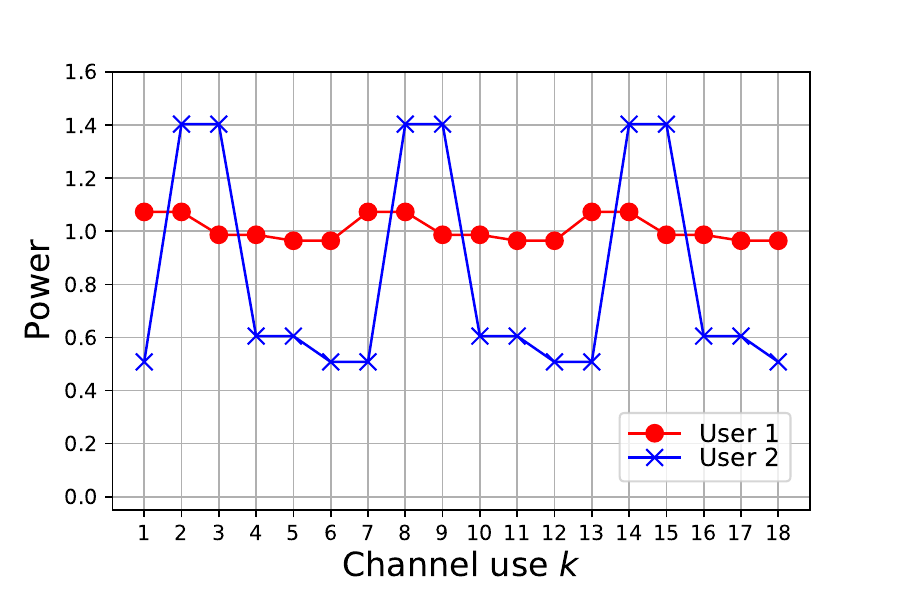}
  \caption{ $K_1=K_2=1$.
  }
  \label{fig:power_linear:B1}
\end{subfigure}
\begin{subfigure}{.49\linewidth}
  \centering
  \includegraphics[width=\linewidth]{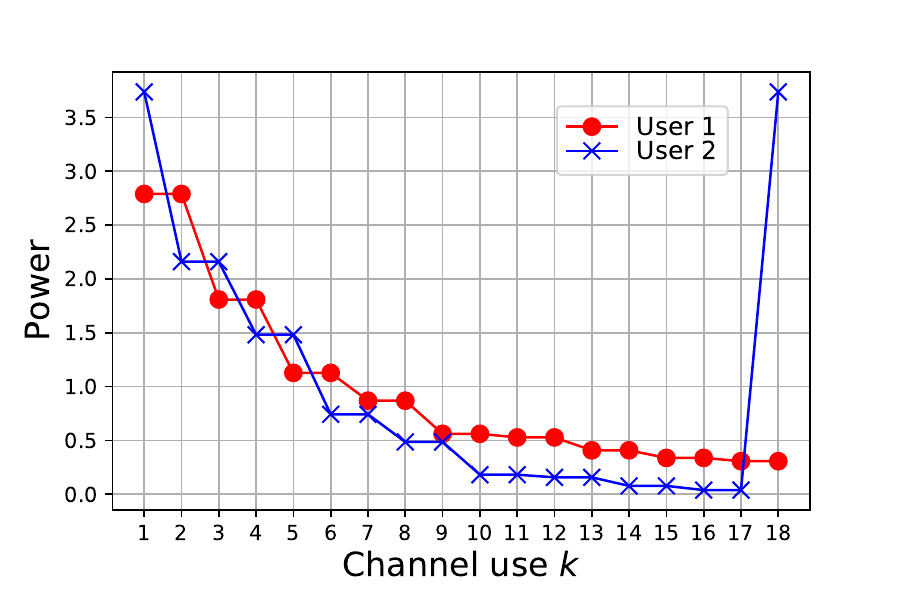}
  \caption{$K_1=K_2=3$.
  }
  \label{fig:power_linear:B3}
\end{subfigure}
  \caption{Power distribution from linear coding along the channel uses when adopting the strategy of alternate channel uses, where $L_1=L_2=6$ and $N_\text{L}=18$. In this way, the users can fully utilize the channel uses.}
  \label{fig:power_linear}
\end{figure}

First, we consider linear coding, where we adopt the strategy of alternate channel uses, discussed in Sec.~\ref{ssec:linear:alternate}.
We consider two different processing bits, $K_1=K_2=1$ in Fig.~\ref{fig:power_linear}(\subref{fig:power_linear:B1}) and $K_1=K_2=3$ in Fig.~\ref{fig:power_linear}(\subref{fig:power_linear:B3}). 
First, for the case of $K_1=K_2=1$ in Fig.~\ref{fig:power_linear}(\subref{fig:power_linear:B1}), each User $i$ has $M_i=L_i/K_i=6$ message symbols, and there are 6 message pairs to be exchanged.
Through the alternate channel use strategy,
every two message pairs are transmitted over 6 channel uses, and the pattern of power distribution during the 6 channel uses are repeated three times to exchange the 6 message pairs.
Specifically, during the first 6 channel uses, to exchange the first message pair, User 1 uses the 1st, 3rd, and 5th channel uses, and User 2 uses the 2nd, 4th, and 6th channel uses, while to exchange the second message pair, User 1 uses the 2nd, 4th, and 6th channel uses, and User 2 uses the  1st, 3rd, and 5th channel uses, as discussed in \eqref{eq:alternate_channel}.
We observe that User 1 slightly decreases its power across the 6 channel uses, while User 2 also decreases its power across the channel uses except when transmitting its message in the 1st and 6th channel uses.
For the case of $K_1=K_2=3$ in Fig.~\ref{fig:power_linear}(\subref{fig:power_linear:B3}), each user has $M_i=L_i/K_i=2$ message symbols, and the two message pairs are transmitted over 18 channel uses via alternate channel uses.
Since the alternate channel use strategy is used, to exchange the first message pair, User 1 uses odd-numbered channel uses, and User 2 uses
the even-numbered channel uses, while to exchange the second message pair, User 1 uses odd-numbered channel uses, and User 2 uses even-numbered channel uses. Similarly, Users 1 and 2 decrease their power across channel uses, with User 2 using the first and last channel uses to transmit its two respective messages.
\begin{figure}[t]
  \centering
\begin{subfigure}{.49\linewidth}
  \centering
  \includegraphics[width=\linewidth]{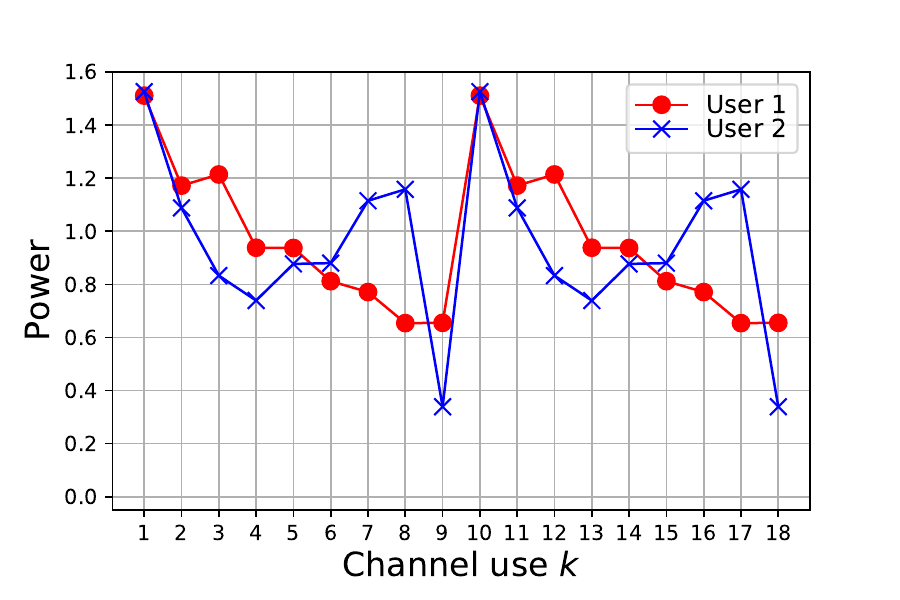}
  \caption{ $K_1=K_2=3$.
  }
  \label{fig:power_RNN:K3:sig}
\end{subfigure}
\begin{subfigure}{.49\linewidth}
  \centering
  \includegraphics[width=\linewidth]{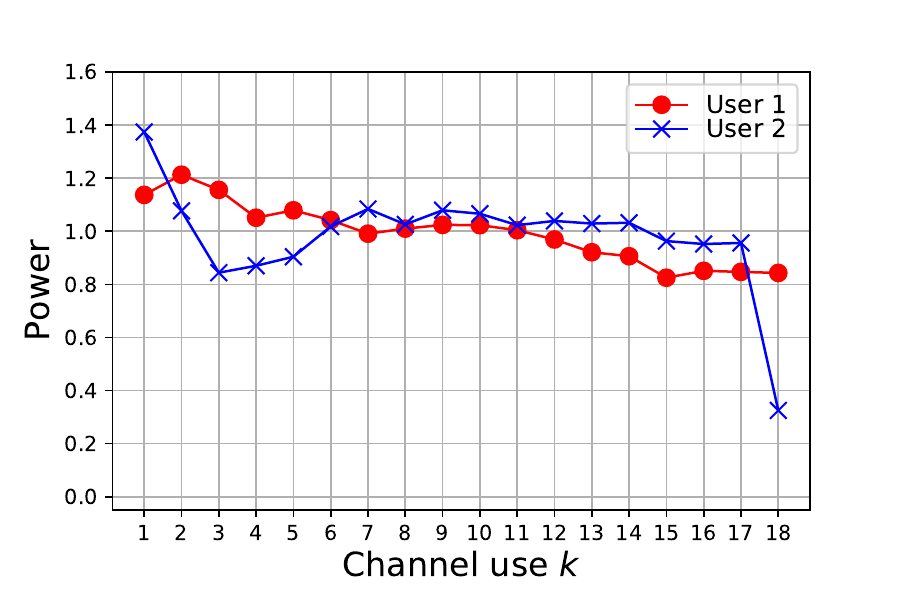}
  \caption{$K_1=K_2=6$.
  }
  \label{fig:power_RNN:K6:soft}
\end{subfigure}
  \caption{Power distribution from RNN-based coding  where $L_1=L_2=6$ and $N_\text{L}=18$. Under this asymmetric channel ($\text{SNR}^{\text{ch}}_1 = 1$dB $<$ $\text{SNR}^{\text{ch}}_2 = 20$dB), User 1 allocates more power at the beginning to fully exploit the feedback information provided by User 2, in order to improve sum-error performance.}
  \label{fig:power_RNN}
\end{figure} 

Next, we consider RNN-based coding, where the softmax function is used for decoding. We consider two different numbers of processing bits, $K_1=K_2=3$ bits in Fig.~\ref{fig:power_RNN}(\subref{fig:power_RNN:K3:sig}) and $K_1=K_2=6$ bits in Fig.~\ref{fig:power_RNN}(\subref{fig:power_RNN:K6:soft}).
For the case of $K_1=K_2=3$ bits in Fig.~\ref{fig:power_RNN}(\subref{fig:power_RNN:K3:sig}), a pair of  3 bits is input to the RNN-based coding architecture at a time to generate transmit symbols for the 9 channel uses, and this process is repeated twice to exchange a total of $L_1=L_2=6$ bits. 
In Fig.~\ref{fig:power_RNN}(\subref{fig:power_RNN:K3:sig}), we observe that during the 9 channel uses, User 1 decreases its power.
The behavior of User 2's power is less interpretable due to the complexity of analyzing neural networks; however, it is clear that User 2 utilizes the 9 channel uses to provide feedback information as well as to transmit its message.
For the case of $K_1=K_2=6$ bits in Fig.~\ref{fig:power_RNN}(\subref{fig:power_RNN:K6:soft}), a pair of 6 bits is input to the RNN-based coding architecture to generate transmit symbols for 18 channel uses. Similarly, we observe that User 1 decreases its power across the channel uses.
The main difference in power distribution between  RNN-based coding and linear coding is the repetition pattern. As an example, we consider RNN-based coding in Fig.~\ref{fig:power_RNN}(\subref{fig:power_RNN:K3:sig}) and linear coding in Fig.~\ref{fig:power_linear}(\subref{fig:power_linear:B3}), both with $K_1=K_2=3$. In RNN-based coding in Fig.~\ref{fig:power_RNN}(\subref{fig:power_RNN:K3:sig}), the power distribution pattern repeats every $N=9$ channel uses. In contrast, in linear coding in Fig.~\ref{fig:power_linear}(\subref{fig:power_linear:B3}), the repeated pattern is observed in alternating channel uses over $2N=18$ channel uses. 
The difference arises because, in RNN-based coding, the two messages are transmitted successively, whereas in linear coding, the two messages are transmitted alternately over the channel uses through the alternate channel use strategy.

The key message conveyed from the power distribution results in Figs.~\ref{fig:power_linear} and \ref{fig:power_RNN} is that the helper concept in two-way channels~\cite{palacio2021achievable} can be related to the transmitter-receiver relationship in feedback-enabled one-way channels~\cite{butman1969general,schalkwijk1966coding, chance2011concatenated,kim2020deepcode,kim2023feedback} in terms of power utilization.  Specifically, from the power distribution of linear and RNN-based coding under this asymmetric SNR scenario ($\text{SNR}^{\text{ch}}_1 < \text{SNR}^{\text{ch}}_2$) in Figs.~\ref{fig:power_linear} and \ref{fig:power_RNN}, 
we commonly observe that User 1 allocates more power at the beginning and decreases its power across the channel uses during the transmission of a message pair (over 6 channel uses in Fig.~\ref{fig:power_linear}(\subref{fig:power_linear:B1}), 9 channel uses in Fig.~\ref{fig:power_RNN}(\subref{fig:power_RNN:K3:sig}), and 18 channel uses in Fig.~\ref{fig:power_linear}(\subref{fig:power_linear:B3}) and Fig.~\ref{fig:power_RNN}(\subref{fig:power_RNN:K6:soft})).
Understanding that a user with lower channel noise acts as a helper~\cite{palacio2021achievable}, we see that User 2 acts as a helper to provide feedback to User 1 to improve User 1's error performance at the expense of sacrificing its own error performance, as demonstrated in Fig.~\ref{fig:threshold}. 
This helper concept is closely related to  feedback-enabled one-way channels~\cite{butman1969general,schalkwijk1966coding, chance2011concatenated,kim2020deepcode,kim2023feedback}, where the receiver helps the transmitter by providing feedback to improve performance.
In this context, User 1 functions similarly to a transmitter as the recipient of help in feedback-enabled one-way channels, while User 2 acts as a receiver providing feedback to User 1. 
It is interesting to see that the power distribution of the transmitter in feedback-enabled one-way channels, as observed in \cite{schalkwijk1966coding, chance2011concatenated, kim2023feedback}, is similar to that of User 1 in two-way channels in terms of decreasing power across the channel uses.
To further understand the behavior of linear and RNN-based coding, we have included additional simulations results on scatter plots of transmit symbols across channel uses in Appendix~\ref{app:scatter:plot}.

\subsection{Block-Length Gain of Two-Way Coding}
\label{ssec:sim:blockgain}

\begin{table}[t]
\captionsetup{justification=centering}
\caption{Block-length gain of linear (two-way) coding along different numbers of processing bits $K=K_1=K_2$ when $\text{SNR}^{\text{ch}}_1=1$dB, $\text{SNR}^{\text{ch}}_2=20$dB, and $r_1=r_2=1/3$.}
\centering
\begin{tabular}{
|c||c|c|c||c|c|c| }
 \hline
 & $\text{BER}_1$ & $\text{BER}_2$ & Sum-BER & $\text{BLER}_{1,K}$ & $\text{BLER}_{2,K}$ & Sum-BLER ($L=60$) \\
 \hline
 $K=6$ & 1.47E-1 & 5.07E-2 & 1.98E-1 & 6.72E-1 & 3.02E-1 & 1.97 \\
 \hline
 $K=5$ & 9.80E-2 & 1.20E-2 & 1.10E-1 & 4.60E-1 & 6.00E-2 & 1.52 \\
 \hline
 $K=4$   & 1.66E-2 & 3.66E-3 & 2.03E-2 & 6.66E-2 & 1.47E-2 & 8.44E-1 \\
 \hline
 $K=3$   & 5.51E-4 & 8.36E-4 & 1.39E-3 & 3.31E-3 & 5.08E-3 & 1.61E-1 \\
 \hline
  $K=2$   & 6.44E-5 & 2.70E-7 & \textbf{6.47E-5} & 1.30E-4 & 4.66E-7 & \textbf{3.90E-3} \\
 \hline
  $K=1$   & 7.76E-4 & $<$1E-12 & 7.76E-4 & 7.76E-4 & $<$1E-12 & 4.55E-2 \\
 \hline
\end{tabular}
\label{table:block_gain:SNR=1dB:linear}
\end{table}

\begin{table}[t]
\captionsetup{justification=centering}
\caption{Block-length gain of RNN-based (two-way) coding along different numbers of processing bits $K=K_1=K_2$ when $\text{SNR}^{\text{ch}}_1=1$dB, $\text{SNR}^{\text{ch}}_2=20$dB, and $r_1=r_2=1/3$.}
\centering
\begin{tabular}{
|c||c|c|c||c|c|c| }
 \hline
 & $\text{BER}_1$ & $\text{BER}_2$ & Sum-BER & $\text{BLER}_{1,K}$ & $\text{BLER}_{2,K}$ & Sum-BLER ($L=60$) \\
 \hline
 $K=6$ & 1.13E-6 & 3.04E-8 & 1.16E-6 & 4.65E-6 & 1.63E-7 & 4.81E-5 \\
 \hline
 $K=5$ & 1.91E-6 & 1.55E-8 & 1.93E-6 & 7.91E-6 & 5.7E-08 & 9.56E-5 \\
 \hline
 $K=4$   & 1.13E-6 & 1.93E-8 & 1.15E-6 & 4.08E-6 & 6.11E-8 & 6.21E-5 \\
 \hline
 $K=3$   & 3.03E-8 & 7.67E-10 & \textbf{3.11E-8} & 8.96E-8 & 1.50E-9 & \textbf{1.82E-6} \\
 \hline
  $K=2$   & 1.18E-7 & 1.50E-9 & 1.20E-7 & 2.35E-7 & 3.00E-9 & 7.14E-6 \\
 \hline
  $K=1$   & 8.00E-4 & 1.02E-6 & 8.01E-4 & 8.00E-4 & 1.02E-6 & 4.69E-2 \\
 \hline
\end{tabular}
\label{table:block_gain:SNR=1dB:RNN}
\end{table}

Next, we investigate the block-length gain of two-way coding for GTWCs, where the block-length gain is measured as the sum-error, either sum-BLER or sum-BER.
For two-way coding, we consider linear coding and RNN-based coding with $r_1=r_2=1/3$. 
We consider two different channel scenarios: $(\text{SNR}_1, \text{SNR}_2) = (1, 20)$dB and $(\text{SNR}_1, \text{SNR}_2) = (-1, 20)$dB.
We have four results summarized in respective tables: 
(i) linear coding under $(\text{SNR}_1, \text{SNR}_2) = (1, 20)$dB in Table~\ref{table:block_gain:SNR=1dB:linear}, 
(ii) RNN-based coding under $(\text{SNR}_1, \text{SNR}_2) = (1, 20)$dB in Table~\ref{table:block_gain:SNR=1dB:RNN}, 
(iii) linear coding under $(\text{SNR}_1, \text{SNR}_2) = (-1, 20)$dB in Table~\ref{table:block_gain:SNR=-1dB:linear}, and 
(iv) RNN-based coding under $(\text{SNR}_1, \text{SNR}_2) = (-1, 20)$dB in Table~\ref{table:block_gain:SNR=-1dB:RNN}.
Overall, we investigate the sum-BER and sum-BLER performance by varying the numbers of processing bits $K=K_1=K_2$. 
We introduce the measure ``Sum-BLER ($L=60$)" or ``Sum-BLER ($L=120$)" to fairly compare the BLER performances with different values of $K$, which is calculated as $1-(1-\text{BLER}_{1,K})^{L/K} + 1-(1-\text{BLER}_{2,K})^{L/K}$, where $\text{BLER}_{i,K}$ is the BLER of $K$ bits of User $i$.
For RNN-based coding, we consider the sigmoid function in (ii) and the softmax function in (iv).

%

For linear coding, both sum-BER and sum-BLER peak at $K=2$ in both  channel SNR scenarios, $(\text{SNR}_1, \text{SNR}_2) = (1, 20)$dB in Table~\ref{table:block_gain:SNR=1dB:linear} and $(\text{SNR}_1, \text{SNR}_2) = (-1, 20)$dB in Table~\ref{table:block_gain:SNR=-1dB:linear}. This indicates that, under these channel SNRs, up to $K=2$, the feedback benefit from an increasing number of channel uses $N=K_i/r_i$ outweighs the performance loss from increasing modulation order $2^K$. However, beyond $K=2$, the  performance loss by increasing modulation order outweighs the feedback benefit. It is important to note that the best choice of $K$  varies with different channel SNRs. For example, while the best performing $K$ is $K=2$ when $(\text{SNR}_1, \text{SNR}_2) = (1, 20)$dB as shown in Table~\ref{table:block_gain:SNR=1dB:linear}, the best performing $K$ when $(\text{SNR}_1, \text{SNR}_2) = (1, 30)$dB is $K=3$, as shown in Fig.~\ref{fig:short}(\subref{fig:short:SNR1=1}).

For RNN-based coding, in the case of $(\text{SNR}_1, \text{SNR}_2) = (1, 20)$dB in Table~\ref{table:block_gain:SNR=1dB:RNN}, both sum-BER and sum-BLER peak at $K=3$, implying that our RNN-based coding exploits the block-length gain up to $K=3$.
In the case of $(\text{SNR}_1, \text{SNR}_2) = (-1, 20)$dB, 
in Table~\ref{table:block_gain:SNR=-1dB:RNN}, 
sum-BER peaks at $K=4$, while sum-BLER  peaks at $K=6$.
For RNN-based coding, we observe that in the low SNR scenario, i.e., $\text{SNR}_1=-1$dB in Table~\ref{table:block_gain:SNR=-1dB:RNN}, more numbers of bits for coding yields better sum-error performance than the case with the higher SNR, i.e., $\text{SNR}_1=1$dB in Table~\ref{table:block_gain:SNR=1dB:RNN}. 
This is because two-way coding potentially benefits from noise-averaging effect by utilizing more numbers of channel uses under low SNR scenarios. 


\begin{table}[t]
\captionsetup{justification=centering}
\caption{Block-length gain of linear (two-way) coding along different numbers of processing bits $K=K_1=K_2$ when $\text{SNR}^{\text{ch}}_1=-1$dB, $\text{SNR}^{\text{ch}}_2=20$dB, and $r_1=r_2=1/3$.}
\centering
\begin{tabular}{
|c||c|c|c||c|c|c| }
 \hline
 & $\text{BER}_1$ & $\text{BER}_2$ & Sum-BER & $\text{BLER}_{1,K}$ & $\text{BLER}_{2,K}$ & Sum-BLER ($L=120$) \\
 \hline
 $K=8$ & 2.40E-1 & 1.41E-1 & 3.81E-1 & 9.17E-1 & 7.60E-1 & 2.00 \\
 \hline
 $K=7$ & 2.12E-1 & 9.38E-2 & 3.06E-1 & 8.49E-1 & 5.68E-1 & N/A \\
 \hline
 $K=6$ & 1.98E-1 & 4.56E-2 & 2.44E-1 & 7.76E-1 & 2.73E-1 & 2.00 \\
 \hline
 $K=5$ & 1.01E-1 & 1.44E-2 & 1.15E-1 & 5.04E-1 &	7.20E-2 & 1.83 \\
 \hline
 $K=4$   & 3.66E-2 & 8.51E-3 & 4.51E-2 & 1.46E-1 & 3.40E-2 & 1.64 \\
 \hline
 $K=3$   & 1.29E-2	 & 8.46E-4 & 1.37E-2 & 3.86E-2 & 2.54E-3 & 8.90E-1 \\
 \hline
  $K=2$   & 7.59E-3 & 1.56E-4 & \textbf{7.75E-3} & 1.52E-2 & 3.17E-4 & \textbf{6.20E-1} \\
 \hline
  $K=1$   & 1.52E-2 & 2.21E-8 & 1.52E-2  & 1.52E-2 & 2.21E-8  & 8.41E-1 \\
 \hline
\end{tabular}
\label{table:block_gain:SNR=-1dB:linear}
\end{table}

\begin{table}[t]
\captionsetup{justification=centering}
\caption{Block-length gain of RNN-based (two-way) coding along different numbers of processing bits $K=K_1=K_2$ when $\text{SNR}^{\text{ch}}_1=-1$dB, $\text{SNR}^{\text{ch}}_2=20$dB, and $r_1=r_2=1/3$.}
\centering
\begin{tabular}{
|c||c|c|c||c|c|c| }
 \hline
 & $\text{BER}_1$ & $\text{BER}_2$ & Sum-BER & $\text{BLER}_{1,K}$ & $\text{BLER}_{2,K}$ & Sum-BLER ($L=120$) \\
 \hline
 $K=8$ & 4.57E-3 & 2.55E-4 & 4.83E-3 & 1.08E-2 & 6.67E-4 & 1.60E-1 \\
 \hline
 $K=7$ & 1.57E-3 & 1.28E-4 & 1.70E-3 & 3.80E-3 & 3.42E-4 & N/A \\
 \hline
 $K=6$ & 9.96E-4 & 1.20E-4 & 1.12E-3 & 2.14E-3 & 2.60E-4 & \textbf{4.71E-2} \\
 \hline
 $K=5$ & 1.27E-3 &	8.87E-5 & 1.36E-3 & 2.90E-3 &	2.32E-4 & 7.29E-2 \\
 \hline
 $K=4$   & 7.10E-4 & 2.24E-5 & \textbf{7.32E-4} & 1.87E-3 & 5.89E-5 & 5.64E-2 \\
 \hline
 $K=3$   & 2.26E-3	 & 3.53E-5 & 2.30E-3 & 4.45E-3 & 6.42E-5 & 1.66E-1 \\
 \hline
  $K=2$   & 4.54E-3 & 2.20E-5 & 4.56E-3 & 8.86E-3 & 3.26E-5 & 4.16E-1 \\
 \hline
  $K=1$   & 1.94E-2 & 2.92E-6 & 1.94E-2  & 1.94E-2 & 2.92E-6  & 9.05E-1 \\
 \hline
\end{tabular}
\label{table:block_gain:SNR=-1dB:RNN}
\end{table}

It is important to note that, for RNN-based coding, at specific SNR scenarios, the best sum-error performance is attained  through (i) the appropriate selection of $K$ and (ii) a proper choice between sigmoid and softmax.
As a result, given a certain channel environment, we can select the best-performing coding architecture from a set of pre-trained architectures that have been trained with various $K$ and activation functions.








\subsection{Experimental Validations for Linear Coding}
\label{ssec:validation_linear_coding}


\begin{figure}[t]
    \centering
    \includegraphics[width=.5\linewidth]{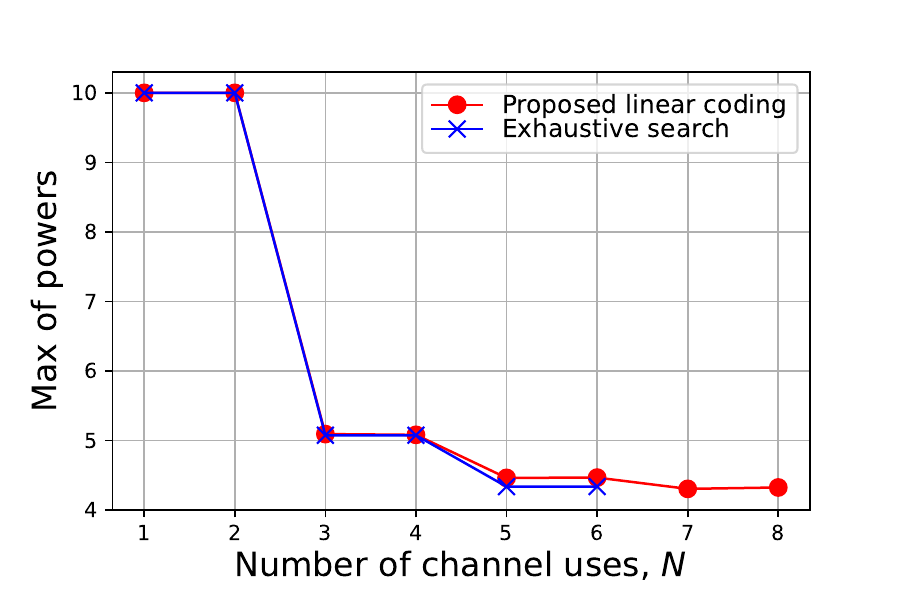}
    \caption{\rev{The max of  power values across various numbers of channel uses $N$ when  $\eta_1=\eta_2=10$, $\text{SNR}^{\text{ch}}_1 = 0$dB, and $\text{SNR}^{\text{ch}}_2 = 10$dB. Assuming that the exhaustive search yields optimal solutions, the optimality loss of our approach (by solving \eqref{opt:linear:weighted-sum} instead of \eqref{opt:linear:max-of-powers:SNR}) is observed to be less than 3\% up to $N=6$.}}
    \label{fig:max_power}
\end{figure}

\rev{We first discuss the potential of optimality loss of our linear coding strategy in Sec.~\ref{ssec:linear:opt:weighted-sum} when solving \eqref{opt:linear:weighted-sum}. This loss may arise due to (i) the use of the weighted sum
method for multi-objective optimization, (ii) the choice of the simplified ${\bf F}_2 \in \mathcal{F}_2$ in \eqref{eq:F2}, and (iii) the use of the iterative optimization strategy in Algorithm~\ref{al:linear:weighted_sum_power}.
To demonstrate how much the performance of our coding strategy deviates from optimal performance, we adopt an exhaustive search method as our benchmark that solves \eqref{opt:linear:max-of-powers:SNR}.
In the exhaustive search, we conduct an exhaustive search for ${\bf g}_1$, ${\bf F}_1$, ${\bf g}_2$, and ${\bf F}_2$, to minimize the max of powers in \eqref{opt:linear:max-of-powers:SNR} while satisfying the SNR constraints of Users 1 and 2.
Fig. \ref{fig:max_power} demonstrates the objective value in \eqref{opt:linear:max-of-powers:SNR}, i.e., max of powers, across various numbers of channel uses, $N$, when $\eta_1=\eta_2=10$, $\text{SNR}^{\text{ch}}_1 = 0$dB, and $\text{SNR}^{\text{ch}}_2 = 10$dB. 
Since the exhaustive search incurs huge computational complexity, we could obtain results only for small values of $N$, specifically up to $N=6$.
For $N=1,2$, the two users do not use any cooperation. When $N\ge 3$, the two users  cooperate, thereby reducing the max of powers. We find that the performance gap between our coding strategy and the exhaustive search is less than 3\% at each $N$. 
This implies that the benefit of analytical tractability from solving \eqref{opt:linear:weighted-sum} instead of \eqref{opt:linear:max-of-powers:SNR} might outweigh the low optimality loss in performance.}

\begin{figure}[t]
    \centering
    \includegraphics[width=.5\linewidth]{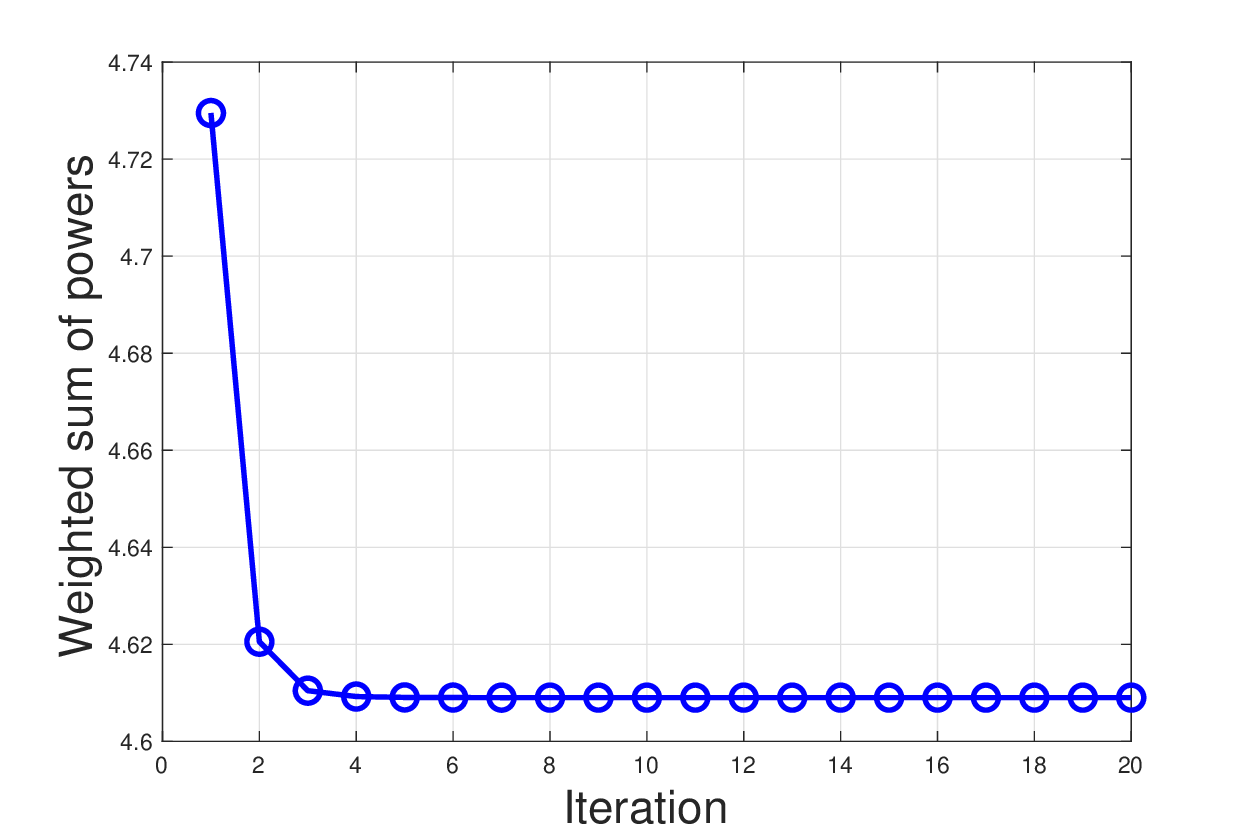}
    \caption{Convergence of the weighted sum power value through the iterations in our iterative optimization for linear coding, with $N=6$, $\alpha=0.7$, $\eta_1=\eta_2=10$, $\text{SNR}^{\text{ch}}_1 = 0$dB, and $\text{SNR}^{\text{ch}}_2 = 10$dB. The objective value decreases and converges to a stable value across the iterations.}
    \label{fig:convergence}
\end{figure}

Next, we discuss the convergence of the iterative optimization for linear coding in Algorithm~\ref{al:linear:weighted_sum_power}, discussed in Sec.~\ref{ssec:linear:opt:weighted-sum:sol}. 
While this iterative process does not guarantee convergence, which is challenging to prove theoretically, we empirically demonstrate convergence through numerical experiments shown in Fig.~\ref{fig:convergence}.
For simulations, we consider $N=6$, $\alpha=0.7$, $\eta_1=\eta_2=10$, $\text{SNR}^{\text{ch}}_1 = 0$dB, and $\text{SNR}^{\text{ch}}_2 = 10$dB
We conduct 30 Monte Carlo simulations and average the objective value (the weighted sum of power values in~\eqref{opt:linear:weighted-sum}) at each iteration. 
Fig.~\ref{fig:convergence} demonstrates that the objective value decreases and converges to a stable value across the iterations of the first and second sub-problems, respectively, in Sec.~\ref{sssec:problem1} and \ref{sssec:problem2}.

\begin{figure}[t]
    \centering
    \includegraphics[width=.5\linewidth]{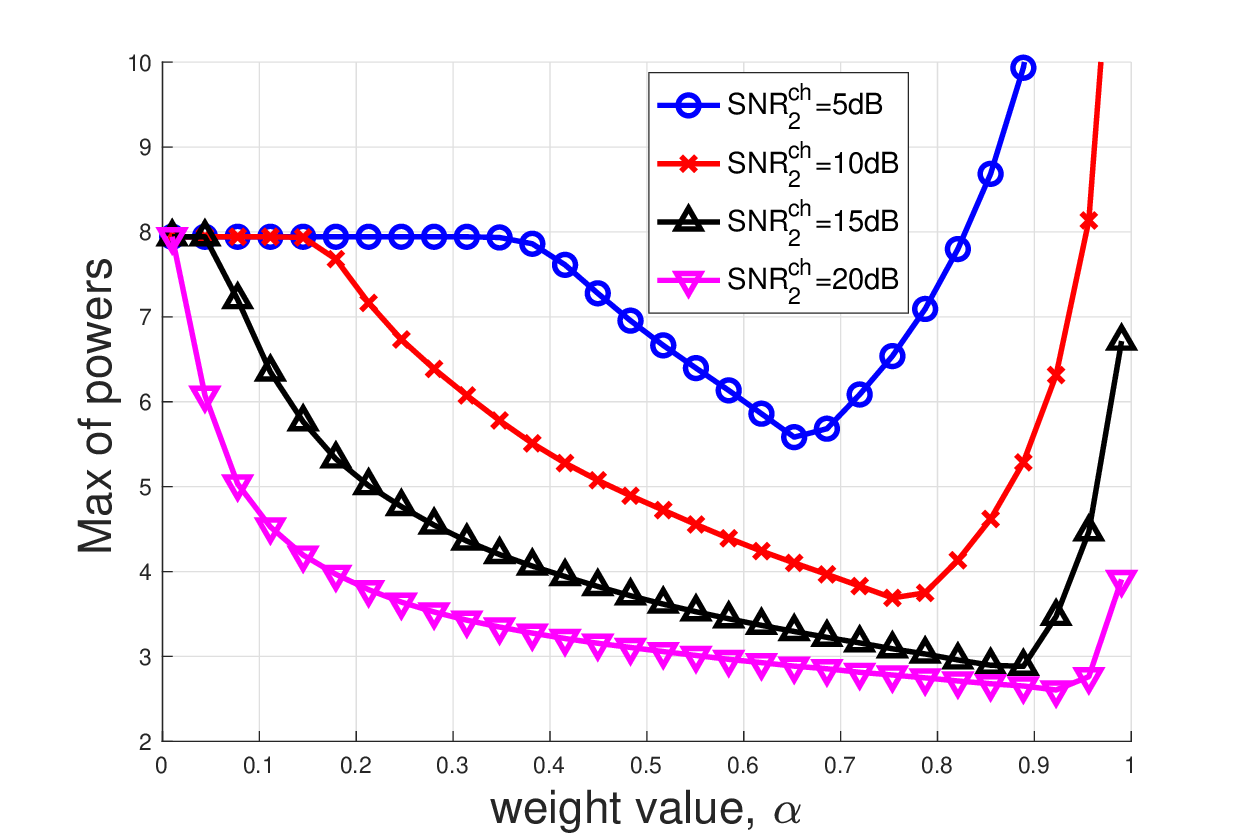}
    \caption{The max-power value, ${\max} \big\{ \mathbb{E}_{\rev{m_1,m_2,{\bf n}_1, {\bf n}_2}}\big[\|{\bf x}_1 \|^2\big], \mathbb{E}_{\rev{m_1,m_2,{\bf n}_1, {\bf n}_2}}\big[\|{\bf x}_2 \|^2\big] \big\}$, obtained by solving the weighted sum problem in~\eqref{opt:linear:weighted-sum} across various $\alpha \in (0, 1)$. The max-power value is shown to be unimodal across different values of $\text{SNR}^{\text{ch}}_2$, which experimentally supports the validity of the unimodality assumption.}
    \label{fig:unimodality}
\end{figure}

Finally, we discuss the unimodality assumption made in Sec.~\ref{sssec:alpha}.
We have conducted numerical simulations in Fig.~\ref{fig:unimodality} to experimentally demonstrate the validity of the unimodality assumption.
Fig.~\ref{fig:unimodality} depicts the objective value, ${\max} \big\{ \mathbb{E}_{\rev{m_1,m_2,{\bf n}_1, {\bf n}_2}}\big[\|{\bf x}_1 \|^2\big], \mathbb{E}_{\rev{m_1,m_2,{\bf n}_1, {\bf n}_2}}\big[\|{\bf x}_2 \|^2\big] \big\}$ in~\eqref{opt:linear:max-of-powers:SNR}, that is obtained by solving the weighted sum problem in~\eqref{opt:linear:weighted-sum} across various $\alpha \in (0, 1)$.
For simulations, we consider $\eta_1=10$, $\eta_2=10$,  $\text{SNR}^{\text{ch}}_1=1$dB, and $N=10$.
In Fig.~\ref{fig:unimodality}, we observe that the max-power value is unimodal, i.e., it decreases monotonically up to a certain point (the minimizer) and then increases monotonically, across different values of $\text{SNR}^{\text{ch}}_2$, which experimentally supports the validity of the unimodality assumption.



\section{Conclusion}

In this work, we focused on balancing the communication reliability between the two users in GTWCs by
minimizing the sum of error probabilities of the users through the design of encoders/decoders for the users. 
We first provided general encoding/decoding functions, and formulated an optimization problem, aiming to minimize the sum-error of the users subject to users' power constraints.
We proposed two coding strategies -- linear coding and learning-based coding -- to address  the challenges of (i) the encoders' coupling effect, (ii) the requirement for effective decoding, and (iii) the need for efficient power management.

For linear coding, we adopted a linear processing for encoding/decoding, which mitigates the complex coupling effect in challenge (i). 
We then derived an optimal form of decoding as a function of encoding schemes, which addresses challenge (ii). 
Next, we solved the sum-error minimization problem subject to power constraints, which addresses challenge (iii). Furthermore, we provided new insights on user cooperation by analyzing the relationship between the channel noise ratio and the weight imposed in weighted sum-power minimization.
%
For learning-based coding, we proposed an 
RNN-based coding architecture composed of multiple novel components. For encoding, we proposed interactive RNNs for addressing challenge (i) and a power control layer for addressing challenge (iii), while for decoding we incorporated bi-directional RNNs with an attention mechanism for addressing challenge (ii). 
To jointly address these challenges, we trained the encoder/decoders via auto-encoder.
We then analyzed the computational complexity of both our linear and non-linear coding schemes.

Through simulations, we demonstrated that our two-way coding
strategies outperform traditional channel coding schemes by wide margins in terms of sum-error performance.
In addition, we examined our two-way coding schemes in terms of their power distribution, two-way coding benefit, different coding rates, and block-length gain.
Our results highlighted the potential of our two-way coding methodologies  to improve/balance the communication reliability in GTWCs.



\bibliographystyle{IEEEtran}
\bibliography{ref}

\begin{thebibliography}{10}
\providecommand{\url}[1]{#1}
\csname url@samestyle\endcsname
\providecommand{\newblock}{\relax}
\providecommand{\bibinfo}[2]{#2}
\providecommand{\BIBentrySTDinterwordspacing}{\spaceskip=0pt\relax}
\providecommand{\BIBentryALTinterwordstretchfactor}{4}
\providecommand{\BIBentryALTinterwordspacing}{\spaceskip=\fontdimen2\font plus
\BIBentryALTinterwordstretchfactor\fontdimen3\font minus
  \fontdimen4\font\relax}
\providecommand{\BIBforeignlanguage}[2]{{%
\expandafter\ifx\csname l@#1\endcsname\relax
\typeout{** WARNING: IEEEtran.bst: No hyphenation pattern has been}%
\typeout{** loaded for the language `#1'. Using the pattern for}%
\typeout{** the default language instead.}%
\else
\language=\csname l@#1\endcsname
\fi
#2}}
\providecommand{\BIBdecl}{\relax}
\BIBdecl

\bibitem{kim2022linear}
J.~Kim, S.~Hosseinalipour, T.~Kim, D.~J. Love, and C.~G. Brinton, ``Linear
  coding for {Gaussian} two-way channels,'' in \emph{58th Annual Allerton
  Conference on Communication, Control, and Computing (Allerton)}, Sep. 2022.

\bibitem{zhang2016full}
Z.~Zhang, K.~Long, A.~V. Vasilakos, and L.~Hanzo, ``Full-duplex wireless
  communications: Challenges, solutions, and future research directions,''
  \emph{Proceedings of the IEEE}, vol. 104, no.~7, pp. 1369--1409, Feb. 2016.

\bibitem{kim2015survey}
D.~Kim, H.~Lee, and D.~Hong, ``A survey of in-band full-duplex transmission:
  From the perspective of {PHY} and {MAC} layers,'' \emph{IEEE Communications
  Surveys \& Tutorials}, vol.~17, no.~4, pp. 2017--2046, Feb. 2015.

\bibitem{shannon1961two}
C.~E. Shannon, ``Two-way communication channels,'' in \emph{Proceedings of the
  Fourth Berkeley Symposium on Mathematical Statistics and Probability, Volume
  1: Contributions to the Theory of Statistics}.\hskip 1em plus 0.5em minus
  0.4em\relax University of California Press, Jan. 1961, pp. 611--644.

\bibitem{han1984general}
T.~Han, ``A general coding scheme for the two-way channel,'' \emph{IEEE
  Transactions on Information Theory}, vol.~30, no.~1, pp. 35--44, Jan. 1984.

\bibitem{palacio2021achievable}
K.~S. Palacio-Baus and N.~Devroye, ``Achievable error exponents of one-way and
  two-way {AWGN} channels,'' \emph{IEEE Transactions on Information Theory},
  vol.~67, no.~5, pp. 2693--2715, May 2021.

\bibitem{vasaldynamic}
D.~Vasal, ``A dynamic program for linear sequential coding for two way
  {Gaussian} channel,'' \emph{ResearchGate}, Dec. 2021.

\bibitem{schalkwijk1966coding}
J.~Schalkwijk and T.~Kailath, ``A coding scheme for additive noise channels
  with feedback--{I}: No bandwidth constraint,'' \emph{IEEE Transactions on
  Information Theory}, vol.~12, no.~2, pp. 172--182, Apr. 1966.

\bibitem{butman1969general}
S.~Butman, ``A general formulation of linear feedback communication systems
  with solutions,'' \emph{IEEE Transactions on Information Theory}, vol.~15,
  no.~3, pp. 392--400, May 1969.

\bibitem{chance2011concatenated}
Z.~Chance and D.~J. Love, ``Concatenated coding for the {AWGN} channel with
  noisy feedback,'' \emph{IEEE Transactions on Information Theory}, vol.~57,
  no.~10, pp. 6633--6649, Oct. 2011.

\bibitem{agrawal2011iteratively}
M.~Agrawal, D.~J. Love, and V.~Balakrishnan, ``An iteratively optimized linear
  coding scheme for correlated {Gaussian} channels with noisy feedback,'' in
  \emph{IEEE 49th Annual Allerton Conference on Communication, Control, and
  Computing}, Sep. 2011, pp. 1012--1018.

\bibitem{elia2004bode}
N.~Elia, ``When {Bode} meets {Shannon}: Control-oriented feedback communication
  schemes,'' \emph{IEEE Transactions on Automatic Control}, vol.~49, no.~9, pp.
  1477--1488, Sep. 2004.

\bibitem{mishra2023linear}
R.~Mishra, D.~Vasal, and H.~Kim, ``Linear coding for {AWGN} channels with noisy
  output feedback via dynamic programming,'' \emph{IEEE Transactions on
  Information Theory}, Mar. 2023.

\bibitem{ben2017interactive}
A.~Ben-Yishai and O.~Shayevitz, ``Interactive schemes for the awgn channel with
  noisy feedback,'' \emph{IEEE Transactions on Information Theory}, vol.~63,
  no.~4, pp. 2409--2427, 2017.

\bibitem{kim2020deepcode}
H.~Kim, Y.~Jiang, S.~Kannan, S.~Oh, and P.~Viswanath, ``Deepcode: Feedback
  codes via deep learning,'' \emph{IEEE Journal on Selected Areas in
  Information Theory}, vol.~1, no.~1, pp. 194--206, Apr. 2020.

\bibitem{safavi2021deep}
A.~R. Safavi, A.~G. Perotti, B.~M. Popovic, M.~B. Mashhadi, and D.~Gunduz,
  ``Deep extended feedback codes,'' \emph{arXiv preprint arXiv:2105.01365}, May
  2021.

\bibitem{ozfatura2022all}
E.~Ozfatura, Y.~Shao, A.~G. Perotti, B.~M. Popovi{\'c}, and D.~G{\"u}nd{\"u}z,
  ``All you need is feedback: Communication with block attention feedback
  codes,'' \emph{IEEE Journal on Selected Areas in Information Theory}, vol.~3,
  no.~3, pp. 587--602, Sep. 2022.

\bibitem{kim2023feedback}
J.~Kim, T.~Kim, D.~Love, and C.~Brinton, ``Robust non-linear feedback coding
  via power-constrained deep learning,'' \emph{To appear at International
  Conference on Machine Learning (ICML)}, Jul. 2023.

\bibitem{proakis2008digital}
J.~G. Proakis, \emph{Digital communications}.\hskip 1em plus 0.5em minus
  0.4em\relax McGraw-Hill, Higher Education, Jan. 2008.

\bibitem{tse2005fundamentals}
D.~Tse and P.~Viswanath, \emph{Fundamentals of wireless communication}.\hskip
  1em plus 0.5em minus 0.4em\relax Cambridge university press, Jul. 2005.

\bibitem{deb2011multi}
K.~Deb, \emph{Multi-objective optimisation using evolutionary algorithms: an
  introduction}.\hskip 1em plus 0.5em minus 0.4em\relax Springer, Jan. 2011.

\bibitem{marler2010weighted}
R.~T. Marler and J.~S. Arora, ``The weighted sum method for multi-objective
  optimization: new insights,'' \emph{Structural and multidisciplinary
  optimization}, vol.~41, pp. 853--862, 2010.

\bibitem{freund2001solving}
R.~W. Freund and F.~Jarre, ``Solving the sum-of-ratios problem by an
  interior-point method,'' \emph{Journal of Global Optimization}, vol.~19,
  no.~1, pp. 83--102, Jan. 2001.

\bibitem{MatlabOTB}
\BIBentryALTinterwordspacing
MathWorks, \emph{MATLAB Optimization Toolbox}, Natick, MA, USA, 2021. [Online].
  Available: \url{https://www.mathworks.com/help/optim/}
\BIBentrySTDinterwordspacing

\bibitem{chong2013introduction}
E.~K. Chong and S.~H. Zak, \emph{An introduction to optimization}.\hskip 1em
  plus 0.5em minus 0.4em\relax John Wiley \& Sons, Jan. 2013, vol.~75.

\bibitem{ferziger1981numerical}
J.~H. Ferziger, ``Numerical methods for engineering applications,'' \emph{New
  York, Wiley-Interscience}, Apr. 1998.

\bibitem{goodfellow2016deep}
I.~Goodfellow, Y.~Bengio, and A.~Courville, \emph{Deep learning}.\hskip 1em
  plus 0.5em minus 0.4em\relax MIT press, 2016.

\bibitem{bengio1993problem}
Y.~Bengio, P.~Frasconi, and P.~Simard, ``The problem of learning long-term
  dependencies in recurrent networks,'' in \emph{IEEE International Conference
  on Neural Networks}.\hskip 1em plus 0.5em minus 0.4em\relax IEEE, Mar. 1993,
  pp. 1183--1188.

\bibitem{bahdanau2014neural}
D.~Bahdanau, K.~Cho, and Y.~Bengio, ``Neural machine translation by jointly
  learning to align and translate,'' \emph{arXiv preprint arXiv:1409.0473},
  Sep. 2014.

\bibitem{ma1986tail}
H.~Ma and J.~Wolf, ``On tail biting convolutional codes,'' \emph{IEEE
  Transactions on Communications}, vol.~34, no.~2, pp. 104--111, Feb. 1986.

\bibitem{LTEstandard}
{3GPP TS 36.212}, ``{LTE}: Evolved universal terrestrial radio access
  ({E-UTRA}): Multiplexing and channel coding,'' vol. V8.8.0 Release 8, 2010.

\bibitem{polyanskiy2010channel}
Y.~Polyanskiy, H.~V. Poor, and S.~Verd{\'u}, ``Channel coding rate in the
  finite blocklength regime,'' \emph{IEEE Transactions on Information Theory},
  vol.~56, no.~5, pp. 2307--2359, 2010.

\bibitem{yavas2024third}
R.~C. Yavas, V.~Kostina, and M.~Effros, ``Third-order analysis of channel
  coding in the small-to-moderate deviations regime,'' \emph{IEEE Transactions
  on Information Theory}, 2024.

\bibitem{berrou1996near}
C.~Berrou and A.~Glavieux, ``Near optimum error correcting coding and decoding:
  Turbo-codes,'' \emph{IEEE Transactions on communications}, vol.~44, no.~10,
  pp. 1261--1271, Oct. 1996.

\bibitem{ungerboeck1982channel}
G.~Ungerboeck, ``Channel coding with multilevel/phase signals,'' \emph{IEEE
  Transactions on Information Theory}, vol.~28, no.~1, pp. 55--67, 1982.

\bibitem{billingsley2012probability}
P.~Billingsley, \emph{Probability and measure}.\hskip 1em plus 0.5em minus
  0.4em\relax John Wiley \& Sons, Feb. 2012.

\end{thebibliography}

\clearpage

\appendices
\section{Conversion between linear signal models and derivation of power expressions}
\label{app:linear:general_signal}


In the literature on GTWCs~\cite{palacio2021achievable, vasaldynamic}, transmit signals are represented as a  function of the message and the receive signals, i.e., ${x}_i[k] = f(m_i, \{{y}_i[k']\}_{k'=1}^{k-1})$. The linear form of this signal model is ${\bf x}_i = {\bf g}_i^{(A)} m_i + {\bf F}_i^{(A)} {\bf y}_i$, which we call Model A, where ${\bf F}_i^{(A)}$ is strictly lower triangular, $i=1,2$.
Besides Model A, we might consider the signal model chosen in this paper as in
    \eqref{eq:x1:tilde}-\eqref{eq:x2:tilde} or
    another form, ${\bf x}_i = {\bf g}_i^{(B)} m_i +  {\bf F}_i^{(B)} {\bf y}_i +  {\bf H}_i^{(B)} {\bf x}_i$, which we call Model B, where 
    ${\bf F}_i^{(B)}$ and ${\bf H}_i^{(B)}$ are strictly lower triangular, $i=1,2$.
    It is important to note that our model in \eqref{eq:x1:tilde}-\eqref{eq:x2:tilde}  and Model B can be converted to the general form, Model A, through changes of variables. 
    The specific derivation for the conversion is provided as below in (i) and (ii). We then provide the power expressions when using Model A and Model B as linear signal models, respectively, in (iii) and (iv).

\textbf{(i) Conversion from our model in \eqref{eq:x1:tilde}-\eqref{eq:x2:tilde} to Model A.}
We first show that our signal models that we consider in this paper in \eqref{eq:x1:tilde}-\eqref{eq:x2:tilde}, ${\bf x}_1 = \tilde {\bf g}_1 m_1 + \tilde {\bf F}_1 ({\bf y}_1 - \tilde {\bf F}_2 {\bf x}_1)$ and ${\bf x}_2 = \tilde {\bf g}_2 m_2 + \tilde {\bf F}_2 ({\bf y}_2 - \tilde {\bf F}_1 {\bf x}_2)$, 
can be converted to the form ${\bf x}_i = {\bf g}_i^{(A)} m_i + {\bf F}_i^{(A)} {\bf y}_i$ through changes of variables. 
Note that $\tilde {\bf F}_i$ is strictly lower triangular, $i=1,2$.
We show the case for ${\bf x}_1$, while the case for ${\bf x}_2$ can be shown in the same way. Note that
\begin{align}
    {\bf x}_1 &= \tilde {\bf g}_1 m_1 + \tilde {\bf F}_1 ({\bf y}_1 - \tilde {\bf F}_2 {\bf x}_1)
    \label{eq:app:x1}
    \\
    ({\bf I} - \tilde {\bf F}_1 \tilde {\bf F}_2) {\bf x}_1 &= \tilde {\bf g}_1 m_1  + \tilde {\bf F}_1 {\bf y}_1
    \label{eq:app:x1:derivation1}
    \\
    {\bf x}_1 &= ({\bf I} - \tilde {\bf F}_1 \tilde {\bf F}_2)^{-1} \tilde {\bf g}_1 m_1  +  ({\bf I} - \tilde {\bf F}_1 \tilde {\bf F}_2)^{-1}  \tilde {\bf F}_1 {\bf y}_1.
    \label{eq:app:x1:derivation2}
\end{align}
We derive \eqref{eq:app:x1:derivation1} by moving the term including ${\bf x}_1$ to the left-hand side in \eqref{eq:app:x1}. 
In \eqref{eq:app:x1:derivation1}, $({\bf I} - \tilde {\bf F}_1 \tilde {\bf F}_2)$ is always invertible, since both $\tilde {\bf F}_1$ and $\tilde {\bf F}_2$ are strictly lower triangular, and $({\bf I} - \tilde {\bf F}_1 \tilde {\bf F}_2)$ has 1's on all diagonal entries.
We then obtain  \eqref{eq:app:x1:derivation2} by taking the inverse of $({\bf I} - \tilde {\bf F}_1 \tilde {\bf F}_2)$ on both sides in  \eqref{eq:app:x1:derivation1}.
In \eqref{eq:app:x1:derivation2}, by defining ${\bf g}_1^{(A)} = ({\bf I} - \tilde {\bf F}_1 \tilde {\bf F}_2)^{-1} \tilde {\bf g}_1$ and ${\bf F}_1^{(A)} = ({\bf I} - \tilde {\bf F}_1 \tilde {\bf F}_2)^{-1}  \tilde {\bf F}_1$, we arrive at 
\begin{equation}
    {\bf x}_1 = {\bf g}_1^{(A)} m_1 + {\bf F}_1^{(A)} {\bf y}_1.
\end{equation}
Note that ${\bf F}_1^{(A)}$ is strictly lower triangular, since the multiplication of the lower triangular matrix $({\bf I} - \tilde {\bf F}_1 \tilde {\bf F}_2)^{-1}$ and the strictly lower triangular matrix $\tilde {\bf F}_1$ is strictly lower triangular.

\textbf{(ii) Conversion from Model B to Model A.}
Now, we show that
Model B, ${\bf x}_i =  {\bf g}_i^{(B)} m_i +  {\bf F}_i^{(B)} {\bf y}_i +  {\bf H}_i^{(B)} {\bf x}_i$, can be converted to Model A, ${\bf x}_i = {\bf g}_i^{(A)} m_i + {\bf F}_i^{(A)} {\bf y}_i$, through changes of variables. 
Note that ${\bf F}_i^{(B)}$ and ${\bf H}_i^{(B)}$ are strictly lower triangular, $i=1,2$.
We have
\begin{align}
    {\bf x}_i &=  {\bf g}_i^{(B)} m_i +  {\bf F}_i^{(B)} {\bf y}_i +  {\bf H}_i^{(B)} {\bf x}_i
    \label{eq:app:xi}
    \\
    ({\bf I} -  {\bf H}_i^{(B)}) {\bf x}_i &=  {\bf g}_i^{(B)} m_i  +  {\bf F}_i^{(B)} {\bf y}_i
    \label{eq:app:xi:derivation1}
    \\
    {\bf x}_i &= ({\bf I} -  {\bf H}_i^{(B)})^{-1}  {\bf g}_i^{(B)} m_i  + ({\bf I} -  {\bf H}_i^{(B)})^{-1}   {\bf F}_i^{(B)} {\bf y}_i.
    \label{eq:app:xi:derivation2}
\end{align}
We obtain \eqref{eq:app:xi:derivation1} by moving the term including ${\bf x}_i$ to the left-hand side in \eqref{eq:app:xi}. 
In \eqref{eq:app:xi:derivation1}, $({\bf I} - {\bf H}_i^{(B)})$ is always invertible since ${\bf H}_i^{(B)}$ is strictly lower triangular, and $({\bf I} - {\bf H}_i^{(B)})$ has 1's on all diagonal entries.
We then derive \eqref{eq:app:xi:derivation2} by taking the inverse of $({\bf I} - {\bf H}_i^{(B)})$ on both sides in  \eqref{eq:app:xi:derivation1}.
In \eqref{eq:app:xi:derivation2}, by defining ${\bf g}_i^{(A)} = ({\bf I} -  {\bf H}_i^{(B)})^{-1}   {\bf g}_i^{(B)}$ and ${\bf F}_i^{(A)} = ({\bf I} -  {\bf H}_i^{(B)})^{-1}   {\bf F}_i^{(B)}$, we derive
\begin{equation}
    {\bf x}_i = {\bf g}_i^{(A)} m_i + {\bf F}_i^{(A)} {\bf y}_i, \; i=1,2.
\end{equation}
Note that $ {\bf F}_i^{(A)}$ is strictly lower triangular, since the multiplication of the lower triangular matrix $ ({\bf I} -  {\bf H}_i^{(B)})^{-1}$ and the strictly lower triangular matrix ${\bf F}_i^{(B)}$ is strictly lower triangular.

\textbf{(iii) Power expression for Model A.}
We show that Model A, ${\bf x}_i = {\bf g}_i^{(A)} m_i + {\bf F}_i^{(A)} {\bf y}_i$, $i=1,2$, leads to the introduction of an inverse matrix form in the power expressions. Note that ${\bf F}_i^{(A)}$ is strictly lower triangular, $i=1,2$.
We consider the case for $i=1$, and can repeat for $i=2$ in the same way. We now have
\begin{align}
    {\bf x}_1 & = {\bf g}_1^{(A)} m_1 + {\bf F}_1^{(A)} {\bf y}_1
    \label{eq:app:power:signal1:derivation0}
    \\
    & = {\bf g}_1^{(A)} m_1 + {\bf F}_1^{(A)} ({\bf g}_2^{(A)} m_2 + {\bf F}_2^{(A)} ({\bf x}_1 + {\bf n}_1) + {\bf n}_2)
    \label{eq:app:power:signal1:derivation1}
    \\
    & = ({\bf I} - {\bf F}_1^{(A)} {\bf F}_2^{(A)})^{-1} ({\bf g}_1^{(A)} m_1 + {\bf F}_1^{(A)} {\bf g}_2^{(A)} m_2 + {\bf F}_1^{(A)} {\bf F}_2^{(A)} {\bf n}_1 + {\bf F}_1^{(A)} {\bf n}_2).
\label{eq:app:power:signal1:derivation2}
\end{align}
We derive \eqref{eq:app:power:signal1:derivation1} by putting ${\bf y}_1 = {\bf x}_2 + {\bf n}_2 = {\bf g}_2^{(A)} m_2 + {\bf F}_2^{(A)} ({\bf x}_1 + {\bf n}_1) + {\bf n}_2$ into \eqref{eq:app:power:signal1:derivation0}, where ${\bf x}_2 = {\bf g}_2^{(A)} m_2 + {\bf F}_2^{(A)} {\bf y}_2  = {\bf g}_2^{(A)} m_2 + {\bf F}_2^{(A)} ({\bf x}_1 + {\bf n}_1)$.
We obtain \eqref{eq:app:power:signal1:derivation2} by moving the term including ${\bf x}_1$ to the left-hand side in \eqref{eq:app:power:signal1:derivation1} and taking the inverse of $({\bf I} - {\bf F}_1^{(A)} {\bf F}_2^{(A)})$ on both sides. 
Note that $({\bf I} - {\bf F}_1^{(A)} {\bf F}_2^{(A)})$ is always invertible, since
${\bf F}_1^{(A)}$ and ${\bf F}_2^{(A)}$ are strictly lower triangular, and $({\bf I} - {\bf F}_1^{(A)} {\bf F}_2^{(A)})$  has 1's on all diagonal entries.
From \eqref{eq:app:power:signal1:derivation2}, the power expression for ${\bf x}_1$ is given by
\begin{align}
    \mathbb{E}_{\rev{m_1,m_2,{\bf n}_1, {\bf n}_2}}\big[\|{\bf x}_1\|^2\big] 
    & = \|({\bf I} - {\bf F}_1^{(A)} {\bf F}_2^{(A)})^{-1} {\bf g}_1^{(A)}\|^2 
    + \|({\bf I} - {\bf F}_1^{(A)} {\bf F}_2^{(A)})^{-1} {\bf F}_1^{(A)} {\bf g}_2^{(A)}\|^2
    \nonumber
    \\
    & \hspace{1cm} + \|({\bf I} - {\bf F}_1^{(A)} {\bf F}_2^{(A)})^{-1} {\bf F}_1^{(A)} {\bf F}_2^{(A)}\|_F^2 \sigma_1^2
    + \|({\bf I} - {\bf F}_1^{(A)} {\bf F}_2^{(A)})^{-1} {\bf F}_1^{(A)} \|_F^2 \sigma_2^2.
    \label{eq:power1:signal1}
\end{align}
In \eqref{eq:power1:signal1}, the inverse matrix $({\bf I} - {\bf F}_1^{(A)} {\bf F}_2^{(A)})^{-1}$ appears in every term, which is challenging to handle in optimization.

\textbf{(iv) Power expression for Model B.}
We show that Model B, ${\bf x}_i =  {\bf g}_i^{(B)} m_i +  {\bf F}_i^{(B)} {\bf y}_i +  {\bf H}_i^{(B)} {\bf x}_i$, lead to the introduction of an inverse matrix form in the power expressions. Note that ${\bf F}_i^{(B)}$ and ${\bf H}_i^{(B)}$ are strictly lower triangular, $i=1,2$.
We consider the case for $i=1$, and can repeat for $i=2$ in the same way. We have
\begin{align}
    {\bf x}_1 & = {\bf g}_1^{(B)} m_1 + {\bf F}_1^{(B)} {\bf y}_1 + {\bf H}_1^{(B)} {\bf x}_1
    \\
    & = {\bf g}_1^{(B)} m_1 + {\bf F}_1^{(B)} ( {\bf x}_2 + {\bf n}_2) + {\bf H}_1^{(B)} {\bf x}_1
    \label{eq:app:power:signal2:derivation1}
    \\
    & = {\bf g}_1^{(B)} m_1 + {\bf F}_1^{(B)} ( ({\bf I} - {\bf H}_2^{(B)})^{-1} ({\bf g}_2^{(B)} m_2 + {\bf F}_2^{(B)} {\bf y}_2) + {\bf n}_2) + {\bf H}_1^{(B)} {\bf x}_1
    \label{eq:app:power:signal2:derivation2}
    \\
    & = {\bf g}_1^{(B)} m_1 + {\bf F}_1^{(B)} ( ({\bf I} - {\bf H}_2^{(B)})^{-1} ({\bf g}_2^{(B)} m_2 + {\bf F}_2^{(B)} ({\bf x}_1 + {\bf n}_1)) + {\bf n}_2) + {\bf H}_1^{(B)} {\bf x}_1
    \label{eq:app:power:signal2:derivation3}
    \\
    & = \big({\bf I} -  {\bf F}_1^{(B)} ({\bf I}- {\bf H}_2^{(B)})^{-1}  {\bf F}_2^{(B)} -  {\bf H}_1^{(B)} \big)^{-1} \big( {\bf g}_1^{(B)} m_1 +  {\bf F}_1^{(B)} ({\bf I} -  {\bf H}_2^{(B)})^{-1}  {\bf g}_2^{(B)} m_2 
    \nonumber
    \\
    & \hspace{1cm} +  {\bf F}_1^{(B)} ({\bf I} -  {\bf H}_2^{(B)})^{-1}  {\bf F}_2^{(B)} {\bf n}_1 +   {\bf F}_1^{(B)} {\bf n}_2 \big).
    \label{eq:app:power:signal2:derivation4}
\end{align}
In \eqref{eq:app:power:signal2:derivation2}, we put the term ${\bf x}_2 = ({\bf I} - {\bf H}_2^{(B)})^{-1} ({\bf g}_2^{(B)} m_2 + {\bf F}_2^{(B)} {\bf y}_2)$ in \eqref{eq:app:power:signal2:derivation1}, which is obtained by moving the term including ${\bf x}_2$ to the left-hand side in
${\bf x}_2 = {\bf g}_2^{(B)} m_2 + {\bf F}_2^{(B)} {\bf y}_2 + {\bf H}_2^{(B)} {\bf x}_2$ and taking the inverse of $({\bf I} - {\bf H}_2^{(B)})$ on both sides.
We note that $({\bf I} - {\bf H}_2^{(B)})$ is always invertible since $({\bf I} - {\bf H}_2^{(B)})$ has 1's on all diagonal entries.
We derive \eqref{eq:app:power:signal2:derivation4} by moving all terms including ${\bf x}_1$ to the left-hand side in \eqref{eq:app:power:signal2:derivation3} and taking the inverse of $\big({\bf I} -  {\bf F}_1^{(B)} ({\bf I}- {\bf H}_2^{(B)})^{-1}  {\bf F}_2^{(B)} -  {\bf H}_1^{(B)} \big)$ on both sides. 
We note that $\big({\bf I} -  {\bf F}_1^{(B)} ({\bf I}- {\bf H}_2^{(B)})^{-1}  {\bf F}_2^{(B)} -  {\bf H}_1^{(B)} \big)$ is always invertible, since ${\bf F}_i^{(B)}$ and ${\bf H}_i^{(B)}$ are strictly lower triangular and $\big({\bf I} -  {\bf F}_1^{(B)} ({\bf I}- {\bf H}_2^{(B)})^{-1}  {\bf F}_2^{(B)} -  {\bf H}_1^{(B)} \big)$ has 1's on all diagonal entries.
From \eqref{eq:app:power:signal2:derivation4}, the power expression for ${\bf x}_1$ is given by
\begin{align}
    \mathbb{E}_{\rev{m_1,m_2,{\bf n}_1, {\bf n}_2}}\big[\|{\bf x}_1\|^2\big] 
    & = \|\big({\bf I} -  {\bf F}_1^{(B)} ({\bf I}- {\bf H}_2^{(B)})^{-1}  {\bf F}_2^{(B)} -  {\bf H}_1^{(B)} \big)^{-1} {\bf g}_1^{(B)}\|^2 
    \nonumber
    \\
    & \hspace{.5cm} + \|\big({\bf I} -  {\bf F}_1^{(B)} ({\bf I}- {\bf H}_2^{(B)})^{-1}  {\bf F}_2^{(B)} -  {\bf H}_1^{(B)} \big)^{-1} {\bf F}_1^{(B)} ({\bf I}- {\bf H}_2^{(B)})^{-1}  {\bf g}_2^{(B)}\|^2
    \nonumber
    \\
    & \hspace{.5cm} + \|\big({\bf I} -  {\bf F}_1^{(B)} ({\bf I}- {\bf H}_2^{(B)})^{-1}  {\bf F}_2^{(B)} -  {\bf H}_1^{(B)} \big)^{-1} {\bf F}_1^{(B)} ({\bf I}- {\bf H}_2^{(B)})^{-1}  {\bf F}_2^{(B)} \|_F^2 \sigma_1^2
    \nonumber
    \\
    & \hspace{.5cm} + \|\big({\bf I} -  {\bf F}_1^{(B)} ({\bf I}- {\bf H}_2^{(B)})^{-1}  {\bf F}_2^{(B)} -  {\bf H}_1^{(B)} \big)^{-1} {\bf F}_1^{(B)} \|_F^2 \sigma_2^2.
    \label{eq:power1:signal2}
\end{align}
We observe that inverse matrix appears in every term in \eqref{eq:power1:signal2}, which is difficult to deal with in optimization.


\section{Derivation for ${\bf g}_1$, ${\bf F}_1$, ${\bf g}_2$, and ${\bf F}_2$ as functions of $\tilde {\bf g}_1$, $\tilde  {\bf F}_1$, $\tilde  {\bf g}_2$, and $\tilde  {\bf F}_2$}
\label{app:system:equivalence}

First, we will  derive the functional forms of ${\bf g}_2$ and ${\bf F}_2$ in \eqref{eq:x2} by starting from \eqref{eq:x2:tilde}.
We move all terms including 
${\bf x}_2$ in \eqref{eq:x2:tilde} to the left-hand side and obtain
    $({\bf I} + \tilde {\bf F}_2 \tilde {\bf F}_1) {\bf x}_2 = \tilde {\bf g}_2 m_2 + \tilde {\bf F}_2 {\bf y}_2.$
We take the inverse of $({\bf I} + \tilde {\bf F}_2 \tilde {\bf F}_1)$ on both sides and obtain
\begin{align}
    {\bf x}_2 = ({\bf I} + \tilde {\bf F}_2 \tilde {\bf F}_1)^{-1}\tilde {\bf g}_2 m_2 + ({\bf I} + \tilde {\bf F}_2 \tilde {\bf F}_1)^{-1}\tilde {\bf F}_2 {\bf y}_2.
    \label{eq:x2:reform}
\end{align}
By comparing \eqref{eq:x2} and \eqref{eq:x2:reform}, we  find ${\bf g}_2 = ({\bf I} + \tilde {\bf F}_2 \tilde {\bf F}_1)^{-1} \tilde {\bf g}_2$ and ${\bf F}_2 = ({\bf I} + \tilde {\bf F}_2 \tilde {\bf F}_1)^{-1}\tilde {\bf F}_2$. 

We will then derive the functional expressions of ${\bf g}_1$ and ${\bf F}_1$ in \eqref{eq:x1} by starting from \eqref{eq:x1:tilde}.
As a first step,
we rewrite  \eqref{eq:x1:tilde} as
\begin{align}
    {\bf x}_1 & =  \tilde {\bf g}_1 m_1 + \tilde {\bf F}_1 ({\bf y}_1 - \tilde {\bf F}_2 {\bf x}_1 + {\bf F}_2 {\bf x}_1 -  {\bf F}_2 {\bf x}_1)
    \nonumber
    \\
    & = \big({\bf I} - \tilde {\bf F}_1 ( {\bf F}_2 - \tilde {\bf F}_2) \big)^{-1} \tilde {\bf g}_1 m_1
    + \big({\bf I} - \tilde {\bf F}_1 ( {\bf F}_2 - \tilde {\bf F}_2) \big)^{-1} \tilde {\bf F}_1 ({\bf y}_1  - {\bf F}_2 {\bf x}_1).
    \label{eq:x1:reform}
\end{align}
To obtain the equality in the first line in \eqref{eq:x1:reform}, we add the term $ {\bf F}_2 {\bf x}_1 -{\bf F}_2 {\bf x}_1$, which is a zero vector, inside the parenthesis in \eqref{eq:x1:tilde}. 
To obtain the equality in the second line in \eqref{eq:x1:reform},
we first move the term of
$ \tilde {\bf F}_1 (-\tilde {\bf F}_2 {\bf x}_1 + {\bf F}_2 {\bf x}_1)$ (in the first line) to the left-hand side, and accordingly obtain
$\big({\bf I} - \tilde {\bf F}_1 ( {\bf F}_2 - \tilde {\bf F}_2) \big) {\bf x}_1 = \tilde {\bf g}_1 m_1 + \tilde {\bf F}_1({\bf y}_1  - {\bf F}_2 {\bf x}_1)$. 
We take the inverse of $\big({\bf I} - \tilde {\bf F}_1 ( {\bf F}_2 - \tilde {\bf F}_2) \big)$ on both sides and finally obtain the equality in the second line in \eqref{eq:x1:reform}.
By comparing  \eqref{eq:x1} and \eqref{eq:x1:reform}, we  find ${\bf g}_1 = {\bf A}^{-1} \tilde {\bf g}_1$ and ${\bf F}_1 = {\bf A}^{-1} \tilde {\bf F}_1$, where ${\bf A} 
= {\bf I} - \tilde {\bf F}_1 ( {\bf F}_2 - \tilde {\bf F}_2)$.
Here, we put the obtained result of ${\bf F}_2 = ({\bf I} + \tilde {\bf F}_2 \tilde {\bf F}_1)^{-1}\tilde {\bf F}_2$ in 
${\bf A}$, and obtain 
$ {\bf A}  = {\bf I} - \tilde {\bf F}_1 ( ({\bf I} + \tilde {\bf F}_2 \tilde {\bf F}_1)^{-1} - {\bf I}) \tilde {\bf F}_2$. 

In summary, we have 
\begin{align}
    {\bf g}_2 & = ({\bf I} + \tilde {\bf F}_2 \tilde {\bf F}_1)^{-1} \tilde {\bf g}_2,
    \label{eq:app:g2} 
    \\
    {\bf F}_2 & = ({\bf I} + \tilde {\bf F}_2 \tilde {\bf F}_1)^{-1} \tilde {\bf F}_2,
    \label{eq:app:F2}
    \\
    {\bf g}_1 & = \big( {\bf I} - \tilde {\bf F}_1 ( ({\bf I} + \tilde {\bf F}_2 \tilde {\bf F}_1)^{-1} - {\bf I}) \tilde {\bf F}_2 \big)^{-1} \tilde {\bf g}_1, 
    \label{eq:app:g1}
    \\
    {\bf F}_1 & = \big( {\bf I} - \tilde {\bf F}_1 ( ({\bf I} + \tilde {\bf F}_2 \tilde {\bf F}_1)^{-1} - {\bf I}) \tilde {\bf F}_2 \big)^{-1} \tilde {\bf F}_1.
    \label{eq:app:F1}
\end{align}

We observe that ${\bf F}_1$ and ${\bf F}_2$ in \eqref{eq:app:F1} and \eqref{eq:app:F2} are strictly lower triangular matrices, since $\tilde {\bf F}_1$ and $\tilde {\bf F}_2$ are strictly lower triangular matrices.

\section{\rev{Derivation for $\tilde {\bf g}_1$, $\tilde  {\bf F}_1$, $\tilde  {\bf g}_2$, and $\tilde  {\bf F}_2$ as functions of ${\bf g}_1$, ${\bf F}_1$, ${\bf g}_2$, and ${\bf F}_2$}}
\label{app:system:equivalence2}

\rev{We show how to obtain $\tilde \bg_1$, $\tilde \bg_2$, $\tilde \bF_1$, and $\tilde \bF_2$ in  \eqref{eq:x1:tilde}-\eqref{eq:x2:tilde}, given $\bg_1$, $\bg_2$, $\bF_1$, and $\bF_2$ in \eqref{eq:x1}-\eqref{eq:x2}  as follows.
From \eqref{eq:x2}, we obtain
\begin{align}
    \bx_2 &= \bg_2 m_2 + \bF_2\by_2
    \label{eq:x2:before}
    \\
    \bx_2 &= \bg_2 m_2 + \bF_2\by_2 - \bF_2 \bA \bx_2 + \bF_2 \bA \bx_2
    \label{eq:x2:after}
    \\
    \bx_2 - \bF_2 \bA \bx_2 &= \bg_2 m_2 + \bF_2\by_2  - \bF_2 \bA \bx_2
    \\
    (\bI - \bF_2 \bA)\bx_2  &= \bg_2 m_2 + \bF_2\by_2 - \bF_2 \bA \bx_2
    \\
    \bx_2  &= (\bI - \bF_2 \bA)^{-1} \bg_2 m_2 + (\bI - \bF_2 \bA)^{-1} \bF_2 (\by_2 -  \bA \bx_2).
    \label{eq:appc:x2}
\end{align}
In \eqref{eq:x2:after}, we have added the zero vector, $- \bF_2 \bA \bx_2 + \bF_2 \bA \bx_2$, to the right-hand side of \eqref{eq:x2:before} to facilitate further derivation.
By matching \eqref{eq:appc:x2} 
 with \eqref{eq:x2:tilde}, we can define 
\begin{align}
    \tilde \bg_2 & = (\bI - \bF_2 \bA)^{-1} \bg_2,
    \label{eq:deriv1:tilg2}
    \\
    \tilde \bF_2 & = (\bI - \bF_2 \bA)^{-1} \bF_2,
    \label{eq:deriv1:tilF2}
    \\
    \tilde \bF_1 & = \bA.
    \label{eq:deriv1:tilF1}
\end{align}
From \eqref{eq:deriv1:tilF1}, $\bA$ is  a strictly lower triangular matrix, since $\tilde \bF_1$ is a strictly lower triangular matrix. We note that $\bI - \bF_2 \bA$ is invertible, since  $\bF_2 \bA$ is strictly lower triangular.}

\rev{We next look into equation \eqref{eq:x1}, which leads to
\begin{align}
    \bx_1 & = \bg_1 m_1 + \bF_1 (\by_1 - \bF_2 \bx_1) 
    \label{eq:x1:before}
    \\
    \bx_1 & = \bg_1 m_1 + \bF_1 (\by_1 - \bF_2 \bx_1) - \bF_1 \bB \bx_1 + \bF_1 \bB \bx_1  
    \label{eq:x1:after}
    \\
    \bx_1 & = \bg_1 m_1 + \bF_1 (\by_1 - (\bF_2 + \bB) \bx_1) + \bF_1 \bB \bx_1
    \\
    \bx_1 - \bF_1 \bB \bx_1  & = \bg_1 m_1 + \bF_1 (\by_1 - (\bF_2 + \bB) \bx_1) 
    \\
    (\bI - \bF_1 \bB) \bx_1  & = \bg_1 m_1 + \bF_1 (\by_1 - (\bF_2 + \bB) \bx_1) 
    \\
    \bx_1  & = (\bI - \bF_1 \bB)^{-1} \bg_1 m_1 + (\bI - \bF_1 \bB)^{-1} \bF_1 (\by_1 - (\bF_2 + \bB) \bx_1).
    \label{eq:appc:x1}
\end{align}
In \eqref{eq:x1:after}, we have added the zero vector, $- \bF_1 \bB \bx_1 + \bF_1 \bB \bx_1$, to the right-hand side of \eqref{eq:x1:before} to facilitate further derivation.
By matching \eqref{eq:appc:x1} with 
\eqref{eq:x1:tilde}, we can define
\begin{align}
    \tilde \bg_1 & = (\bI - \bF_1 \bB)^{-1} \bg_1,
    \label{eq:deriv2:tilg1}
    \\
    \tilde \bF_1 & = (\bI - \bF_1 \bB)^{-1} \bF_1,
    \label{eq:deriv2:tilF1}
    \\
    \tilde \bF_2 & = \bF_2 + \bB.
    \label{eq:deriv2:tilF2}
\end{align}
From \eqref{eq:deriv2:tilF2},  $\bB$ is a strictly lower triangular matrix, since $\tilde \bF_2$ and $\bF_2$ are strictly lower triangular matrices.
We also note that $\bI - \bF_1 \bB$  is invertible, since $\bF_1 \bB$  is strictly lower triangular.}

\rev{In terms of $\tilde \bF_1$, equations \eqref{eq:deriv1:tilF1} and 
\eqref{eq:deriv2:tilF1} should be the same, which leads to
\begin{align}
    \bA = (\bI - \bF_1 \bB)^{-1} \bF_1.
    \label{eq:A}
\end{align}
Similarly, in terms of $\tilde \bF_2$, equations \eqref{eq:deriv1:tilF2} and 
\eqref{eq:deriv2:tilF2} should be the same, which leads to
\begin{align}
    \bB = (\bI - \bF_2 \bA)^{-1} \bF_2 - \bF_2.
    \label{eq:B}
\end{align}
Ideally,  $\bA$ and $\bB$ would be explicitly expressed in terms of $\bg_1$, $\bg_2$, $\bF_1$, and $\bF_2$. However, the coupled and interdependent structure of equations \eqref{eq:A} and \eqref{eq:B}, along with their inverse forms, renders the derivation of a closed-form solution intractable.
As an alternative, an iterative approach can be employed to find $\bA$ and $\bB$, once $\bF_1$ and $\bF_2$ are given.
The iterations proceed until convergence, ensuring that equations \eqref{eq:A} and \eqref{eq:B} are satisfied. The pseudo code for the detailed procedure is outlined in Algorithm~\ref{al:AB}.}

 \begin{algorithm}[t]
 \caption{\rev{The pseudo code for finding $\bA$ and $\bB$ from $\bF_1$ and $\bF_2$}}
 \label{al:AB}
 \begin{algorithmic}[1]
 \footnotesize
\State \textbf{Input.} 
Matrices $\bF_1$ and $\bF_2$, and the tolerance acceptance value $\epsilon$.
\State \textbf{Output.} 
Matrices $\bA$ and $\bB$.
\State \textbf{Initialization.} Set $\bA = \bO$ and $\bB = \bO$ (zero matrices), and initialize $c_1=c_2=\epsilon_0$, where $\epsilon_0$ is any value such that $\epsilon_0>\epsilon$.
  \While {$c_1 > \epsilon$ or $c_2 > \epsilon$}
    \State Obtain $\bA'$ using equation \eqref{eq:A}: $\bA' = (\bI - \bF_1 \bB)^{-1} \bF_1 $.
    \State Obtain $\bB'$ using equation \eqref{eq:B}: $\bB' = (\bI - \bF_2 \bA')^{-1} \bF_2 - \bF_2 $.
    \State Calculate the convergence errors using the Frobenius norm: $c_1 = \| \bA' - \bA \|_F^2 $ and $c_2 = \| \bB' - \bB \|_F^2 $.
    \State Update $\bA$ and $\bB$: $\bA = \bA'$ and $\bB = \bB'$.
    \EndWhile 
 \end{algorithmic}
 \end{algorithm}

\rev{In summary, given $\bg_1$, $\bg_2$, $\bF_1$, and $\bF_2$, we can determine $\tilde \bg_1$, $\tilde \bg_2$, $\tilde \bF_1$, and $\tilde \bF_2$. More specifically, we obtain $\bA$ and $\bB$ 
from Algorithm~\ref{al:AB}, and subsequently, compute 
$\tilde \bg_1$ from \eqref{eq:deriv2:tilg1}, $\tilde \bg_2$ from \eqref{eq:deriv1:tilg2}, $\tilde \bF_1$ either from \eqref{eq:deriv1:tilF1} or \eqref{eq:deriv2:tilF1}, and $\tilde \bF_2$ either from \eqref{eq:deriv1:tilF2} or \eqref{eq:deriv2:tilF2}.}
\section{Validity of Remark \ref{lemma1} within our linear coding framework}
\label{app:linear:lemma1}

Let (${\bf g}_1$, ${\bf F}_1$, ${\bf g}_2$,  ${\bf F}_2$) be the solution tuple of the optimization in \eqref{opt:linear:sum-error}.
Let's consider $\tilde P > P $.
We can always find $\tilde {\bf g}_1 = (1+\epsilon){\bf g}_1$, where $\epsilon>0$ can be chosen to satisfy both $\mathbb{E}_{\rev{m_1,m_2,{\bf n}_1, {\bf n}_2}}\big[\|{\bf x}_1 \|^2\big] \le N \tilde P$ and $\mathbb{E}_{\rev{m_1,m_2,{\bf n}_1, {\bf n}_2}}\big[\|{\bf x}_2 \|^2\big] \le  N \tilde P$ in \eqref{eq:power1}-\eqref{eq:power2}. Then, the solution tuple ($\tilde {\bf g}_1$, ${\bf F}_1$, ${\bf g}_2$,  ${\bf F}_2$) yields larger $\text{SNR}_1$ in \eqref{eq:SNR_opt}, leading to smaller  $\mathcal{E}_1(\text{SNR}_1)$. However, $\mathcal{E}_2(\text{SNR}_2)$ remains the same here since $\text{SNR}_2$ is not a function of $\tilde {\bf g}_1$. Thus, $\mathcal{E}_1(\text{SNR}_1) + \mathcal{E}_2(\text{SNR}_2)$ decreases.
\section{Proof of Lemma \ref{lemma_optimization_conversion}}
\label{app:linear:lemma:optimality}

\begin{proof} 
Denote a tuple of variables as $s=({\bf g}_1, {\bf F}_1, {\bf g}_2, {\bf F}_2)$. To indicate the dependency on $s$, we denote the SNR values as $\text{SNR}_1(s)$ and $\text{SNR}_2(s)$, and on the power value as $P_i(s)= \mathbb{E}_{\rev{m_1,m_2,{\bf n}_1, {\bf n}_2}}\big[\|{\bf x}_i \|^2\big]$, for $i=1,2$. 
We first define the set of optimal solutions of \eqref{opt:linear:sum-error} as $\mathcal{S}^\star$. For any optimal solution $s^\star \in \mathcal{S}^\star$ for \eqref{opt:linear:sum-error}, the minimum objective value is $\mathcal{E}_1(\text{SNR}_1(s^\star)) + \mathcal{E}_2(\text{SNR}_2(s^\star)) = w^\star$, 
and the power constraint is satisfied as $P_i(s^\star) \le NP$ for $i=1,2$ or equivalently as $\max \{ P_1(s^\star), P_2(s^\star) \} \le NP$.
We define the set of optimal solutions in \eqref{opt:linear:SVEM} as $\mathcal{\bar S}$.
For any optimal solution $\bar s \in \mathcal{\bar S}$, the minimum  objective value in \eqref{opt:linear:SVEM} is $\mathcal{E}_1(\text{SNR}_1(\bar s)) + \mathcal{E}_2(\text{SNR}_2(\bar s)) = \bar w$.
Recall that $\mathcal{S}$ is the feasible solution set of \eqref{opt:linear:SVEM}, and any feasible solution $s \in \mathcal{S}$ satisfies the power constraint, $\max \{ P_1(s), P_2(s) \} \le NP$, by its definition. Therefore,
 any optimal solution $\bar s \in \mathcal{\bar S} \subset \mathcal{S}$ satisfies the power constraint.


\textbf{Claim.}
We claim that $\mathcal{S}^\star = \mathcal{\bar S}$. We prove this by showing the two following statements: (i) if $s_0 \in \mathcal{S}^\star$, then $s_0 \in \mathcal{\bar S}$, and (ii) if $s_0 \in \mathcal{\bar S}$, then $s_0 \in \mathcal{S}^\star$.

\textbf{Proof of statement (i) by contrapositive.}
The contrapositive of statement (i) is that, if $s_0 \notin \mathcal{\bar S}$, then $s_0 \notin \mathcal{S}^\star$. Suppose $s_0 \notin \mathcal{\bar S}$. Then, we have either of the two following cases.

Case (i-a). $s_0$ does not satisfy the power constraint, i.e., $P_i(s_0) > NP$ for some $i \in \{1, 2\}$.

Case (i-b). $s_0$ does not minimize the sum-error value in \eqref{opt:linear:SVEM} while satisfying the power constraints, 
i.e., $\mathcal{E}_1(\text{SNR}_1(s_0)) + \mathcal{E}_2(\text{SNR}_2(s_0)) = w^o > \bar w$ and $P_i(s_0) \le NP$, $i=1,2$.

First, we consider Case (i-a). Since $s_0$ does not satisfy the power constraint, it is not a solution of \eqref{opt:linear:sum-error}. Therefore, it cannot be an optimal solution for \eqref{opt:linear:sum-error}, i.e., $s_0 \notin \mathcal{S}^\star$.
Next, we consider Case (i-b). We first note that $\bar s \in \mathcal{\bar S}$ is a feasible solution of \eqref{opt:linear:sum-error}, since $\bar s$ satisfies the power constraints.
It is obvious that $\bar w \ge w^\star$ since the feasible solution space of \eqref{opt:linear:SVEM} is a subset of that of  \eqref{opt:linear:sum-error}.
From Case (i-b), we have
$w^o > \bar w$. 
Combining the previous two results, we arrive at $w^o > w^\star$. 
This shows that $s_0$ does not yield the lowest sum-error value in \eqref{opt:linear:sum-error}, and thus it is not an optimal solution of \eqref{opt:linear:sum-error}, i.e., $s_0 \notin \mathcal{S}^\star$.
Therefore, for either of the cases, $s_0 \notin \mathcal{S}^\star$.

\textbf{Proof of statement (ii) by contrapositive.} 
The contrapositive of statement (ii) is that, if $s_0 \notin \mathcal{S}^\star$, then $s_0 \notin \mathcal{\bar S}$. Suppose $s_0 \notin \mathcal{S}^\star$.
Then, we have either of the two following cases.

Case (ii-a). $s_0$ does not satisfy the power constraint, i.e., $P_i(s_0) > NP$ for some $i \in \{1,2\}$.

Case (ii-b). $s_0$ does not minimize the sum-error value in \eqref{opt:linear:sum-error} while satisfying the power constraints, 
i.e., $\mathcal{E}_1(\text{SNR}_1(s_0)) + \mathcal{E}_2(\text{SNR}_2(s_0)) = w^o > w^\star$ and $P_i(s_0) \le NP$, $i=1,2$.

First, we consider Case (ii-a). Since $s_0$ does not satisfy the power constraint, it is not a solution of \eqref{opt:linear:SVEM},
i.e., $s_0 \notin \mathcal{\bar S}$.
Next, we consider Case (ii-b). 
We note that $s^\star$ is a feasible solution of \eqref{opt:linear:max-of-powers:SNR} when $\eta_1 = \text{SNR}_1(s^\star)$ and $\eta_2 = \text{SNR}_2(s^\star)$. Also, $s^\star$ satisfies the power constraints, $P_i(s^\star) \le NP$, $i=1,2$.
Therefore, $s^\star$ is a feasible solution of \eqref{opt:linear:SVEM}, i.e.,  $s^\star \in \mathcal{S}$.
From Case (ii-b), we have
$w^o > w^\star$.
This means that  $s_0$ yields a higher objective value than an existing feasible solution $s^\star$ of \eqref{opt:linear:SVEM}, and thus $s_0$ cannot be an optimal solution of \eqref{opt:linear:SVEM}, i.e., $s_0 \notin \mathcal{\bar S}$.
For either of the cases, $s_0 \notin \mathcal{\bar S}$.
\end{proof}
\section{Proof of Lemma \ref{lemma:max_inc}}
\label{app:linear:lemma2}

\begin{proof}
We first define the optimal value of ${\max}\big\{$$\mathbb{E}_{\rev{m_1,m_2,{\bf n}_1, {\bf n}_2}}\big[\|{\bf x}_1 \|^2\big],$ $\mathbb{E}_{\rev{m_1,m_2,{\bf n}_1, {\bf n}_2}}\big[\|{\bf x}_2 \|^2\big] \big\}$ in \eqref{opt:linear:max-of-powers:SNR} with the constraints, $\text{SNR}_1=\delta$ and $\text{SNR}_2 = \eta_2$, as $h(\delta, \eta_2)$.
We want to show that $h(\delta, \eta_2)$ is an increasing function of $\delta$.

We show this by contradiction. 
Suppose that $h(\delta, \eta_2)$ is not an increasing function of $\delta$. 
This implies that there exist $\delta_1$ and $\delta_2$, such that $\delta_1 < \delta_2$ and $h(\delta_1, \eta_2) \ge h(\delta_2, \eta_2)$.
Let us denote the optimal solution tuple that yields $h(\delta_2, \eta_2)$ and satisfies the constraints, $\text{SNR}_1=\delta_2$ and $\text{SNR}_2 = \eta_2$, as (${\bf g}_1$, ${\bf F}_1$, ${\bf g}_2$,  ${\bf F}_2$).
With fixed ${\bf F}_1$, ${\bf g}_2$, and ${\bf F}_2$, we can find $\tilde {\bf g}_1 = (1-\epsilon){\bf g}_1$, where $0<\epsilon<1$ is chosen to satisfy $\text{SNR}_1=\delta_1 < \delta_2$ in \eqref{eq:SNR_opt}.
The tuple ($\tilde {\bf g}_1$, ${\bf F}_1$, ${\bf g}_2$,  ${\bf F}_2$) satisfies $\text{SNR}_2=\eta_2$
since $\text{SNR}_2$ is not a function of ${\bf g}_1$.
Furthermore, this tuple ($\tilde {\bf g}_1$, ${\bf F}_1$, ${\bf g}_2$,  ${\bf F}_2$) yields smaller $\mathbb{E}_{\rev{m_1,m_2,{\bf n}_1, {\bf n}_2}}\big[\|{\bf x}_1 \|^2\big]$ in \eqref{eq:power1} and $\mathbb{E}_{\rev{m_1,m_2,{\bf n}_1, {\bf n}_2}}\big[\|{\bf x}_2 \|^2\big]$ in \eqref{eq:power2}, since $\|\tilde {\bf g}_1\|^2 < \|{\bf g}_1\|^2$ and $\| {\bf F}_2 \tilde {\bf g}_1\|^2 < \|{\bf F}_2 {\bf g}_1\|^2$.
That is, the tuple ($\tilde {\bf g}_1$, ${\bf F}_1$, ${\bf g}_2$,  ${\bf F}_2$) yields  a smaller objective value of ${\max} \big\{ \mathbb{E}_{\rev{m_1,m_2,{\bf n}_1, {\bf n}_2}}\big[\|{\bf x}_1 \|^2\big], \mathbb{E}_{\rev{m_1,m_2,{\bf n}_1, {\bf n}_2}}\big[\|{\bf x}_2 \|^2\big] \big\}$, say $\nu$, than the one obtained by (${\bf g}_1$, ${\bf F}_1$, ${\bf g}_2$,  ${\bf F}_2$), i.e., $\nu < h(\delta_2, \eta_2)$.


We observe that the tuple ($\tilde {\bf g}_1$, ${\bf F}_1$, ${\bf g}_2$, ${\bf F}_2$) is a feasible solution for the problem in \eqref{opt:linear:max-of-powers:SNR} with the constraints, $\text{SNR}_1=\delta_1$ and $\text{SNR}_2=\eta_2$, yielding an objective value of $\nu$. 
As $h(\delta_1, \eta_2)$ defines the optimal (or smallest) objective value for this problem, $h(\delta_1, \eta_2)$ must be less than or equal to $\nu$, i.e., $h(\delta_1, \eta_2) \le \nu$.
Combining with the previous result, $\nu < h(\delta_2, \eta_2)$, we then have $h(\delta_1, \eta_2) \le \nu < h(\delta_2, \eta_2)$. This contradicts the statement of $h(\delta_1, \eta_2) \ge h(\delta_2, \eta_2)$. Thus, $h(\delta, \eta_2)$ is an increasing function of $\delta$.
\end{proof}
\section{Proof of Proposition \ref{pro:f2n}}
\label{app:linear:pro:f2n}

\begin{proof}
%
%
We let (${\bf g}_1$, ${\bf F}_1$, ${\bf g}_2$, ${\bf F}_2$) be any feasible solution of \eqref{opt:linear:weighted-sum} with ${\bf F}_2 \in \mathcal{F}_2$. We want to show that 
it is optimal that $f_{2,N}=0$ in ${\bf F}_2 \in \mathcal{F}_2$ for any feasible solution (${\bf g}_1$, ${\bf F}_1$, ${\bf g}_2$, ${\bf F}_2$). That is, we will show that the objective value obtained from the solution with $f_{2,N}=0$ is always smaller than or equal to that with any $f_{2,N}$.

%
First, we let $\bar {\bf F}_2$ be equal to ${\bf F}_2$, except that the last entry of $\bar {\bf F}_2$ is zero, i.e.,  $\bar f_{2,N}=0$. 
We will show that  (i) the solution (${\bf g}_1$, ${\bf F}_1$, $\bar {\bf g}_2$, $\bar {\bf F}_2$)  is a feasible solution where $\bar {\bf g}_2 = (1-\epsilon){\bf g}_2$ with some $\epsilon \in [0,1)$, and 
(ii) the solution (${\bf g}_1$, ${\bf F}_1$, $\bar {\bf g}_2$, $\bar {\bf F}_2$) results in an objective value smaller than or equal to that obtained by (${\bf g}_1$, ${\bf F}_1$, ${\bf g}_2$, ${\bf F}_2$).
By showing the two statements above, we show that the feasible solution (${\bf g}_1$, ${\bf F}_1$, $\bar {\bf g}_2$, $\bar {\bf F}_2$) with $\bar f_{2,N}=0$ always yields a smaller objective value than that obtained by any feasible solution (${\bf g}_1$, ${\bf F}_1$, ${\bf g}_2$, ${\bf F}_2$), implying that  it is optimal that $f_{2,N}=0$ in ${\bf F}_2 \in \mathcal{F}_2$ in \eqref{opt:linear:weighted-sum}.

{\bf Proof of Statement (i):}
Since (${\bf g}_1$, ${\bf F}_1$, ${\bf g}_2$, ${\bf F}_2$) is a feasible solution, it 
satisfies the constraints for $\text{SNR}_1$ and $\text{SNR}_2$ in \eqref{opt:linear:weighted-sum}.
First, for $\text{SNR}_2$, using \eqref{eq:SNR_opt} and \eqref{eq:Q2}, we get
\begin{align}
    \text{SNR}_2 = \eta_2 &= {\bf g}_2^\top ({\bf F}_2 {\bf F}_2^\top \sigma_1^2 + \sigma_2^2{\bf I} )^{-1} {\bf g}_2 
  \nonumber   \\&\le {\bf g}_2^\top (\bar {\bf F}_2 \bar {\bf F}_2^\top \sigma_1^2 + \sigma_2^2{\bf I} )^{-1} {\bf g}_2.
  \label{lemma2:eq:SNR2}
\end{align}
In \eqref{lemma2:eq:SNR2},
we can always choose $\bar {\bf g}_2 = (1-\epsilon){\bf g}_2$ with $\epsilon \in [0,1)$ that satisfies  $\bar {\bf g}_2^\top (\bar {\bf F}_2 \bar {\bf F}_2^\top \sigma_1^2 + \sigma_2^2{\bf I} )^{-1} \bar {\bf g}_2 = \eta_2$. 
This implies that (${\bf g}_1$, ${\bf F}_1$, $\bar {\bf g}_2$, $\bar {\bf F}_2$) satisfies the constraint for ${\rm SNR}_2$.
The constraint for ${\rm SNR}_1$ is also satisfied with (${\bf g}_1$, ${\bf F}_1$, $\bar {\bf g}_2$, $\bar {\bf F}_2$)
since ${\rm SNR}_1$ relies on ${\bf Q}_1 $ in \eqref{eq:Q1} and we have ${\bf F}_1 {\bf F}_2 = {\bf F}_1 \bar {\bf F}_2$.
Therefore, (${\bf g}_1$, ${\bf F}_1$, $\bar {\bf g}_2$, $\bar {\bf F}_2$) is a feasible solution of \eqref{opt:linear:weighted-sum}.

{\bf Proof of Statement (ii):}
First, (${\bf g}_1$, ${\bf F}_1$, $\bar {\bf g}_2$, $\bar {\bf F}_2$) yields
a smaller or an equal transmit power of $\mathbb{E}_{\rev{m_1,m_2,{\bf n}_1, {\bf n}_2}}\big[\|{\bf x}_2 \|^2\big]$  
since
\begin{align}
    \mathbb{E}_{\rev{m_1,m_2,{\bf n}_1, {\bf n}_2}}\big[\|{\bf x}_2 \|^2\big] & = \| ({\bf I} + {\bf F}_2 {\bf F}_1 ) {\bf g}_2\|^2 + \|{\bf F}_2{\bf g}_1\|^2 + 
    \| {\bf F}_2( {\bf I} + {\bf F}_1{\bf F}_2 ) \|_F^2 \sigma_1^2
    + \| {\bf F}_2{\bf F}_1 \|_F^2 \sigma_2^2 
      \nonumber  \\
    & 
    \ge \| ({\bf I} + \bar {\bf F}_2 {\bf F}_1 ) \bar {\bf g}_2\|^2 + \|\bar {\bf F}_2{\bf g}_1\|^2 
    +
    \| \bar {\bf F}_2( {\bf I} + {\bf F}_1 \bar {\bf F}_2 ) \|_F^2 \sigma_1^2
    + \| \bar {\bf F}_2{\bf F}_1 \|_F^2 \sigma_2^2  .
\end{align}
Note that $\mathbb{E}_{\rev{m_1,m_2,{\bf n}_1, {\bf n}_2}}\big[\|{\bf x}_1 \|^2\big]$ in \eqref{eq:power1} are not dependent on $f_{2,N}$ since ${\bf F}_1 {\bf F}_2$ does not include $f_{2,N}$.

Therefore, when $f_{2,N}=0$,
we can always obtain a smaller or an equal objective value of $\alpha \mathbb{E}_{\rev{m_1,m_2,{\bf n}_1, {\bf n}_2}}$ $\big[\|{\bf x}_1 \|^2\big] + (1-\alpha) \mathbb{E}_{\rev{m_1,m_2,{\bf n}_1, {\bf n}_2}}\big[\|{\bf x}_2 \|^2\big]$ in \eqref{opt:linear:weighted-sum},
while satisfying the SNR constraints. It is thus optimal to set $f_{2,N}=0$ in \eqref{opt:linear:weighted-sum} with ${\bf F}_2 \in \mathcal{F}_2$.
\end{proof}
\section{Proof of Proposition~\ref{pro:B_lower_upper}}
\label{app:linear:pro:B_lower_upper}

\begin{proof}
We first prove the upper bound, $\nu_{\min}[{\bf B}]  \le (1-\alpha) \sigma_2^2$.
We rewrite ${\bf B}$ in \eqref{eq:linear:B} as 
${\bf B} = \alpha {\bf Q}_{2,\text{sqrt}}  {\bf F}_1^\top  {\bf F}_1$ 
${\bf Q}_{2,\text{sqrt}} + (1-\alpha){\bf Q}_{2} + (1-\alpha){\bf Q}_{2,\text{sqrt}} \big(
     {\bf F}_2 {\bf F}_1 + {\bf F}_1^\top {\bf F}_2^\top + {\bf F}_1^\top {\bf F}_2^\top {\bf F}_2 {\bf F}_1
    \big){\bf Q}_{2,\text{sqrt}}$.
Then, we obtain  ${\bf B} = (1-\alpha) \sigma_2^2 {\bf I} + {\bf C}$ by using ${\bf Q}_{2} = {\bf F}_2 {\bf F}_2^\top \sigma_1^2 + \sigma_2^2 {\bf I}$ from \eqref{eq:Q2} and defining ${\bf C} = 
 (1-\alpha) \sigma_1^2 {\bf F}_2 {\bf F}_2^\top 
 + {\bf Q}_{2,\text{sqrt}} \big( 
    \alpha {\bf F}_1^\top {\bf F}_1 
    + (1-\alpha) 
    ( {\bf F}_2 {\bf F}_1 + {\bf F}_1^\top {\bf F}_2^\top + {\bf F}_1^\top {\bf F}_2^\top {\bf F}_2 {\bf F}_1
    )
    \big)
 {\bf Q}_{2,\text{sqrt}}$.
By the Weyl's inequality, $\nu_{\min}[{\bf B}] \le (1-\alpha) \sigma_2^2 + \nu_{\min}[{\bf C}]$.
Showing $\nu_{\min}[{\bf B}]  \le (1-\alpha) \sigma_2^2$ is equivalent to showing $\nu_{\min}[{\bf C}] \le 0$.
We note that all the entries at the last column and row of ${\bf C}$ are zeros for any strictly lower triangular matrix ${\bf F}_1$ and any ${\bf F}_2 \in \mathcal{F}_2$.
Since the last ($N$-th) column of ${\bf C}$ is a zero vector, ${\bf C}$ has an eigenvalue of 0 with the corresponding eigenvector $[0,...,0,1]^\top$, i.e., ${\bf C} [0,...,0,1]^\top = 0 \cdot [0,...,0,1]^\top $.
Since ${\bf C}$ has an eigenvalue of 0, the smallest eigenvalue of ${\bf C}$ should be less than or equal to 0, i.e., $\nu_{\min}[{\bf C}] \le 0$. Then, $\nu_{\min}[{\bf B}]  \le (1-\alpha) \sigma_2^2$.

Next, we prove the lower bound, $\nu_{\min}[{\bf B}] \ge 0$. We first define the terms of ${\bf B}$ in \eqref{eq:linear:B} as ${\bf B}_1 = \alpha {\bf Q}_{2,\text{sqrt}}  {\bf F}_1^\top  {\bf F}_1 {\bf Q}_{2,\text{sqrt}}$ and ${\bf B}_2 = (1-\alpha) {\bf Q}_{2,\text{sqrt}}  ({\bf I} + {\bf F}_2 {\bf F}_1 )^\top ({\bf I} + {\bf F}_2 {\bf F}_1 ) {\bf Q}_{2,\text{sqrt}}$, so that ${\bf B} = {\bf B}_1 + {\bf B}_2$.
Note that ${\bf B}_1 $ and ${\bf B}_2$, are positive semidefinite (PSD), respectively. Therefore, their sum, ${\bf B}$, is also PSD. This implies that $\nu_{\min}[{\bf B}] \ge 0$.
\end{proof}
\section{Proof of Conjecture \ref{conj:B} When $N=3$}
\label{app:linear:conj:B}

\begin{proofN3}
We will show $\min \{\alpha \sigma_1^2, (1-\alpha) \sigma_2^2 \} \le \nu_{\min}[{\bf B}]$.
By understanding that $\min \{\alpha \sigma_1^2, (1-\alpha) \sigma_2^2 \} \le (1-\alpha) \sigma_2^2$, in the special case with $N=3$,
we will instead show that $\nu_{\min}[{\bf B}]  = (1-\alpha) \sigma_2^2$ for any strictly lower triangular ${\bf F}_1$ and any ${\bf F}_2 \in \mathcal{F}_2$.
 We first rewrite ${\bf B}  = (1-\alpha) \sigma_2^2 {\bf I} + {\bf C}$ where ${\bf C} = 
 (1-\alpha) \sigma_1^2 {\bf F}_2 {\bf F}_2^\top 
 + {\bf Q}_{2,\text{sqrt}} \big( 
    \alpha {\bf F}_1^\top {\bf F}_1 
    + (1-\alpha) 
    ( {\bf F}_2 {\bf F}_1 + {\bf F}_1^\top {\bf F}_2^\top + {\bf F}_1^\top {\bf F}_2^\top {\bf F}_2 {\bf F}_1
    )
    \big)
 {\bf Q}_{2,\text{sqrt}}$. 
 Then,
 showing $\nu_{\min}[{\bf B}] = (1-\alpha) \sigma_2^2$ is equivalent to showing $\nu_{\min}[{\bf C}] = 0$.
 First, since ${\bf F}_2 {\bf F}_1={\bf 0}$ due to $f_{2,3}=0$ from Proposition \ref{pro:f2n}, we have 
 \begin{equation}
    {\bf C} = (1-\alpha) \sigma_1^2 {\bf F}_2 {\bf F}_2^\top 
    + \alpha {\bf Q}_{2,\text{sqrt}}  
     {\bf F}_1^\top {\bf F}_1 {\bf Q}_{2,\text{sqrt}}.
     \label{eq:C:appendixD}
 \end{equation}
In \eqref{eq:C:appendixD}, ${\bf F}_2 {\bf F}_2^\top$ and ${\bf Q}_{2,\text{sqrt}} {\bf F}_1^\top {\bf F}_1 {\bf Q}_{2,\text{sqrt}}$ are positive semidefinite (PSD), respectively, since each of them is a form of a matrix multiplied with its own transpose. Since $(1-\alpha)\sigma_1^2>0$ and $\alpha>0$, $(1-\alpha) \sigma_1^2 {\bf F}_2 {\bf F}_2^\top$ and $\alpha {\bf Q}_{2,\text{sqrt}}  {\bf F}_1^\top {\bf F}_1 {\bf Q}_{2,\text{sqrt}}$ in \eqref{eq:C:appendixD} are PSD, respectively. Thus, the summation of the two PSD matrices is also PSD, i.e., ${\bf C}$ is PSD.
This implies that $\nu_{\min}[{\bf C}] \ge 0$.
The remaining part to claim $\nu_{\min}[{\bf C}] = 0$ is to show that ${\bf C}$ has an eigenvalue of 0.
Since the last ($N$-th) column of ${\bf C}$ is a zero vector, ${\bf C}$ has an eigenvalue of 0 with the corresponding eigenvector $[0,0,1]^\top$, i.e., ${\bf C} [0,0,1]^\top = 0 \cdot [0,0,1]^\top $.
We then have $\nu_{\min}[{\bf C}] = 0$, which leads to $\nu_{\min}[{\bf B}] = (1-\alpha)\sigma_2^2$.
%
\end{proofN3}
\section{Proof of Proposition \ref{pro:g:optimal}}
\label{app:linear:pro:g:optimal}

\begin{proof}
We have a lower bound of the objective function in \eqref{opt:q2} as
${\bf q}_2^\top {\bf B} {\bf q}_2 
    \ge \nu_{\min}[{\bf B}] \| {\bf q}_2 \|^2.$
From Conjecture~\ref{conj:B}, we have $\nu_{\min}[{\bf B}]
= (1-\alpha) \sigma_2^2$
when $\alpha \ge  \frac{\sigma_2^2}{\sigma_1^2 + \sigma_2^2}$.
Then, we have the lower bound as ${\bf q}_2^\top {\bf B} {\bf q}_2 
    \ge (1-\alpha) \sigma_2^2 \| {\bf q}_2 \|^2.$
Here, ${\bf q}_2^\star  = [0, ...,0, \sqrt{\eta_2}]^\top$ satisfies the lower bound 
with $\|{\bf q}^\star_2\|^2 = \eta_2$, 
which can be easily shown by the fact that
all the entries in the last column and row of ${\bf B}$ are  zeros except the last diagonal entry is $(1-\alpha) \sigma_2^2$ due to $f_{2,N}=0$ from Proposition~\ref{pro:f2n}.
In other words, ${\bf q}_2^\star$ is an optimal solution of 
\eqref{opt:q2}.
We then have the optimal solution for \eqref{opt:g2} as
${\bf g}_2^\star = {\bf Q}_{2,\text{sqrt}} {\bf q}_2^\star = [0, ...,0,  \sqrt{\eta_2} \sigma_2]^\top $,
since 
$f_{2,N}=0$ from Proposition 1.
For the specific case of $N=3$, we have $\nu_{\min}[{\bf B}]
= (1-\alpha) \sigma_2^2$ for any 
$\alpha$ (refer to Appendix~\ref{app:linear:conj:B} for the proof). Therefore, with the same analysis as above, ${\bf q}_2^\star  = [0, 0, \sqrt{\eta_2}]^\top$ (or equivalently ${\bf g}_2^\star = {\bf Q}_{2,\text{sqrt}} {\bf q}_2^\star = [0,0,  \sqrt{\eta_2} \sigma_2]^\top $) is optimal in \eqref{opt:q2}.
\end{proof}
\section{Proof of Corollary~\ref{cor1}}
\label{app:linear:cor1}

\begin{proof}
    We first note that, if we choose ${\bf g}_2^\star = [0,...,0, \sqrt{\eta}\sigma_2]^\top$ (or equivalently, ${\bf q}_2^\star = {\bf Q}_{2,\text{sqrt}}^{-1} {\bf g}_2^\star = [0,...,0,\sqrt{\eta_2}]^\top$), 
    we have ${{\bf q}_2^\star}^\top {\bf B} {\bf q}_2^\star = (1-\alpha)\sigma_2^2 \eta_2$ in \eqref{opt:q2} since all the entries in the last column and row of ${\bf B}$ are zeros except the last diagonal entry is $(1-\alpha)\sigma_2^2$.
    This choice of ${\bf q}_2$ may not be optimal when $\alpha < \frac{\sigma_2^2}{\sigma_1^2 + \sigma_2^2}$.
    When $\alpha < \frac{\sigma_2^2}{\sigma_1^2 + \sigma_2^2}$,
    the lower bound in Conjecture~\ref{conj:B} is $\alpha \sigma_1^2$. That is, we have $\alpha \sigma_1^2 \le \nu_{\min}[{\bf B}]$ for any ${\bf F}_1$ and ${\bf F}_2 \in \mathcal{F}_2$.
    This means that the lower bound of the objective value ${\bf  q}_2^\top {\bf B} {\bf q}_2$ in \eqref{opt:q2} (while satisfying the constraint $\|{\bf q}_2\|^2 = \eta_2$) is $\alpha \sigma_1^2 \eta_2$. Here, we can set the direction of ${\bf q}_2$ to align with the eigenvector corresponding to the smallest eigenvalue of ${\bf B}$.
    Therefore, in the worst case, i.e., when 
    $\nu_{\min}[{\bf B}] = \alpha \sigma_1^2$ (with specific ${\bf F}_1$ and  ${\bf F}_2$), the optimality loss in \eqref{opt:q2} will be $(1-\alpha)\sigma_2^2 \eta_2 - \alpha \sigma_1^2 \eta_2 = \eta_2 (\sigma_1^2 + \sigma_2^2) ( \frac{\sigma_2^2}{\sigma_1^2+\sigma_2^2} - \alpha )$.
\end{proof}
\section{Proof of Lemma \ref{lemma:RNN:power}}
\label{app:RNN:lemma3}

\begin{proof}
    We extend the proof of the one-way coding architecture in \cite{kim2023feedback} to the two-way coding architecture.
    Define the training data tuples as $\{\mathcal{T}_j\}_{j=1}^J$, where $\mathcal{T}_j = \{{\bf b}_1^{(j)}, {\bf b}_2^{(j)}, {\bf n}_1^{(j)}, {\bf n}_2^{(j)} \}$ is the $j$-th tuple of the training data.
    Let us denote $\tilde \eta_i^{(j)}[k]$ as the output at timestep $k$ generated by data $j$, $\mathcal{T}_j$, through the User $i$'s encoding process of \eqref{eq:s1}-\eqref{eq:enc:non-linear}, $k=1,...,N$.
    It is obvious that $\tilde \eta_i^{(j)}[k]$ is independent and identically distributed (IID) over $j$ assuming the data tuples $\{\mathcal{T}_j\}_{j=1}^J$ are IID from each other. We define the mean and variance of $\tilde \eta_i^{(j)}[k]$ as $\mu_{i,k}$ and $\delta^2_{i,k}$,  respectively.
    With the sample mean 
    $m_{i,k}(J) = \frac{1}{J}\sum_{j=1}^{J} \tilde \eta_i^{(j)}[k]$ and the sample variance 
    $d^2_{i,k}(J) = \frac{1}{J} \sum_{j=1}^{J} (\tilde \eta_i^{(j)}[k] - m_{i,k}(J))^2$, we define the normalization function 
    as 
    $\gamma_{i,k}^{(J)}(x) = (x-m_{i,k}(J))/d_{i,k}(J)$.
    
    Let us define $\tilde x_i[k]$ as the output at timestep $k$ at User $i$, generated by the data tuple for inference, $ \{{\bf b}_1, {\bf b}_2, {\bf n}_1,  {\bf n}_2 \}$.
    Assuming the training and inference data tuples are extracted from the same distribution, the mean and variance of $\tilde x_i[k]$ are then $\mu_{i,k}$ and $\delta^2_{i,k}$, respectively. 
    We then have
    $\mathbb{E}_{{\bf b}_1, {\bf b}_2, {\bf n}_1,  {\bf n}_2} \big[ (  \gamma^{(J)}_{i,k}(\tilde {x}_i[k]) )^2   \big] = \frac{ \delta_{i,k}^2 + (m_{i,k}(J)-\mu_{i,k})^2}{d^2_{i,k}(J)}$. 
    By the strong law of large number (SLLN)~\cite{billingsley2012probability}, $m_{i,k}(J) \rightarrow \mu_{i,k}$ and $d^2_{i,k}(J) \rightarrow \delta^2_{i,k}$ almost surely (a.s.) as $J \rightarrow \infty$. 
    Then, by continuous mapping theorem~\cite{billingsley2012probability}, $\mathbb{E}_{{\bf b}_1, {\bf b}_2, {\bf n}_1,  {\bf n}_2} \big[ (  \gamma^{(J)}_{i,k}(\tilde {x}_i[k]) )^2   \big] \rightarrow 1$ a.s. as $J \rightarrow \infty$.
    Then, $\mathbb{E}_{{\bf b}_1, {\bf b}_2, {\bf n}_1,  {\bf n}_2} \big[ \sum_{k=1}^N   x_i^2[k] \big] 
    = \sum_{k=1}^N   w^2_{i}[k]  \mathbb{E}_{{\bf b}_1, {\bf b}_2, {\bf n}_1,  {\bf n}_2} \big[ ( \gamma_{i,k}^{(J)} (\tilde {x}_i[k]) )^2 \big]  \rightarrow N P$ a.s. as $J \rightarrow \infty$.
\end{proof}
\section{Derivation for $\Phi_i({\bf f}_{1,i})$ in \eqref{eq:power1_hf}}
\label{app:linear:phi}

We express each of the terms of $\mathbb{E}_{\rev{m_1,m_2,{\bf n}_1, {\bf n}_2}}[\|{\bf x}_1\|^2] $ in \eqref{eq:power1:q1} as follows:
\begin{equation}
    \|{\bf q}^\top_1 ({\bf I} + {\bf F}_1{\bf F}_2) \|^2 \sigma_1^2 
    = \sigma_1^2 \bigg( 
    \sum_{i=2}^{N-1} (q_{1,i-1} + f_{2,i} {\bf h}_i^\top {\bf f}_{1,i}  )^2 
    + q_{1,N-1}^2 + q_{1,N}^2,
    \bigg)
    \label{eq:append:Ex1:1}
\end{equation}
where ${\bf q}^\top_1 ({\bf I} + {\bf F}_1{\bf F}_2) = [q_{1,1} + f_{2,2} {\bf h}_2^\top {\bf f}_{1,2},  \;
q_{1,2} + f_{2,3} {\bf h}_3^\top {\bf f}_{1,3}, \;
... , \;  q_{1,N-2} + f_{2,N-1} {\bf h}_{N-1}^\top {\bf f}_{1,N-1}, $ $q_{1,N-1}, \; q_{1,N}]$,
\begin{align}
    \| {\bf q}^\top_1 {\bf F}_1 \|^2 \sigma_2^2
    &= \sigma_2^2 \sum_{i=1}^{N-1} ({\bf h}_{i}^\top {\bf f}_{1,i})^2,
    \\
    \|{\bf F}_1 {\bf F}_2 \|_F^2 \sigma_1^2
    &= \sigma_1^2 \text{tr}( ({\bf F}_1 {\bf F}_2)^\top {\bf F}_1 {\bf F}_2 ) 
    = \sigma_1^2 \sum_{i=2}^{N-1} f_{2,i}^2 {\bf f}_{1,i}^\top {\bf f}_{1,i},
    \\
    \| {\bf F}_1 \|_F^2 \sigma_2^2
    &= \sigma_2^2 \text{tr}({\bf F}_1^\top {\bf F}_1) 
    = \sigma_2^2 \sum_{i=1}^{N-1} {\bf f}_{1,i}^\top {\bf f}_{1,i}.
    \label{eq:append:Ex1:4}
\end{align}
%
%
With \eqref{eq:append:Ex1:1}-\eqref{eq:append:Ex1:4}, we represent $\mathbb{E}_{\rev{m_1,m_2,{\bf n}_1, {\bf n}_2}}[\|{\bf x}_1\|^2]$ as 
\begin{align}
    \mathbb{E}_{\rev{m_1,m_2,{\bf n}_1, {\bf n}_2}}[\|{\bf x}_1\|^2] & =  
    \sigma_1^2 \bigg( 
    \sum_{i=2}^{N-1}  (q_{1,i-1} + f_{2,i} {\bf h}_i^\top {\bf f}_{1,i}  )^2 
    + q_{1,N-1}^2 + q_{1,N}^2
    \bigg) 
    \nonumber
    \\
    & \hspace{1cm} 
    +\sigma_2^2 \sum_{i=1}^{N-1} ({\bf h}_{i}^\top {\bf f}_{1,i})^2
    + \sigma_1^2 \sum_{i=2}^{N-1} f_{2,i}^2 {\bf f}_{1,i}^\top {\bf f}_{1,i}
    + \sigma_2^2 \sum_{i=1}^{N-1} {\bf f}_{1,i}^\top {\bf f}_{1,i}
    \\
    & = ({\bf h}_1^\top {\bf f}_{1,1} )^2 \sigma_2^2  +  {\bf f}_{1,1}^\top {\bf f}_{1,1} \sigma_2^2
    +
    \sum_{i=2}^{N-1} \bigg( \big( q_{1,i-1} + f_{2,i} {\bf h}_{i}^\top {\bf f}_{1,i} \big)^2 \sigma_1^2 
    + ( {\bf h}_{i}^\top {\bf f}_{1,i} )^2 \sigma_2^2  
    \nonumber
    \\
    & \hspace{1cm} +{\bf f}_{1,i}^\top {\bf f}_{1,i}  (f_{2,i}^2 \sigma_1^2 + \sigma_2^2) \bigg)
    + \sigma_1^2 (q_{1,N-1}^2 + q_{1,N}^2).
    \label{eq:append:phi}
\end{align}
We note that equation \eqref{eq:append:phi} is equivalent to equation \eqref{eq:power1_hf}.
\section{Derivation of \eqref{eq:f1i_opt} from  \eqref{eq:f1i_deriv}}
\label{app:linear:f1i}

To satisfy the equality in \eqref{eq:f1i_deriv}, we must have
\begin{align}
    (f_{2,i}^2 \sigma_1^2 + \sigma_2^2)({\bf h}_{i} {\bf h}_{i}^\top + {\bf I}){\bf f}_{1,i} = - q_{1,i-1} f_{2,i} \sigma_1^2 {\bf h}_{i}.
\end{align}
Finally, the optimal solution form of ${\bf f}_{1,i}$, $i\in\{2,...,N-1\}$, is given in terms of the entries of ${\bf q}_1$ (encapsulated in ${\bf h}_{i}$ according to \eqref{eq:h_i}) as follows:
\begin{align}
    {\bf f}_{1,i} &= - \frac{q_{1,i-1} f_{2,i} \sigma_1^2}{f_{2,i}^2\sigma_1^2 + \sigma_2^2}
    \big( {\bf h}_{i} {\bf h}_{i}^\top + {\bf I} \big)^{-1} {\bf h}_{i}
   \nonumber  \\
    &\overset{(i)}{=} - \frac{q_{1,i-1} f_{2,i} \sigma_1^2}{f_{2,i}^2 \sigma_1^2 + \sigma_2^2}
    \bigg({\bf I} - \frac{{\bf h}_{i}{\bf h}_{i}^\top}
    {1+ \| {\bf h}_{i}\|^2 } \bigg) {\bf h}_{i}
    \nonumber
\\
    &= - \frac{ f_{2,i}\sigma_1^2}{f_{2,i}^2 \sigma_1^2 + \sigma_2^2}
    \frac{q_{1,i-1}}
    {1+ \| {\bf h}_{i}\|^2 }  {\bf h}_{i},
    \label{eq:app:f1i_opt}
\end{align}
where the Sherman–Morrison formula is used to obtain equality (i) in \eqref{eq:app:f1i_opt}.
\section{Derivation for $\alpha \frac{\partial \mathbb{E}_{\rev{m_1,m_2,{\bf n}_1, {\bf n}_2}}\big[\|{\bf x}_1 \|^2\big]}{\partial f_{2,i}}  + (1-\alpha) \frac{\partial \mathbb{E}_{\rev{m_1,m_2,{\bf n}_1, {\bf n}_2}}\big[\|{\bf x}_2 \|^2\big]}{\partial f_{2,i}} $ in \eqref{eq:2ndsub:deriv}}
\label{app:linear:deriv_f2}

From the equation of $\mathbb{E}_{\rev{m_1,m_2,{\bf n}_1, {\bf n}_2}}\big[\|{\bf x}_1 \|^2\big]$ in \eqref{eq:power1_hf}, we can readily obtain 
\begin{equation}
    \frac{\partial \mathbb{E}_{\rev{m_1,m_2,{\bf n}_1, {\bf n}_2}}\big[\|{\bf x}_1 \|^2\big]}{\partial f_{2,i}} 
    = 2 q_{1,i-1} {\bf h}_{i}^\top {\bf f}_{1,i}\sigma_1^2 
    + 2 \sigma_1^2 \big( 
    \vert{\bf h}_{i}^\top {\bf f}_{1,i}\vert^2 + \| {\bf f}_{1,i} \|^2
    \big) f_{2,i}, \quad i \in \{2,...,N-1\}.
    \label{eq:append:deriv:eq1}
\end{equation}

For $\mathbb{E}_{\rev{m_1,m_2,{\bf n}_1, {\bf n}_2}}\big[\|{\bf x}_2 \|^2\big]$, 
we first 
express each of the terms including ${\bf F}_2$ in $\mathbb{E}_{\rev{m_1,m_2,{\bf n}_1, {\bf n}_2}}[\|{\bf x}_2\|^2]$ of \eqref{eq:power2:q1} as a sum of entries of ${\bf F}_2$, i.e., $\{f_{2,i}\}_{i =2 }^{N-1}$. First, revisiting the second term in \eqref{eq:power2:q1}, we obtain
\begin{align}
     \|{\bf F}_2 {\bf Q}_{1,\text{sqrt}} {\bf q}_1\|^2 & = {\bf q}_1^\top {\bf Q}_{1,\text{sqrt}} {\bf F}_2^\top {\bf F}_2 {\bf Q}_{1,\text{sqrt}} {\bf q}_1
     = {\bf p}^\top {\bf F}_2^\top {\bf F}_2 {\bf p}
      = \sum_{i=2}^N p_{i-1}^2 f_{2,i}^2,
     \label{eq:derv_f2_term1}
\end{align}
where we assumed that ${\bf p} \triangleq {\bf Q}_{1,\text{sqrt}} {\bf q}_1 = [p_1, ..., p_N]^\top$ is fixed for tractability although ${\bf Q}_{1,\text{sqrt}}$ depends on ${\bf F}_2$.
We then express the third term in \eqref{eq:power2:q1} as
\begin{align}
     & \|{\bf F}_2({\bf I} + {\bf F}_1{\bf F}_2 ) \|_F^2 \sigma_1^2  
     = \sigma_1^2 \sum_{i=2}^N f_{2,i}^2 + \sigma_1^2 \sum_{i=2}^{N-2} \sum_{j=i+1}^{N-1} f_{1,j,i}^2 f_{2,i}^2 f_{2,j+1}^2.
     \label{eq:derv_f2_term2}
\end{align}
Also, the last term in \eqref{eq:power2:q1} can be expressed as
\begin{align}
    \| {\bf F}_2 {\bf F}_1 \|_F^2 \sigma_2^2 
    & = \sigma_2^2 \sum_{i=3}^{N}  f_{2,i}^2 \sum_{j=1}^{i-2} f_{1,i-1,j}^2.
    \label{eq:derv_f2_term3}
\end{align}


We take the derivative to the terms \eqref{eq:derv_f2_term1}-\eqref{eq:derv_f2_term3} with respect to $f_{2,i}$. Accordingly, we obtain
\begin{equation}
    \frac{\partial \mathbb{E}_{\rev{m_1,m_2,{\bf n}_1, {\bf n}_2}}\big[\|{\bf x}_2 \|^2\big]}{\partial f_{2,i}}
    = 2 p_{i-1}^2 f_{2,i} + 2 \sigma_1^2 f_{2,i}
    + 2 \sigma_1^2 \bigg( \sum_{j=i+1}^{N-1} f_{1,j,i}^2 f_{2,j+1}^2 + \sum_{k=2}^{i-2} f_{1,i-1,k}^2 f_{2,k}^2
    \bigg)
    + 2 \sigma_2^2 \sum_{j=1}^{i-2} f_{1,i-1,j}^2,
    \label{eq:append:deriv:eq2}
\end{equation}
for $i \in \{2,...,N-1\}$.
Using \eqref{eq:append:deriv:eq1} and \eqref{eq:append:deriv:eq2}, we obtain 
$\alpha \frac{\partial \mathbb{E}_{\rev{m_1,m_2,{\bf n}_1, {\bf n}_2}}\big[\|{\bf x}_1 \|^2\big]}{\partial f_{2,i}}  + (1-\alpha) \frac{\partial \mathbb{E}_{\rev{m_1,m_2,{\bf n}_1, {\bf n}_2}}\big[\|{\bf x}_2 \|^2\big]}{\partial f_{2,i}}$ in \eqref{eq:2ndsub:deriv}.
\section{Scatter plots of transmit symbols for linear and RNN-based coding}
\label{app:scatter:plot}

\begin{figure}[t]
  \centering
\begin{subfigure}{.8\linewidth}
  \centering
  \includegraphics[width=\linewidth]{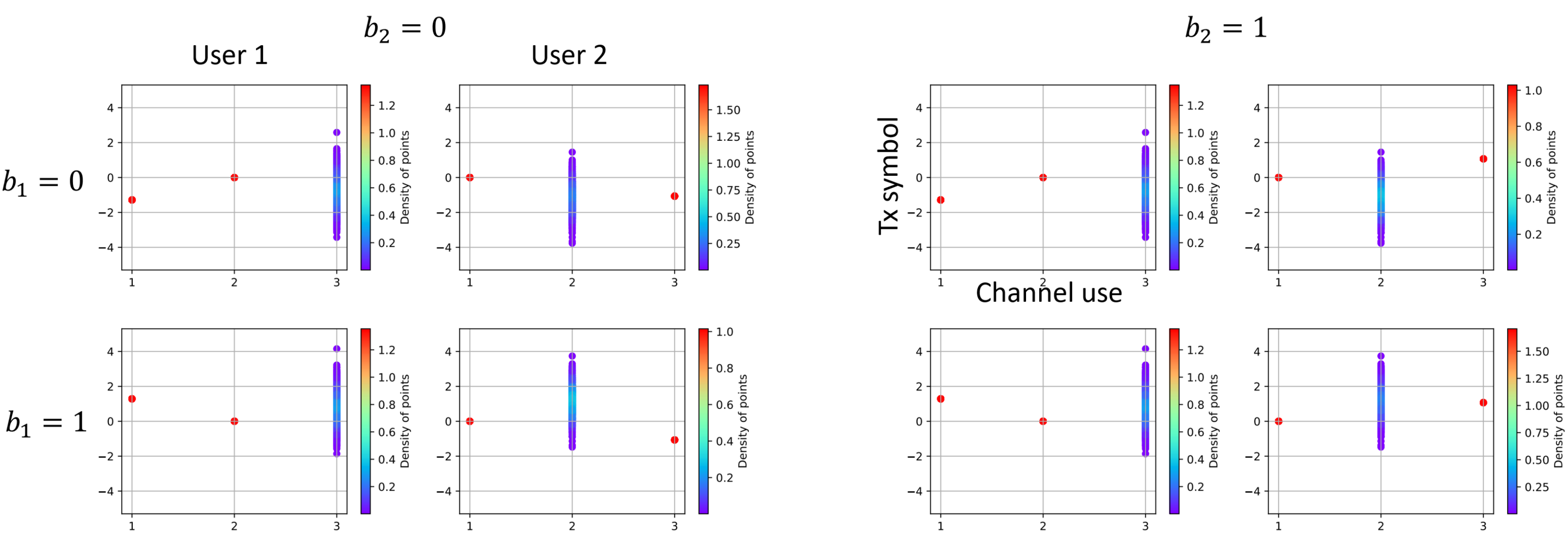}
  \caption{ Linear coding.
  }
  \label{fig:linear_K1N3}
\end{subfigure}
\begin{subfigure}{.8\linewidth}
  \centering
  \includegraphics[width=\linewidth]{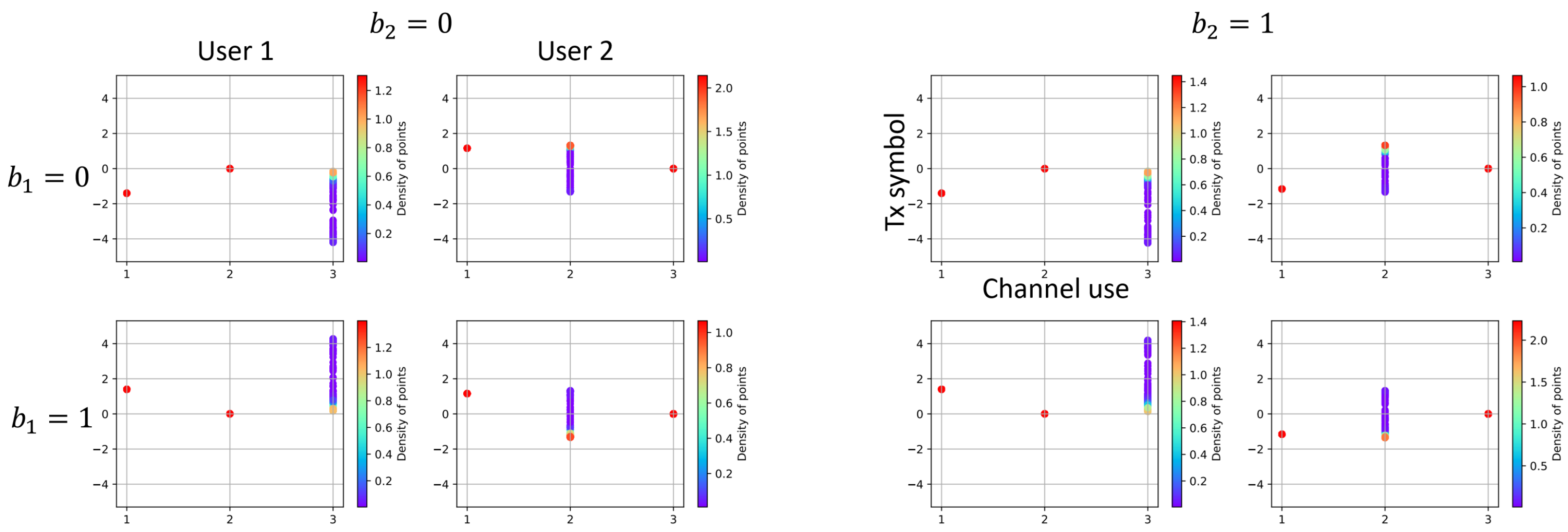}
  \caption{RNN-based coding.
  }
  \label{fig:RNN_K1N3}
\end{subfigure}
  \caption{Scatter plots of transmit symbols across channel uses, when $K=1$, $N=3$, $\text{SNR}^{\text{ch}}_1 =1$dB, and $\text{SNR}^{\text{ch}}_2=10$dB. The non-scattered point (the red point) indicates  the transmit symbol used only for message transmission without containing any feedback information. It is interesting to see that some channel uses are not utilized in RNN-based coding, similar to linear coding. This may suggest that concentrating power on specific channel uses would be more effective for the scenario of $K=1$ and $N=3$.}
  \label{fig:scatter:K1N3}
\end{figure}

\begin{figure}[t]
  \centering
\begin{subfigure}{.99\linewidth}
  \centering
  \includegraphics[width=\linewidth]{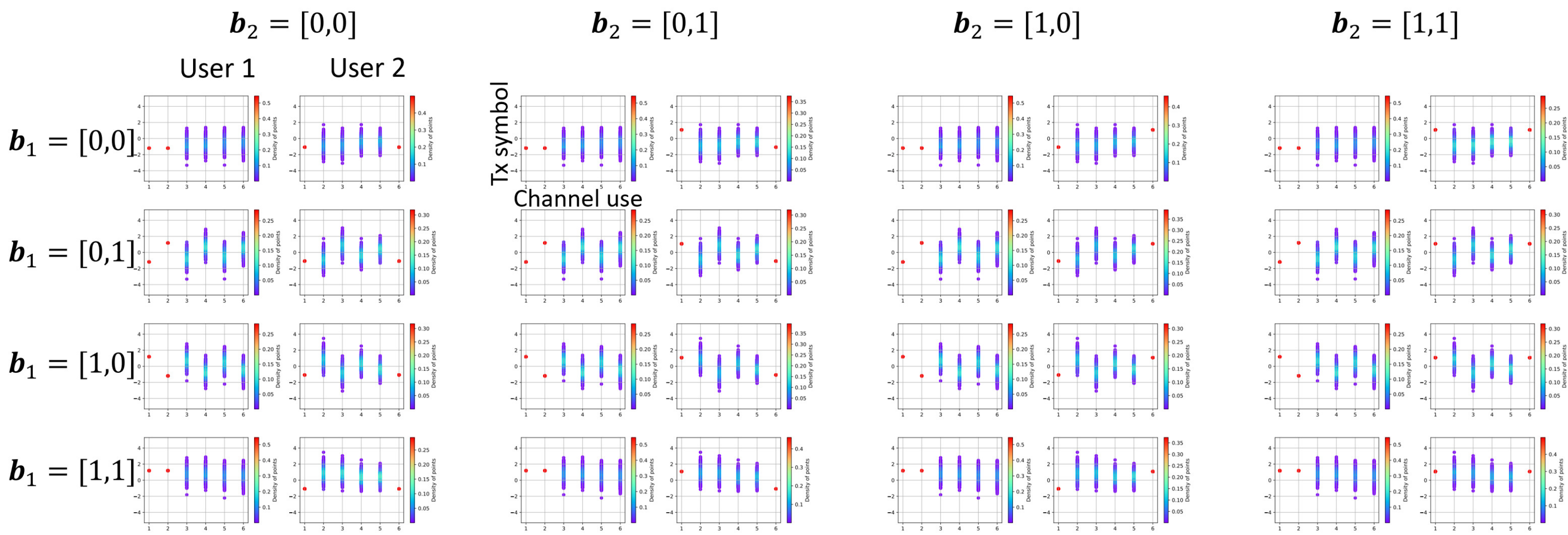}
  \caption{ Linear coding.
  }
  \label{fig:linear_K2N6}
\end{subfigure}
\begin{subfigure}{.99\linewidth}
  \centering
  \includegraphics[width=\linewidth]{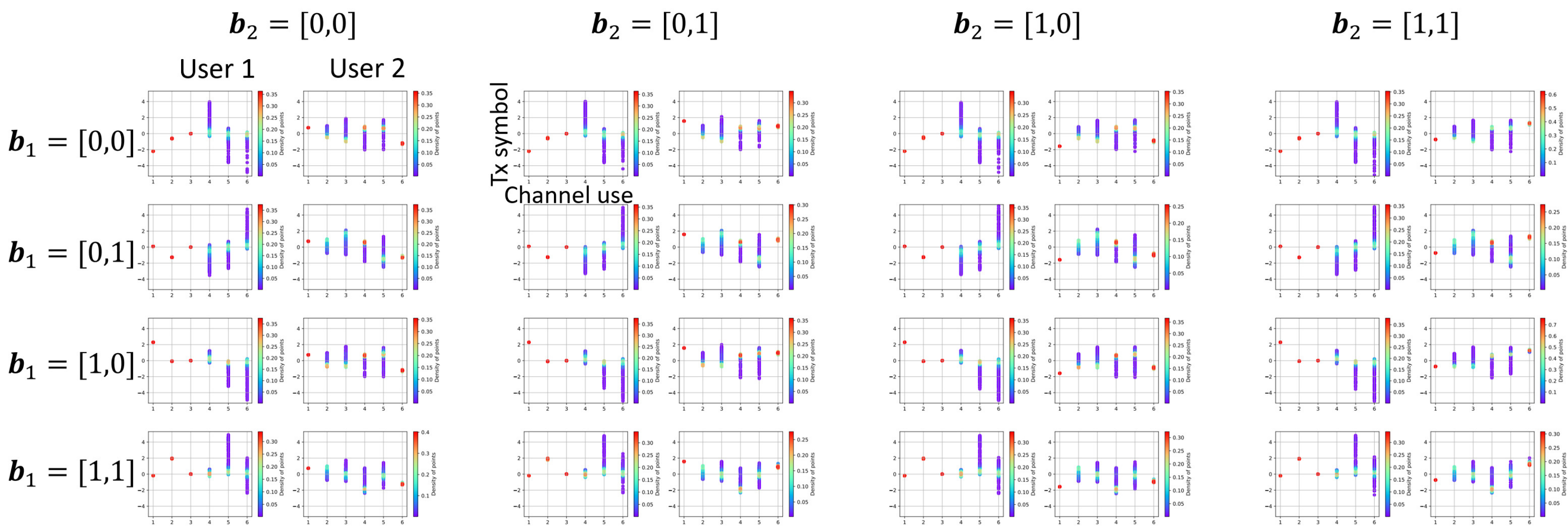}
  \caption{RNN-based coding.
  }
  \label{fig:RNN_K2N6}
\end{subfigure}
  \caption{Scatter plots of transmit symbols across channel uses, when $K=2$, $N=6$, $\text{SNR}^{\text{ch}}_1 =1$dB, and $\text{SNR}^{\text{ch}}_2=10$dB. For RNN-based coding in (b), the transmit symbols on the first three channel uses at User 1 are deterministic, and thus they contain only the transmit messages of ${\bf b}_1$ without any feedback information. User 2 uses the first and last channel uses to transmit its message of ${\bf b}_2$, which is a similar pattern to linear coding in (a). This pattern of the transmit symbols on the first and last channel uses at User 2 may be interpreted to be modulating two bits in ${\bf b}_2$ into transmit symbols.}
  \label{fig:scatter:K2N6}
\end{figure}

We compare the behavior of linear coding and RNN-based coding with scatter plots.
We analyzed both coding schemes under different parameters: (i) $K=1$ and $N=3$ for linear coding in Fig.~\ref{fig:scatter:K1N3}(\subref{fig:linear_K1N3}) and for RNN-based coding in Fig.~\ref{fig:scatter:K1N3}(\subref{fig:RNN_K1N3}), and (ii) $K=2$ and $N=6$ for linear coding in Fig.~\ref{fig:scatter:K2N6}(\subref{fig:linear_K2N6}) and for RNN-based coding in Fig.~\ref{fig:scatter:K2N6}(\subref{fig:RNN_K2N6}). 
For RNN-based coding, we used the softmax function for decoding.
We consider $\text{SNR}^{\text{ch}}_1 =1$dB and $\text{SNR}^{\text{ch}}_2=10$dB.
Each plot represents the scatter points of transmit symbols, ${\bf x}_1$ for User 1 and ${\bf x}_2$ for User 2, across the $N$ channel uses. For each bit-pair representation, we consider 1000 samples where each sample represents the randomness of noises ${\bf n}_1 \sim \mathcal{CN}({\bf 0}, 10^{-\text{SNR}^{\text{ch}}_1/10} {\bf I})$ and ${\bf n}_2 \sim \mathcal{CN}({\bf 0}, 10^{-\text{SNR}^{\text{ch}}_2/10} {\bf I} )$, where we set the average power per channel use as $P=1$.

We first explain the basic understanding of the scatter plots.
In each plot, the non-scattered point (the red point) indicates no randomness at that channel use, meaning that the transmit symbol at that channel use does not contain any feedback information. Note that feedback information always contain the noise values, so it must be scattered in the scatter plot due to its randomness. Therefore, the non-scattered point represents message transmission only.

We first look at the case when  $K=1$ and $N=3$ in Fig.~\ref{fig:scatter:K1N3}. 
%
We consider that each bit $b_i \in \{0,1\}$ at User $i$ is modulated to $m_i=2b_i-1 \in \{-1,1\}$ through pulse amplitude modulation (PAM).
In Fig.~\ref{fig:scatter:K1N3}(\subref{fig:linear_K1N3}), we observe that User 1 uses the first and third channel uses, while User 2 uses the second and third channel uses. By Proposition 2, User 2 only transmits its message during the last channel use. User 1 transmits its message on the first channel use, receives feedback from User 2 on the second channel use, and then transmits a linear mixture of the message and feedback to User 2 on the third channel use.
In Fig.~\ref{fig:scatter:K1N3}(\subref{fig:RNN_K1N3}), for RNN-based coding, we observe that User 1 uses the first channel use to send its message and changes the sign of the transmit symbol based on the value of  $b_1$: a negative value when $b_1=0$ and a positive value when $b_1=1$.
Also, User 1 does not use the second channel use. 
These behaviors at User 1 are similar to those in linear coding.
However, we observe that User 2's behavior differs between linear and RNN-based coding.
In RNN-based coding, User 2 uses the first channel use for its message transmission, while in linear coding, it uses the third channel use.
Overall, it is interesting that some channel uses are not utilized in RNN-based coding, similar to linear coding. This may suggest that concentrating power on specific channel uses would be more effective for the scenario of $K=1$ and $N=3$

Next, we look at the case when $K=2$ and $N=6$ Fig.~\ref{fig:scatter:K2N6}.
For linear coding, the alternate channel use strategy is employed. The first message at User $i$ is given by $m_i= 2b_i[1]-1$, while the second message is $m_i' = 2b_i[2]-1$. For exchanging the first message pair $(m_1, m_2)$, User 1 uses odd-numbered channel uses, and User 2 uses even-numbered channel uses. For exchanging the second message pair $(m_1', m_2')$, User 1 uses even-numbered channel uses, and User 2 uses odd-numbered channel uses. 
In RNN-based coding in  Fig.~\ref{fig:scatter:K2N6}(\subref{fig:RNN_K2N6}), it is interesting to observe that the transmit symbols on the first three channel uses at User 1 are deterministic and do not exhibit any randomness. This indicates that these channel uses are only used for User 1 to transmit messages to User 2. It is also notable that User 2 uses the first and last channel uses to transmit its message, which is a similar pattern to linear coding. The signs of the transmit symbols on the first and last channel uses at User 2 are (+,-), (+,+), (-,-), (-,+) for each case of ${\bf b}_2 =[0,0]$, $[0,1]$, $[1,0]$, $[1,1]$, respectively. 
Since RNN-based coding includes a modulation operation in its feedback coding, this pattern may be interpreted to be modulating two bits of ${\bf b}_2$ into transmit symbols.

\end{document}